\documentclass{IEEEtran}
\usepackage[T1]{fontenc}
\usepackage{color}
\usepackage{float}
\usepackage{calc}
\usepackage{bm}
\usepackage{amsmath}
\usepackage{amsthm}
\usepackage{amssymb}
\usepackage{graphicx}
\PassOptionsToPackage{normalem}{ulem}
\usepackage{ulem}
\usepackage[unicode=true,
 bookmarks=true,bookmarksnumbered=true,bookmarksopen=true,bookmarksopenlevel=1,
 breaklinks=false,pdfborder={0 0 0},pdfborderstyle={},backref=false,colorlinks=false]
 {hyperref}
\hypersetup{pdftitle={Your Title},
 pdfauthor={Your Name},
 pdfpagelayout=OneColumn, pdfnewwindow=true, pdfstartview=XYZ, plainpages=false}

\makeatletter

\floatstyle{ruled}
\newfloat{algorithm}{tbp}{loa}
\providecommand{\algorithmname}{Algorithm}
\floatname{algorithm}{\protect\algorithmname}

\let\oldforeign@language\foreign@language
\DeclareRobustCommand{\foreign@language}[1]{%
  \lowercase{\oldforeign@language{#1}}}
\theoremstyle{plain}
\newtheorem{thm}{\protect\theoremname}
\theoremstyle{plain}
\newtheorem{prop}[thm]{\protect\propositionname}
\theoremstyle{plain}
\newtheorem{lem}[thm]{\protect\lemmaname}
\theoremstyle{remark}
\newtheorem{rem}[thm]{\protect\remarkname}

\usepackage[caption=false,font=footnotesize]{subfig}
\usepackage{caption}
\usepackage{algorithm}
\usepackage{algpseudocode}
\usepackage{cite}
\usepackage{multirow}
\usepackage[table]{xcolor}
\usepackage{url}
\usepackage{stfloats}
\makeatother

\providecommand{\lemmaname}{Lemma}
\providecommand{\propositionname}{Proposition}
\providecommand{\remarkname}{Remark}
\providecommand{\theoremname}{Theorem}

\begin{document}
\title{Rate Maximizations for Reconfigurable Intelligent Surface-Aided Wireless
Networks: A Unified Framework via Block Minorization-Maximization }
\author{Zepeng~Zhang,~\IEEEmembership{Student Member,~IEEE}, and Ziping~Zhao,~\IEEEmembership{Member,~IEEE}\thanks{This work was supported in part by the National Nature Science Foundation
of China (NSFC) under Grant 62001295 and in part by the Shanghai Sailing
Program under Grant 20YF1430800. Part of the paper was preliminary
presented at the 22nd IEEE International Workshop on Signal Processing
Advances in Wireless Communications (SPAWC), Lucca, Italy, September
2021 and was selected as a Finalist for the Best Student Paper Award
\cite{zhang2021weighted}. \emph{(Corresponding author: Ziping Zhao.)}}\thanks{The authors are with the School of Information Science and Technology,
ShanghaiTech University, Shanghai 201210, China. (e-mail: \protect\href{mailto:zhangzp1@shanghaitech.edu.cn}{zhangzp1@shanghaitech.edu.cn};
\protect\href{mailto:zhaoziping@shanghaitech.edu.cn}{zhaoziping@shanghaitech.edu.cn}).}}
\markboth{\MakeLowercase{}}{ZHANG AND ZHAO: Rate Maximizations for RIS-Aided Wireless Networks}
\maketitle
\begin{abstract}
The reconfigurable intelligent surface (RIS) has arose an upsurging
research interest recently due to its promising outlook in 5G-and-beyond
wireless networks. With the assistance of RIS, the wireless propagation
environment is no longer static and could be customized to support
diverse service requirements. In this paper, we will approach the
rate maximization problems in RIS-aided wireless networks by considering
the beamforming and reflecting design jointly. Three representative
design problems from different system settings are investigated based
on a proposed unified algorithmic framework via the block minorization-maximization
(BMM) method. Extensions and generalizations of the proposed framework
in dealing with some other related problems are further presented.
Merits of the proposed algorithms are demonstrated through numerical
simulations in comparison with the state-of-the-art methods. 
\end{abstract}


\section{Introduction}

Along with the worldwide deployment of 5G wireless networks, more
stringent requirements (e.g., ultra-high reliability, capacity, and
efficiency, as well as low latency) are anticipated to be fulfilled
in a holistic fashion in the next-generation wireless networks \cite{giordani2020toward,tataria20216g}.
The existing technology trends in 5G networks (e.g., massive multiple-input
multiple-output (MIMO) and millimeter wave communications) \cite{boccardi2014five,shafi20175g},
unfortunately, may be insufficient to meet such daunting demands since
they will generally involve increased hardware costs and power consumptions.
With the theoretical and experimental breakthroughs in micro electromechanical
systems and metamaterials, reconfigurable intelligent surface (RIS),
a.k.a. software-controlled metasurfaces \cite{liaskos2018new} and
intelligent reflecting surfaces \cite{wu2019towards}, has recently
been advocated as a powerful solution to enhance the spectrum efficiency
and energy efficiency of wireless networks in a cost-effective way
\cite{wu2019towards,huang2020holographic,yuan2021reconfigurable}.

With the assistance of RIS, whose properties (e.g., scattering, absorption,
reflection, and diffraction) are reconfigurable rather than static,
the wireless environment is no longer an uncontrollable element, but
can be customized to support diverse service requirements \cite{di2020smart}.
Considering an interference channel in the multi-user communication
systems, where independent data streams are sent to some target receivers
simultaneously, one classical goal for the system design is to suppress
the inter-user interference and thus achieve a high system rate \cite{tan2011maximizing,weeraddana2012weighted}.
As a crucial aspect of the RIS-aided wireless networks, rate maximization
problems have received significant attention in different systems
with interference channels, leading to the problems of designing the
beamformers and configuring the elements of RISs simultaneously \cite{zhou2020intelligent,guo2020weighted,pan2020multicell,zhang2021weighted}.

In this paper, we focus on the rate maximization problems in RIS-aided
wireless networks, and a unified algorithmic framework based on the
block minorization-maximization (BMM) method \cite{razaviyayn2013unified,sun2016majorization}
is proposed. This framework is broadly applicable to diverse RIS-aided
systems, where the specific minorization-maximization (MM) techniques
are problem-tailored. Merits of the BMM algorithms are illustrated
via three different representative system design problems with different
design criteria, namely, weighted sum-rate (WSR) maximization for
multi-hop RIS-aided multi-user multi-input single-output (MISO) cellular
networks, minimum rate (MR) maximization for RIS-aided multi-user
MISO cellular networks, and sum-rate (SR) (i.e., the system capacity)
maximization for RIS-aided MIMO device-to-device (D2D) networks. 

\emph{WSR maximization for RIS-aided multi-user MISO cellular networks.}
Noticing that if no RIS is deployed in the network, this problem reduces
to the classical WSR maximization, for which many algorithms have
been proposed, two approaches were brought up in \cite{guo2020weighted}
based on the block coordinate descent (BCD) method (a.k.a. alternating
minimization or Gauss-Seidel method) \cite{bertsekas1999nonlinear},
where the design variables are partitioned into different blocks and
are updated cyclically with the remaining blocks fixed. The first
approach is to solve the beamforming block via the weighted minimum
mean square error (WMMSE) method \cite{christensen2008weighted,shi2011iteratively}
and the reflecting block via the Riemannian conjugate gradient (RCG)
method \cite{absil2009optimization}. Inheriting a double-loop nature,
this approach may invoke many iterations to converge and result in
high computational complexity. The other approach is through transforming
the problem by the fractional programming (FP) \cite{shen2018fractional}
where the reflecting block is solved by MM with a carefully chosen
stepsize for line search. This approach, however, relies on the manifold
structure of the continuous phase constraint on RISs, making it lame
in dealing with other system design problems like the case with RISs
of discrete phase \cite{di2020hybrid}. 

\emph{MR maximization for RIS-aided multi-user MISO cellular networks.}
Besides WSR, the MR metric, which is able to provide fairness among
the multiple users in the network, is also worth considering. Generally,
the MR objective in this case becomes nonsmooth. In \cite{zhou2020intelligent},
a similar problem under a multi-group multi-cast system setting was
considered, where the authors aimed at solving two approximation problems.
By convexifying the nonconvex unimodulus phase constraint on RISs,
the first problem was tackled with a BMM algorithm where a second-order
cone programming (SOCP) is invoked in each iteration. Besides, another
approximation problem is obtained by smoothing the MR objective, based
on which the per-iteration SOCP could be removed.

\emph{SR maximization for RIS-aided MIMO D2D networks.} Apart from
the MISO systems, the joint beamforming and reflecting design for
SR maximization in MIMO systems is further investigated. We consider
a MIMO D2D network, where a RIS is deployed to alleviate the co-channel
interference among D2D pairs caused by the full frequency reuse \cite{asadi2014survey}.

To make it clear, main contributions of this paper are summarized
in the following.
\begin{itemize}
\item A unified algorithmic framework via BMM for rate maximizations in
RIS-aided networks by joint beamforming and reflecting design is presented.
The proposed algorithms are of low signal processing complexity and
are broadly applicable for a class of system design problems.
\item To showcase the flexibility of the algorithm, three specific design
cases are investigated covering various rate-related design criteria
like WSR, MR, and SR, and diverse wireless system settings including
MISO and MIMO. 
\item Merits of the proposed algorithmic framework are demonstrated both
theoretically and empirically, while extensions and generalizations
of the framework, e.g., in handling more general constraints and dealing
with more general system configurations, are also demonstrated. 
\end{itemize}
The rest of this paper is organized as follows. In Section \ref{sec:BMM},
we present the BMM method. Three system design problems, namely, WSR
maximization for RIS-aided multi-user MISO networks, MR maximization
for RIS-aided multi-user MISO networks, and SR maximization for RIS-aided
MIMO D2D networks, are illustrated in Section \ref{sec:WSR}, Section
\ref{sec:MR}, and Section \ref{sec:SR}, respectively, with their
convergence and complexity analyses given in Section \ref{sec:Convergence-and-Complexity}.
In Section \ref{sec:Extensions-and-Generalizations}, we further discuss
several extensions and generalizations on RIS-aided system designs
under the proposed algorithm. Section \ref{sec:Numerical-Simulations}
provides simulation results, followed by conclusions in Section \ref{sec:Conclusions}.

\emph{Notations:} Boldface upper-case letters denote matrices, boldface
lower-case letters denote column vectors, and italics stand for scalars.
We denote by $\mathbf{1}$ the all-one vectors and by $\mathbf{I}$
the identity matrices respectively. We denote the all-zero vectors
and all-zero matrices uniformly by $\mathbf{0}$. The real (complex)
numbers are denoted by $\mathbb{R}$ ($\mathbb{C}$), the $N$-dimensional
real (complex) vectors are denoted by $\mathbb{R}^{N}$ ($\mathbb{C}^{N}$),
and the $N\times N$-dimensional complex matrices (Hermitian matrices)
are denoted by $\mathbb{C}^{N\times N}$ $(\mathbb{H}^{N})$. Superscripts
$(\cdot)^{*}$, $(\cdot)^{\mathrm{T}}$, $(\cdot)^{\mathrm{H}}$,
and $(\cdot)^{-1}$ denote the matrix conjugate, transpose, Hermitian,
and inverse operations, respectively. $[\mathbf{x}]_{i}$ denotes
the $i$-th element of vector $\mathbf{x}$, and $[\mathbf{x}]_{-i}$
denotes $\mathbf{x}$ with its $i$-th element replaced by zero. $[\mathbf{X}]_{ij}$
denotes the ($i$-th, $j$-th) element of matrix $\mathbf{X}$, and
$[\mathbf{X}]_{:,i}$ denotes the $i$-th column of matrix $\mathbf{X}$.
Given $\mathbf{A}$, $\mathbf{B}\in\mathbb{H}^{N}$, $\mathbf{A}\succeq\mathbf{B}$
($\mathbf{A}\succ\mathbf{B}$) means $\mathbf{A}-\mathbf{B}$ is a
positive semidefinite (definite) matrix. $\mathfrak{j}$ denotes the
imaginary unit satisfying $\mathfrak{j}^{2}=-1$. $\mathrm{Re}(\cdot)$,
$\mathrm{Im}(\cdot)$, $\bigl|\cdot\bigr|$, and $\mathrm{ang}(\cdot)$
denote the real part, the imaginary part, the modulus, and the angle
of a complex number, respectively. $\otimes$ and $\odot$ denote
the Kronecker product and the Hadamard product of two matrices respectively.
$\mathrm{tr}(\cdot)$ and $\left\Vert \cdot\right\Vert _{\mathrm{F}}$
denote the trace and the Frobenius norm of a matrix respectively.
$\mathrm{diag}(\mathbf{x})$ denote a diagonal matrix with entries
of $\mathbf{x}$ being on the diagonal.

\section{Block Minorization-Maximization Method \label{sec:BMM}}

In this section, we present the general scheme of the BMM method \cite{razaviyayn2013unified,sun2016majorization},
which can be regarded as a combination of the BCD method \cite{bertsekas1999nonlinear}
and the MM method \cite{sun2016majorization}. BCD is an optimization
method aiming at finding a local optimum of the problem by optimizing
along one variable block at a time while the other blocks are held
fixed. Instead of solving for an exact variable update as in BCD,
BMM solves a series of simpler surrogate problems w.r.t. one variable
block each time via carrying out an inexact variable update. Specifically,
consider the following maximization problem:
\begin{equation}
\begin{aligned} & \underset{\mathbf{x}_{1},\ldots,\mathbf{x}_{m}}{\mathrm{maximize}} &  & f(\mathbf{x}_{1},\ldots,\mathbf{x}_{m})\\
 & \mathrm{subject}\ \mathrm{to} &  & \mathbf{x}_{i}\in\mathcal{X}_{i},\ i=1,\ldots,m,
\end{aligned}
\label{eq:BMM}
\end{equation}
where $f:\prod_{i=1}^{m}{\cal X}_{i}\rightarrow\mathbb{R}$. To make
the problem well-defined, we assume $f$ is regular at every point
in $\prod_{i=1}^{m}{\cal X}_{i}$ and the level set $\bigl\{\mathbf{x}_{1},\ldots,\mathbf{x}_{m}\mid f(\mathbf{x}_{1},\ldots,\mathbf{x}_{m})\geq f(\mathbf{x}_{1}^{(t)},\ldots,\mathbf{x}_{m}^{(t)})\bigr\}$
is compact \cite{hong2015unified}, where $\mathbf{x}_{i}^{(t)}$'s
denote some given values. In the BMM method, different variable blocks
are updated in a cyclic order. At the $t$-th iteration, the $i$-th
variable block is updated by solving a maximization subproblem as
follows:
\[
\begin{aligned} & \underset{\mathbf{x}_{i}}{\mathrm{maximize}} &  & f_{i}^{\prime}(\mathbf{x}_{i},\mathbf{x}_{-i}^{(t)})\\
 & \mathrm{subject}\ \mathrm{to} &  & \mathbf{x}_{i}\in\mathcal{X}_{i},
\end{aligned}
\]
where $f_{i}^{\prime}(\mathbf{x}_{i},\mathbf{x}_{-i}^{(t)})$ is defined
as a minorizing function of $f$ w.r.t. $\mathbf{x}_{i}$ at iterate
$\mathbf{x}_{-i}^{(t)}\negthinspace=\negthinspace(\mathbf{x}_{1}^{(t)}\negthinspace,\negthinspace\ldots,\mathbf{x}_{i-1}^{(t)},\mathbf{x}_{i}^{(t-1)}\negthinspace,\negthinspace\ldots,\mathbf{x}_{m}^{(t-1)})$.
Suppose $\mathcal{X}_{1},\negthinspace\ldots,\mathcal{X}_{m}$ are
convex, the generated sequence $\{\mathbf{x}_{1}^{(t)},\ldots,\mathbf{x}_{m}^{(t)}\}$
can be proved to converge to a stationary point of Problem \eqref{eq:BMM}
if the following mild assumptions hold for $f_{i}^{\prime}(\mathbf{x}_{i},\mathbf{x}_{-i}^{(t)})$,
$i=1,\ldots,m$ \cite{razaviyayn2013unified}:
\begin{subequations}
\begin{align}
 & f_{i}^{\prime}(\mathbf{x}_{i},\mathbf{x}_{-i}^{(t)})\ \text{is continuous in }(\mathbf{x}_{i},\mathbf{x}_{-i}^{(t)})\\
 & f_{i}^{\prime}(\mathbf{x}_{i}^{(t-1)},\mathbf{x}_{-i}^{(t)})=f(\mathbf{x}_{-i}^{(t)}),\ \forall\mathbf{x}_{-i}^{(t)}\in\mathcal{X}\\
 & f_{i}^{\prime}(\mathbf{x}_{i},\mathbf{x}_{-i}^{(t)})\leq f(\mathbf{x}_{i},\mathbf{x}_{-i}^{(t)}),\ \forall\mathbf{x}_{i}\in\mathcal{X}_{i},\:\forall\mathbf{x}_{-i}^{(t)}\in\mathcal{X}\\
 & \begin{aligned} & \nabla f_{i}^{\prime}(\mathbf{x}_{i},\mathbf{x}_{-i}^{(t)};\mathbf{d})\big|_{\mathbf{x}_{i}=\mathbf{x}_{i}^{(t-1)}}=\nabla f(\mathbf{x}_{-i}^{(t)};\mathbf{d})\big|_{\mathbf{x}_{i}=\mathbf{x}_{i}^{(t-1)}}\\
 & \hspace{1.2cm}\forall\mathbf{d}=(\mathbf{0},\ldots,\mathbf{d}_{i},\ldots,\mathbf{0})\text{ s.t. }\mathbf{x}_{i}^{(t-1)}+\mathbf{d}_{i}\in\mathcal{X}_{i},
\end{aligned}
\label{BMM-Property-Derivative}
\end{align}
\end{subequations}
where $\nabla f(\mathbf{x};\mathbf{d})$ denotes the directional derivative
at $\mathbf{x}$ along $\mathbf{d}$. If $\mathcal{X}_{i}$ is nonconvex,
assumption \eqref{BMM-Property-Derivative} should be changed to
\[
\begin{aligned} & \nabla f_{i}^{\prime}(\mathbf{x}_{i},\mathbf{x}_{-i};\mathbf{d})\big|_{\mathbf{x}_{i}=\mathbf{x}_{i}^{(t-1)}}=\nabla f(\mathbf{x}_{-i};\mathbf{d})\big|_{\mathbf{x}_{i}=\mathbf{x}_{i}^{(t-1)}}\\
 & \hspace{1.1cm}\forall\mathbf{d}=(\mathbf{0},\ldots,\mathbf{d}_{i},\ldots,\mathbf{0})\text{ s.t. }\mathbf{d}_{i}\in\mathcal{T}_{\mathcal{X}_{i}}(\mathbf{x}_{i}^{(t-1)}),
\end{aligned}
\tag{\ensuremath{\text{2d}^{\prime}}}
\]
where $\mathcal{T}_{\mathcal{X}_{i}}(\mathbf{x}_{i}^{(t-1)})$ denotes
the Boulingand tangent cone of $\mathcal{X}_{i}$ at $\mathbf{x}_{i}^{(t-1)}$
\cite{rockafellar2009variational}. For minimization problems, a counterpart
called block majorization-minimization (BMM), a.k.a. block successive
upperbound minimization (BSUM) \cite{razaviyayn2013unified}, can
be applied where the maximization step of a minorizing function is
replaced by a minimization step of a majorizing function. Minorizing
and majorizing functions in BMM can be chosen in a flexible way \cite{sun2016majorization}
while a properly chosen one can make the updates easy and may lead
to a faster convergence over iterations. In practice, the subproblems
in BMM are applaudable if they are convex or have closed-form solutions. 

\section{Weighted Sum-Rate Maximization for\protect \\
 RIS-Aided Multi-User MISO Cellular Networks \label{sec:WSR} }

\subsection{System Model and Problem Formulation}

We consider a multi-hop RIS-aided multi-user MISO downlink communication
system, where the base station (BS) equipped with $M$ antennas communicates
with $K$ users with single antenna in a circular region. We assume
there are $L$ cascaded RISs deployed in the system and the transmitted
signal experiences $I_{k}$ $(I_{k}\leq L)$ hops on the RISs to arrive
the $k$-th user. Denote $\mathbf{W}=[\mathbf{w}_{1},\ldots,\mathbf{w}_{K}]\in\mathbb{C}^{M\times K}$
as the beamforming matrix with $\mathbf{w}_{k}$, $k=1,\ldots,K$
being the beamforming vector for the $k$-th user. Denote $\mathbf{G}_{0,1}\in\mathbb{C}^{N_{1}\times M}$,
$\mathbf{G}_{i-1,i}\in\mathbb{C}^{N_{i}\times N_{i-1}}$, and $\boldsymbol{\Theta}_{i}\triangleq\mathrm{diag}(\boldsymbol{\theta}_{i})\in\mathbb{C}^{N_{i}\times N_{i}}$,
$i=1,\ldots,L$ as the channel matrix from the BS to the first RIS,
the channel matrix from the $(i-1)$-th RIS to the $i$-th RIS, and
the phase shift matrix of the $i$-th RIS, respectively. Denote $\mathbf{h}_{k}^{\mathsf{r}}\in\mathbb{C}^{N_{I_{k}}}$
and $\mathbf{h}_{k}^{\mathsf{d}}\in\mathbb{C}^{M}$ as the channel
from the last RIS in the reflection channel to the $k$-th user and
the direct channel from the BS to the $k$-th user, respectively.
Then the received signal at the $k$-th user is given by
\begin{align*}
y_{k}= & \underbrace{\mathbf{w}_{k}^{\mathrm{H}}\Bigl(\prod_{i=1}^{I_{k}}\bigl(\mathbf{G}_{i-1,i}\boldsymbol{\Theta}_{i}\bigr)\mathbf{h}_{k}^{\mathsf{r}}+\mathbf{h}_{k}^{\mathsf{d}}\Bigr)s_{k}}_{\text{desired signal}}\\
 & \hspace{1.2cm}+\underbrace{\mathbf{w}_{k}^{\mathrm{H}}\sum_{j,j\neq k}^{K}\Bigl(\prod_{i=1}^{I_{k}}\bigl(\mathbf{G}_{i-1,i}\boldsymbol{\Theta}_{i}\bigr)\mathbf{h}_{k}^{\mathsf{r}}+\mathbf{h}_{k}^{\mathsf{d}}\Bigr)s_{j}+e_{k}}_{\text{interference plus noise}},
\end{align*}
where $s_{1},\ldots,s_{K}$ are the $K$ independent user symbols
with zero mean and unit variance, and $e_{k}\in\mathcal{CN}(0,\sigma^{2})$
represents the additive white Gaussian noise for the $k$-th user.
In this section, for the illustration simplicity, the system model
is set up into a cascaded multi-hop signal transmission scenario,
where only the direct transmission paths and the transmission paths
through $I_{k}$ RISs from the BS to the users have been considered.
We will show in Section \ref{subsec:WSR-Maximization-for-general-topology}
such a setup can be easily extended to a more general signal transmission
scenario. This multi-hop relaying system model is classical \cite{shen2009multi,gunduz2010multi}
and can be deployed to combat the propagation distance problem and
to improve the coverage range. Specially, when $I_{1}=\cdots=I_{k}=0$,
there only exists direct transmission paths, which reduces to the
traditional system with no RIS deployed. Recently, a deep reinforcement
learning approach was proposed for a similar multi-hop system design
problem \cite{huang2021multi}, whose performance highly relies on
the carefully chosen initializations from some preliminary iterative
algorithms. 

Given the signal model, the signal-to-interference-plus-noise ratio
(SINR) at the $k$-th user is computed as follows:\footnote{Acknowledging that channel estimation is a nontrivial and crucial
problem in RIS-aided system design, while we will assume perfect channel
state information to be available through all this paper.}
\[
\mathsf{SINR}_{k}=\frac{\Big|\mathbf{w}_{k}^{\mathrm{H}}\Bigl(\prod_{i=1}^{I_{k}}\bigl(\mathbf{G}_{i-1,i}\boldsymbol{\Theta}_{i}\bigr)\mathbf{h}_{k}^{\mathsf{r}}+\mathbf{h}_{k}^{\mathsf{d}}\Bigr)\Big|^{2}}{\sum_{j,j\neq k}^{K}\Big|\mathbf{w}_{j}^{\mathrm{H}}\Bigl(\prod_{i=1}^{I_{k}}\bigl(\mathbf{G}_{i-1,i}\boldsymbol{\Theta}_{i}\bigr)\mathbf{h}_{k}^{\mathsf{r}}+\mathbf{h}_{k}^{\mathsf{d}}\Bigr)\Big|^{2}+\sigma^{2}}.
\]
Our interest is to maximize the WSR of the system by jointly designing
the beamforming matrix $\mathbf{W}$ and the phase shift matrices
$\{\boldsymbol{\Theta}_{i}\}$. With the data rate (in nats per second
per Hertz (nps/Hz)) at the $k$-th user defined by $\mathsf{R}_{k}=\log(1+\mathsf{SINR}_{k})$,\footnote{The natural logarithm is chosen since optimal solutions of all the
rate maximization problems given later are irrelevant to bases of
the log-functions.} the WSR maximization problem is defined as follows:
\begin{equation}
\begin{aligned} & \underset{\mathbf{W},\{\boldsymbol{\Theta}_{i}\}}{\mathrm{maximize}} &  & \negthinspace f_{\mathsf{WSR}}(\mathbf{W},\{\boldsymbol{\Theta}_{i}\})=\sum_{k=1}^{K}\omega_{k}\mathsf{R}_{k}\\
 & \mathrm{subject}\ \mathrm{to} &  & \negthinspace\mathbf{W}\in\mathcal{W},\ \boldsymbol{\Theta}_{i}\in\mathcal{C}_{i},\ \forall i=1,\ldots,L,
\end{aligned}
\tag{WSRMax}\label{WSR:Formulation}
\end{equation}
where $\omega_{1},\ldots,\omega_{K}$ are the predefined nonnegative
weights,
\[
\mathcal{W}=\left\{ \mathbf{W}\mid\bigl\Vert\mathbf{W}\bigr\Vert_{\mathrm{F}}^{2}\leq P\right\} 
\]
denotes the transmit power limit constraint of the BS, and
\[
\mathcal{C}_{i}=\negthinspace\left\{ \boldsymbol{\Theta}_{i}\negthinspace\mid\boldsymbol{\Theta}_{i}\negthinspace=\negthinspace\mathrm{diag}(\boldsymbol{\theta}_{i}),\,\boldsymbol{\theta}_{i}\negthinspace\in\negthinspace\mathbb{C}^{N_{i}}\negthinspace,\,\bigl|[\boldsymbol{\theta}_{i}]_{j}\bigr|\negthinspace=\negthinspace1,\,j\negthinspace=\negthinspace1,\negthinspace\ldots,N_{i}\right\} 
\]
represents the constant modulus constraint for the $i$-th RIS indicating
that there is no energy loss for the signal when going through the
RISs. Problem \eqref{WSR:Formulation} is nonconvex and NP-hard \cite{luo2008dynamic}.
In the following, we will develop a globally convergent algorithm
via BMM for problem resolution. 

\subsection{The Update Step of $\mathbf{W}$ \label{WSR:Section-W-Block}}

We take beamforming variables $\{\mathbf{w}_{i}\}$ as one block.
Given iterate $\{\underline{\mathbf{W}},\{\underline{\boldsymbol{\Theta}_{i}}\}\}$\footnote{In this paper, \uline{underlined variables} denote those whose
values are given.}, the objective function $f_{\mathsf{WSR}}$ w.r.t. $\mathbf{W}$
is\footnote{$f_{\mathsf{WSR},\mathbf{W}}(\mathbf{W})$ has been used to represent
$f_{\mathsf{WSR},\mathbf{W}}(\mathbf{W},\underline{\mathbf{W}},\{\underline{\boldsymbol{\Theta}_{i}}\})$
for notational simplicity. Similar simplifications will be adopted
along this paper.}
\begin{equation}
f_{\mathsf{WSR},\mathbf{W}}(\mathbf{W})=\sum_{k=1}^{K}\omega_{k}\log\bigl(1+\frac{\bigl|\mathbf{w}_{k}^{\mathrm{H}}\mathbf{h}_{k}\bigr|^{2}}{\sum_{j,j\neq k}^{K}\bigl|\mathbf{w}_{j}^{\mathrm{H}}\mathbf{h}_{k}\bigr|^{2}+\sigma^{2}}\bigr),\label{WSR: W-Block WSR}
\end{equation}
where $\mathbf{h}_{k}=\prod_{i=1}^{I_{k}}\bigl(\mathbf{G}_{i-1,i}\underline{\boldsymbol{\Theta}_{i}}\bigr)\mathbf{h}_{k}^{\mathsf{r}}+\mathbf{h}_{k}^{\mathsf{d}}$.

\begin{figure}[t]
\includegraphics[width=0.9\columnwidth]{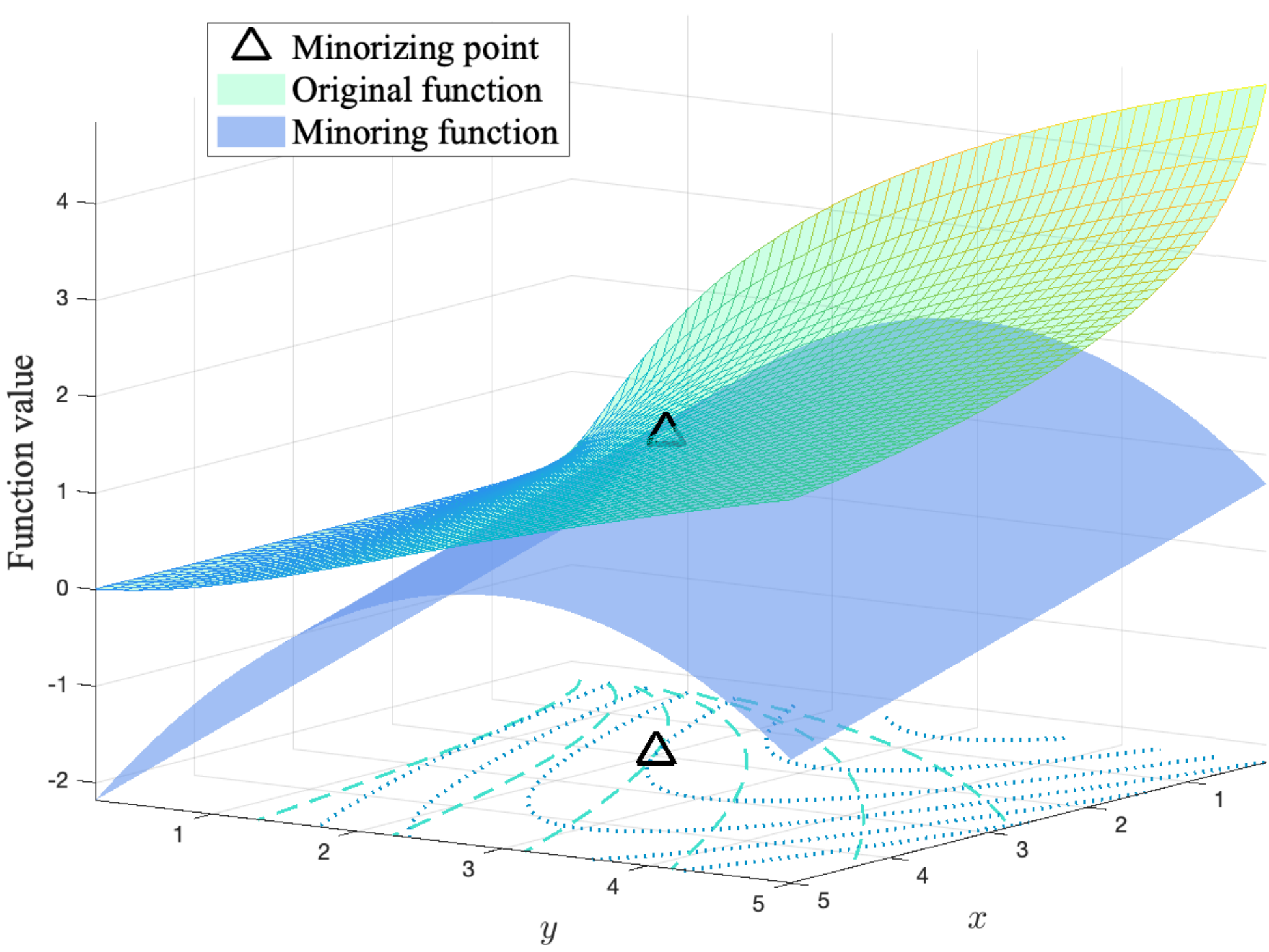}

\caption{\label{Lemma:Scalar MM Figure}A pictorial demonstration of the minorization
procedure for $\log(1+\frac{x^{2}}{y})$ ($x\in\mathbb{R}$ and $y>0$)
at $(x_{\triangle},y_{\triangle})=(2,2)$. }
\end{figure}

Optimization with $f_{\mathsf{WSR},\mathbf{W}}(\mathbf{W})$ reduces
to the classic WSR maximization problem, which is still nonconvex
and NP-hard. We first introduce the following result, with which a
quadratic minorizing function for $f_{\mathsf{WSR},\mathbf{W}}(\mathbf{W})$
can be constructed.
\begin{prop}
\label{Prop: Scalar MM} The $\log(1+\frac{|x|^{2}}{y})$ with $x\in\mathbb{C}$
and $y>0$ is minorized at $(\underline{x},\underline{y})$ as follows:
\begin{align*}
\log(1+\frac{|x|^{2}}{y})\geq & -\frac{|\underline{x}|^{2}}{\underline{y}(\underline{y}+|\underline{x}|^{2})}(y+|x|^{2})+\frac{2}{\underline{y}}\mathrm{Re}(\underline{x}^{*}x)\\
 & \hspace{3cm}+\log(1+\frac{|\underline{x}|^{2}}{\underline{y}})-\frac{|\underline{x}|^{2}}{\underline{y}}.
\end{align*}
\end{prop}
\begin{IEEEproof}
The proof is deferred to Appendix \ref{Appendix: Scalar MM Proof}.
\end{IEEEproof}
A pictorial illustration of the minorization procedure in Lemma \ref{Prop: Scalar MM}
is demonstrated in Fig. \ref{Lemma:Scalar MM Figure}.
 Based on Lemma \ref{Prop: Scalar MM}, taking $\mathbf{w}_{k}^{\mathrm{H}}\mathbf{h}_{k}$
as $x$ and $\sum_{j,j\neq k}^{K}\bigl|\mathbf{w}_{j}^{\mathrm{H}}\mathbf{h}_{k}\bigr|^{2}+\sigma^{2}$
as $y$, a minorizing function for $f_{\mathsf{WSR},\mathbf{W}}$
is constructed as follows:
\begin{equation}
\begin{aligned} & f_{\mathsf{WSR},\mathbf{W}}^{\prime}(\mathbf{W})\\
= & \sum_{k=1}^{K}\omega_{k}\bigl(-\alpha_{k}\sum_{j=1}^{K}\bigl|\mathbf{w}_{j}^{\mathrm{H}}\mathbf{h}_{k}\bigr|^{2}\negthinspace+\negthinspace2\mathrm{Re}(\beta_{k}\mathbf{w}_{k}^{\mathrm{H}}\mathbf{h}_{k})\bigr)\negthinspace+\negthinspace\mathsf{const}_{w},
\end{aligned}
\label{eq:f_WSR_W^prime}
\end{equation}
where
\[
\alpha_{k}=\frac{\underline{\mathsf{SINR}_{k}}}{\sum_{j=1}^{K}\bigl|\underline{\mathbf{w}_{j}^{\mathrm{H}}}\mathbf{h}_{k}\bigr|^{2}+\sigma^{2}},\hspace{1cm}\beta_{k}=\frac{\underline{\mathsf{SINR}_{k}}}{\underline{\mathbf{w}_{k}^{\mathrm{H}}}\mathbf{h}_{k}},
\]
and $\mathsf{const}_{w,k}=\sum_{k=1}^{K}\omega_{k}\left(\underline{\mathsf{R}_{k}}-\underline{\mathsf{SINR}_{k}}-\alpha_{k}\sigma^{2}\right)$,
in which $\underline{\mathsf{R}_{k}}$ and $\underline{\mathsf{SINR}_{k}}$
are calculated with the given $\{\underline{\mathbf{W}},\{\underline{\boldsymbol{\Theta}_{i}}\}\}$.
By rearranging the terms and ignoring the constant terms, we obtain
the resultant convex subproblem for $\mathbf{W}$ given by
\begin{equation}
\begin{aligned} & \underset{\mathbf{W}\in\mathcal{W}}{\mathrm{minimize}} &  & \sum_{k=1}^{K}\bigl(\mathbf{w}_{k}^{\mathrm{H}}\mathbf{R}\mathbf{w}_{k}-2\mathrm{Re}(\mathbf{w}_{k}^{\mathrm{H}}\bigl[\mathbf{Q}\bigr]_{:,k})\bigr),\end{aligned}
\label{WSR:W-Subproblem-vector-form}
\end{equation}
where $\mathbf{R}\negthinspace=\negthinspace\sum_{k=1}^{K}\omega_{k}\alpha_{k}\mathbf{h}_{k}\mathbf{h}_{k}^{\mathrm{H}}$
and $\bigl[\mathbf{Q}\bigr]_{:,k}\negthinspace\negthinspace=\negthinspace\omega_{k}\beta_{k}\mathbf{h}_{k}$.
Note that Problem \eqref{WSR:W-Subproblem-vector-form} can be cast\textcolor{black}{{}
as a constrained weighted sum mean square error minimization problem}
and\textcolor{black}{{} }can be rewritten as
\begin{equation}
\begin{aligned} & \underset{\mathbf{W}\in\mathcal{W}}{\mathrm{minimize}} &  & \mathrm{tr}\left(\mathbf{W}^{\mathrm{H}}\mathbf{RW}\right)-2\mathrm{Re}\left(\mathrm{tr}(\mathbf{W}^{\mathrm{H}}\mathbf{Q})\right).\end{aligned}
\label{WSR:W-Subproblem}
\end{equation}

\begin{lem}
\label{Prop: W-Subproblem-Solution} By solving the Karush-Kuhn-Tucker
(KKT) system, the optimal solution to Problem \eqref{WSR:W-Subproblem}
is given by
\[
\mathbf{W}^{\star}=\left\{ \begin{array}{ll}
\mathbf{R}^{-1}\mathbf{Q} & \hspace{0.3cm}\text{if }\bigl\Vert\mathbf{R}^{-1}\mathbf{Q}\bigr\Vert_{\mathrm{F}}^{2}\leq P\\
\left(\mathbf{R}+\gamma\mathbf{I}\right)^{-1}\mathbf{Q} & \hspace{0.3cm}\text{otherwise, }
\end{array}\right.
\]
where the variable $\gamma$ satisfies
\[
\bigl\Vert(\mathbf{R}+\gamma\mathbf{I})^{-1}\mathbf{Q}\bigr\Vert_{\mathrm{F}}^{2}=P,
\]
and can be readily found via one-dimensional line search methods to
meet
\[
\sum_{n=1}^{N}\frac{\bigl[\mathbf{V}^{\mathrm{H}}\mathbf{Q}\mathbf{Q}^{\mathrm{H}}\mathbf{V}\bigr]_{nn}}{\bigl(\bigl[\bm{\Lambda}\bigr]_{nn}+\gamma\bigr)^{2}}=P,
\]
where $\mathbf{V}$ and $\boldsymbol{\Lambda}$ are obtained from
the eigendecomposition $\mathbf{R}=\mathbf{V}\bm{\Lambda}\mathbf{V}^{\mathrm{H}}$.
\end{lem}

\subsection{The Update Step of $\{\boldsymbol{\Theta}_{i}\}$ \label{WSR:Section-Theta-Block}}

In this section, we choose to update the $L$ phase shift matrices
$\{\boldsymbol{\Theta}_{i}\}$ successively. Given iterate $\{\underline{\mathbf{W}},\{\underline{\boldsymbol{\Theta}_{i}}\}\}$,
$f_{\mathsf{WSR}}$ w.r.t. $\boldsymbol{\Theta}_{l}$ for $l=1,\ldots,L$,\footnote{Note that w.l.o.g. we have assumed $I_{k}\geq l$ holds for $k=1,\ldots,K$.}
is given by
\begin{align}
\begin{aligned} & f_{\mathsf{WSR},\boldsymbol{\Theta}_{l}}\left(\boldsymbol{\Theta}_{l}\right)\\
= & \sum_{k=1}^{K}\omega_{k}\log\bigl(1+\frac{\bigl|\underline{\mathbf{w}_{k}^{\mathrm{H}}}\mathbf{F}_{k,l}\boldsymbol{\theta}_{l}+\underline{\mathbf{w}_{k}^{\mathrm{H}}}\mathbf{h}_{k}^{\mathsf{d}}\bigr|^{2}}{\sum_{j,j\neq k}^{K}\bigl|\underline{\mathbf{w}_{j}^{\mathrm{H}}}\mathbf{F}_{k,l}\boldsymbol{\theta}_{l}+\underline{\mathbf{w}_{j}^{\mathrm{H}}}\mathbf{h}_{k}^{\mathsf{d}}\bigr|^{2}+\sigma^{2}}\bigr),
\end{aligned}
\label{WSR: Theta-Block WSR}
\end{align}
with $\mathbf{F}_{k,l}\negthinspace=\negthinspace\negthinspace\prod_{i=1}^{l-1}(\mathbf{G}_{i\negthinspace-\negthinspace1,i}\,\underline{\boldsymbol{\Theta}}_{i})\mathbf{G}_{l-1,l}\mathrm{diag}\bigl(\prod_{i=l+1}^{I_{k}}\negthinspace\negthinspace\mathbf{G}_{i\negthinspace-\negthinspace1,i}\,\underline{\boldsymbol{\Theta}}_{i}\mathbf{h}_{k}^{\mathsf{r}}\bigr)$.
Based on Lemma \ref{Prop: Scalar MM}, a minorizing function for $f_{\mathsf{WSR},\boldsymbol{\Theta}_{l}}$
is constructed in the following way\footnote{Note that $\alpha_{k}$ and $\beta_{k}$ have the same expressions
as given in Section \ref{WSR:Section-W-Block}, while the iterate
value $\{\underline{\mathbf{W}},\{\underline{\boldsymbol{\Theta}_{i}}\}\}$
may be different.}
\begin{equation}
\begin{aligned}f_{\mathsf{WSR},\boldsymbol{\Theta}_{l}}^{\prime}(\boldsymbol{\Theta}_{l}) & =\sum_{k=1}^{K}\omega_{k}\Bigl(-\alpha_{k}\sum_{j=1}^{K}\Bigl|\underline{\mathbf{w}_{j}^{\mathrm{H}}}\mathbf{F}_{k,l}\boldsymbol{\theta}_{l}+\underline{\mathbf{w}_{j}^{\mathrm{H}}}\mathbf{h}_{k}^{\mathsf{d}}\Bigr|^{2}\\
 & \hspace{-0.4cm}+2\mathrm{Re}\bigl(\beta_{k}\underline{\mathbf{w}_{k}^{\mathrm{H}}}\mathbf{F}_{k,l}\boldsymbol{\theta}_{l}+\beta_{k}\underline{\mathbf{w}_{k}^{\mathrm{H}}}\mathbf{h}_{k}^{\mathsf{d}}\bigr)\Bigr)+\mathsf{const}_{w},
\end{aligned}
\label{eq:f_WSR_theta_l^prime}
\end{equation}
which can be further compactly rewritten as
\begin{align}
\hspace{-5bp}\begin{aligned}f_{\mathsf{WSR},\boldsymbol{\Theta}_{l}}^{\prime}(\boldsymbol{\Theta}_{l}) & =-\boldsymbol{\theta}_{l}^{\mathrm{H}}\mathbf{L}_{l}\boldsymbol{\theta}_{l}+\sum_{k=1}^{K}2\omega_{k}\mathrm{Re}\Bigl(\boldsymbol{\theta}_{l}^{\mathrm{H}}\mathbf{F}_{k,l}^{\mathrm{H}}\bigl(\beta_{k}^{*}\underline{\mathbf{w}_{k}}\\
 & \hspace{1.5cm}-\alpha_{k}\sum_{j=1}^{K}\underline{\mathbf{w}_{j}}\underline{\mathbf{w}_{j}^{\mathrm{H}}}\mathbf{h}_{k}^{\mathsf{d}}\bigr)\Bigr)+\mathsf{const}_{\theta,l},
\end{aligned}
\label{eq:f_WSR_theta_l^prime2}
\end{align}
where $\mathbf{L}_{l}=\sum_{k=1}^{K}\omega_{k}\alpha_{k}\sum_{j=1}^{K}\mathbf{F}_{k,l}^{\mathrm{H}}\underline{\mathbf{w}_{j}}\underline{\mathbf{w}_{j}^{\mathrm{H}}}\mathbf{F}_{k,l}$
and $\mathsf{const}_{\theta,l}=\sum_{k=1}^{K}\omega_{k}\bigl(-\alpha_{k}\sum_{j=1}^{K}\bigl|\underline{\mathbf{w}_{j}^{\mathrm{H}}}\mathbf{h}_{k}^{\mathsf{d}}\bigr|^{2}+2\mathrm{Re}(\beta_{k}\underline{\mathbf{w}_{k}^{\mathrm{H}}}\mathbf{h}_{k}^{\mathsf{d}})\bigr)+\mathsf{const}_{w}$.
Optimizing $f_{\mathsf{WSR},\boldsymbol{\Theta}_{l}}^{\prime}(\boldsymbol{\Theta}_{l})$
under $\mathcal{C}_{l}$ is intricate, hence we further introduce
the following two results to linearize $f_{\mathsf{WSR},\boldsymbol{\Theta}_{l}}^{\prime}$.
\begin{lem}[$\negmedspace$\cite{zhao2018mimo}$\,$]
\label{lem:quadratic_MM}Let $\mathbf{L}$\textup{,} $\mathbf{M}\in\mathbb{H}^{N}$
such that $\mathbf{M}\succeq\mathbf{L}$. Then the function $\mathbf{x}^{\mathrm{H}}\mathbf{L}\mathbf{x}$
with $\mathbf{x}\in\mathbb{C}^{N}$ is majorized at $\underline{\mathbf{x}}$
as follows:
\[
\mathbf{x}^{\mathrm{H}}\mathbf{L}\mathbf{x}\leq\mathbf{x}^{\mathrm{H}}\mathbf{M}\mathbf{x}+2\mathrm{Re}(\mathbf{x}^{\mathrm{H}}(\mathbf{L}-\mathbf{M})\text{\ensuremath{\underline{\mathbf{x}}}})+\underline{\mathbf{x}}^{\mathrm{H}}(\mathbf{M}-\mathbf{L})\underline{\mathbf{x}}.
\]
\end{lem}
\begin{lem}
\label{lem:Matrix Bound} Given $\mathbf{M}=\left\Vert \mathbf{x}\right\Vert _{2}^{2}\mathbf{I}$
and $\mathbf{L}=\mathbf{x}\mathbf{x}^{\mathrm{H}}$ with $\mathbf{x}\in\mathbb{C}^{n}$,
it follows that $\mathbf{M}-\mathbf{L}\succeq\mathbf{0}$.
\end{lem}
\begin{IEEEproof}
For any $\mathbf{y}\in\mathbb{C}^{N}$, we can obtain $\mathbf{y}^{\mathrm{H}}(\mathbf{M}-\mathbf{L})\mathbf{y}=\left\Vert \mathbf{x}\right\Vert _{2}^{2}\left\Vert \mathbf{y}\right\Vert _{2}^{2}-|\mathbf{x}^{\mathrm{H}}\mathbf{y}|^{2}\geq0$,
which completes the proof.
\end{IEEEproof}
Applying Lemma \ref{lem:quadratic_MM} and Lemma \ref{lem:Matrix Bound}
to the first term in \eqref{eq:f_WSR_theta_l^prime2}, a linear minorizing
function for $f_{\mathsf{WSR},\boldsymbol{\Theta}_{l}}^{\prime}$
can be obtained as
\begin{equation}
\!f_{\mathsf{WSR},\boldsymbol{\Theta}_{l}}^{\prime\prime}(\boldsymbol{\Theta}_{l})\!\!=\!-2\mathrm{Re}\bigl(\boldsymbol{\theta}_{l}^{\mathrm{H}}\mathbf{b}_{l}\bigr)-N_{l}\lambda_{l}-\underline{\boldsymbol{\theta}_{l}^{\mathrm{H}}}\bigl(\lambda_{l}\mathbf{I}-\mathbf{L}_{l}\bigr)\underline{\boldsymbol{\theta}_{l}}+\mathsf{const}_{\theta,l},\label{eq:f_WSR_theta_l^primeprime}
\end{equation}
where
\[
\mathbf{b}_{l}=\sum_{k=1}^{K}\omega_{k}\mathbf{F}_{k,l}^{\mathrm{H}}\bigl(\alpha_{k}\sum_{j=1}^{K}\underline{\mathbf{w}_{j}}\underline{\mathbf{w}_{j}^{\mathrm{H}}}\mathbf{h}_{k}^{\mathsf{d}}-\beta_{k}^{*}\underline{\mathbf{w}_{k}}\bigr)+(\mathbf{L}_{l}-\lambda_{l}\mathbf{I})\underline{\boldsymbol{\theta}_{l}}
\]
with $\lambda_{l}=\sum_{k=1}^{K}\omega_{k}\alpha_{k}\sum_{j=1}^{K}\bigl\Vert\underline{\mathbf{w}_{j}^{\mathrm{H}}}\mathbf{F}_{l,k}\bigr\Vert_{2}^{2}$.
Finally, noticing the second term in $f_{\mathsf{WSR},\boldsymbol{\Theta}_{l}}^{\prime\prime}$
is constant over $\mathcal{C}_{l}$ and discarding the constants,
the subproblem for $\boldsymbol{\Theta}_{l}$ is given by
\begin{equation}
\begin{aligned} & \underset{\boldsymbol{\Theta}_{l}\in\mathcal{C}_{l}}{\mathrm{minimize}} &  & \mathrm{Re}\bigl(\boldsymbol{\theta}_{l}^{\mathrm{H}}\mathbf{b}_{l}\bigr).\end{aligned}
\label{WSR:Theta-Subproblem-Parallel}
\end{equation}

\begin{lem}[$\negmedspace$\cite{zhao2018mimo}$\,$]
\label{Prop: Theta-Subproblem-Solution}Optimal solutions to Problem
\eqref{WSR:Theta-Subproblem-Parallel} can be obtained in closed-forms
as $\boldsymbol{\theta}_{l}^{\star}=e^{\mathfrak{j}\mathrm{ang}(-\mathbf{b}_{l})}$.
\end{lem}
In Problem \eqref{WSR:Theta-Subproblem-Parallel}, elements of $\boldsymbol{\theta}_{l}$
are separable in the objective and the constraint, and hence can be
updated in parallel.

In summary, based on BMM the variable blocks $\mathbf{W}$ and $\{\boldsymbol{\Theta}_{i}\}$
will be updated cyclically in closed-forms until some convergence
criterion is met.\footnote{It is sometimes called ``semi-closed'' as line search methods are
invovled.} The overall BMM algorithm is summarized in Algorithm \ref{WSR:Algorithm}
with its convergence and complexity analyses deferred to Section \ref{sec:Convergence-and-Complexity}.
\begin{algorithm}[t]
\caption{The BMM Algorithm for Problem \eqref{WSR:Formulation}. \label{WSR:Algorithm}}

\noindent\begin{minipage}[t]{1\columnwidth}%
\textbf{Input:} $\{\mathbf{h}_{i}^{\mathsf{r}}\}$, $\{\mathbf{h}_{i}^{\mathsf{d}}\}$,
$\{\mathbf{G}_{i-1,i}\}$, $P$, $\{\omega_{k}\}$, $\sigma^{2}$,
initial feasible values of $\mathbf{W}$ and $\{\boldsymbol{\Theta}_{i}\}$.

\textbf{Repeat}

1. $\negthinspace$Update $\mathbf{W}$ by solving Prob. \eqref{WSR:W-Subproblem-vector-form}
via Lemma \ref{Prop: W-Subproblem-Solution};

2. $\negthinspace$Update $\negthinspace\{\boldsymbol{\Theta}_{i}\}\negthinspace$
successively by solving Prob.$\negthinspace$ \eqref{WSR:Theta-Subproblem-Parallel}
$\negthinspace$via Lemma $\negthinspace$\ref{Prop: Theta-Subproblem-Solution};

\textbf{Until} the value of the objective function converges.%
\end{minipage}
\end{algorithm}

\section{Minimum Rate Maximization for \protect \\
RIS-Aided Multi-User MISO Cellular Networks \label{sec:MR}}

\subsection{System Model and Problem Formulation}

The system model considered in this section is quite similar to the
one discussed in Section \ref{sec:WSR}, except that there is only
a single RIS with $N$ reflecting elements. We denote $\mathbf{G}\in\mathbb{C}^{N\times M}$
and $\boldsymbol{\Theta}\triangleq\mathrm{diag}\left(\boldsymbol{\theta}\right)\in\mathbb{C}^{N\times N}$
as the channel matrix from the BS to the RIS and the phase shift matrix
of the RIS, respectively. The received signal at the $k$-th user
is given by
\begin{align*}
y_{k} & =\underset{\text{desired signal}}{\underbrace{\mathbf{w}_{k}^{\mathrm{H}}\bigl(\mathbf{G}\boldsymbol{\Theta}\mathbf{h}_{k}^{\mathsf{r}}+\mathbf{h}_{k}^{\mathsf{d}}\bigr)s_{k}}}+\negthinspace\underset{\text{interference plus noise}}{\underbrace{\sum_{j,j\neq k}^{K}\mathbf{w}_{j}^{\mathrm{H}}\bigl(\mathbf{G}\boldsymbol{\Theta}\mathbf{h}_{k}^{\mathsf{r}}+\mathbf{h}_{k}^{\mathsf{d}}\bigr)s_{j}+e_{k}}},
\end{align*}
and the SINR at the $k$-th user is accordingly computed as
\[
\mathsf{SINR}_{k}=\frac{\bigl|\mathbf{w}_{k}^{\mathrm{H}}\bigl(\mathbf{G}\boldsymbol{\Theta}\mathbf{h}_{k}^{\mathsf{r}}+\mathbf{h}_{k}^{\mathsf{d}}\bigr)\bigr|^{2}}{\sum_{j,j\neq k}^{K}\bigl|\mathbf{w}_{j}^{\mathrm{H}}\bigl(\mathbf{G}\boldsymbol{\Theta}\mathbf{h}_{k}^{\mathsf{r}}+\mathbf{h}_{k}^{\mathsf{d}}\bigr)\bigr|^{2}+\sigma^{2}}.
\]
Our design target is to maximize the MR of the system by jointly designing
the beamforming matrix $\mathbf{W}$ and the phase shift matrix $\boldsymbol{\Theta}$.
With the data rate at the $k$-th user defined by $\mathsf{R}_{k}=\mathsf{log}(1+\mathsf{SINR}_{k})$,
the MR maximization problem is 
\begin{equation}
\begin{aligned} & \underset{\mathbf{W},\boldsymbol{\Theta}}{\mathrm{maximize}} &  & f_{\mathsf{MR}}\left(\mathbf{W},\boldsymbol{\Theta}\right)=\underset{k=1,\ldots,K}{\text{\ensuremath{\min}}}\ \mathsf{R}_{k}\\
 & \mathrm{subject}\ \mathrm{to} &  & \mathbf{W}\in\mathcal{W},\ \boldsymbol{\Theta}\in\mathcal{C},
\end{aligned}
\tag{MRMax}\label{MaxMin:Formulation}
\end{equation}
where $\mathcal{W}$ is defined as before and
\[
\mathcal{C}=\bigl\{\boldsymbol{\Theta}\negthinspace\mid\negthinspace\boldsymbol{\Theta}=\mathrm{diag}(\boldsymbol{\theta}),\:\boldsymbol{\theta}\in\mathbb{C}^{N}\negthinspace,\:\bigl|[\boldsymbol{\theta}]_{j}\bigr|=1,\:\forall j=1,\ldots,N\bigr\}.
\]
Problem \eqref{MaxMin:Formulation} is nonconvex and NP-hard. Like
in the last section, a low-complexity and globally convergent BMM-based
algorithm will be developed for problem resolution. 

\subsection{The Update Step of $\mathbf{W}$}

Given the iterate $\left\{ \underline{\mathbf{W}},\underline{\boldsymbol{\Theta}}\right\} $,
the objective $f_{\mathsf{MR}}$ w.r.t. $\mathbf{W}$ is
\[
f_{\mathsf{MR},\mathbf{W}}(\mathbf{W})=\underset{k}{\text{\ensuremath{\min}}}\,\log\Bigl(1+\frac{\bigl|\mathbf{w}_{k}^{\mathrm{H}}\mathbf{h}_{k}\bigr|^{2}}{\sum_{j,j\neq k}^{K}\bigl|\mathbf{w}_{j}^{\mathrm{H}}\mathbf{h}_{k}\bigr|^{2}+\sigma^{2}}\Bigr),
\]
where $\mathbf{h}_{k}=\mathbf{G}\underline{\boldsymbol{\Theta}}\mathbf{h}_{k}^{\mathsf{r}}+\mathbf{h}_{k}^{\mathsf{d}}$.
For the pointwise minimum function $f_{\mathsf{MR},\mathbf{W}}$,
a minorizing function of it can be obtained based on the minorizing
functions of $\mathsf{R}_{1},\ldots,\mathsf{R}_{k}$ (see a proof
given in \cite{zhao2016maximin}). With Lemma \ref{Prop: Scalar MM},
a minorizing function for $f_{\mathsf{MR},\mathbf{W}}$ can be constructed
as follows:
\[
\begin{aligned} & \hspace{13bp}f_{\mathsf{MR},\mathbf{W}}^{\prime}(\mathbf{W})\\
 & =\underset{k}{\text{\ensuremath{\min}}}-\alpha_{k}\sum_{j=1}^{K}\bigl|\mathbf{w}_{j}^{\mathrm{H}}\mathbf{h}_{k}\bigr|^{2}+2\mathrm{Re}\left(\beta_{k}\mathbf{w}_{k}^{\mathrm{H}}\mathbf{h}_{k}\right)+\mathsf{const}_{w,k},
\end{aligned}
\]
where $\mathsf{const}_{w,k}=\underline{\mathsf{R}_{k}}-\underline{\mathsf{SINR}_{k}}-\alpha_{k}\sigma^{2}$.
Then the subproblem for $\mathbf{W}$ becomes a minimax problem given
by
\begin{equation}
\begin{aligned} & \underset{\mathbf{W}\in\mathcal{W}}{\negthickspace\negthickspace\mathrm{minimize}} &  & \negthickspace\negthickspace\underset{k}{\max}\;\mathrm{tr}\bigl(\mathbf{W}^{\mathrm{H}}\mathbf{R}_{k}\mathbf{W}\bigr)\negthinspace-\negthinspace2\mathrm{Re}(\mathbf{w}_{k}^{\mathrm{H}}\mathbf{q}_{k})\negthinspace-\negthinspace\mathsf{const}_{w,k},\end{aligned}
\negthickspace\negthickspace\negthickspace\negthickspace\negthickspace\label{MaxMin:W-Subproblem}
\end{equation}
where $\mathbf{R}_{k}=\alpha_{k}\mathbf{h}_{k}\mathbf{h}_{k}^{\mathrm{H}}$
and $\mathbf{q}_{k}=\beta_{k}\mathbf{h}_{k}$. The discrete maximum
of the objective can be easily transformed to be a continuous maximum
over a simplex, and we have the following problem
\begin{equation}
\negthinspace\begin{aligned} & \underset{\mathbf{W}\in\mathcal{W}}{\mathrm{minimize}} &  & \negthickspace\negthickspace\negthickspace\underset{\mathbf{s}\in\mathcal{S}}{\max}\negthinspace\sum_{k=1}^{K}\negthinspace s_{k}\negthinspace\bigl(\mathrm{tr}(\mathbf{W}^{\mathrm{H}}\mathbf{R}_{k}\negthinspace\mathbf{W})\negmedspace-\negmedspace2\mathrm{Re}(\mathbf{w}_{k}^{\mathrm{H}}\mathbf{q}_{k})\negmedspace-\negmedspace\mathsf{const}_{w,k}\bigr),\end{aligned}
\label{MR:W-Subproblem-with-s}
\end{equation}
where\textcolor{red}{{} }$\mathcal{S}=\left\{ \mathbf{s}\in\mathbb{R}^{K}\mid\mathbf{1}^{\mathrm{T}}\mathbf{s}=c,\mathbf{s}\geq\mathbf{0}\right\} $
with constant $c>0$. Then solutions to Problem \eqref{MaxMin:W-Subproblem}
can be obtained by solving Problem \eqref{MR:W-Subproblem-with-s}.
The objective of Problem \eqref{MR:W-Subproblem-with-s} is convex-concave
in $\mathbf{W}$ and $\mathbf{s}$, and the constraint sets $\mathcal{W}$
and $\mathcal{S}$ are both nonempty compact and convex. Hence, a
saddle point always exists for Problem \eqref{MR:W-Subproblem-with-s}
and then it can be swapped to be a maximin problem without affecting
its solutions \cite{rockafellar1970convex} as
\begin{equation}
\negthinspace\begin{aligned} & \underset{\mathbf{s}\in\mathcal{S}}{\mathrm{maximize}} &  & \negthickspace\negthickspace\negthickspace\negthinspace\underset{\mathbf{W}\in\mathcal{W}}{\min}\negthinspace\sum_{k=1}^{K}\negthinspace s_{k}\negthinspace\bigl(\mathrm{tr}(\mathbf{W}^{\mathrm{H}}\mathbf{R}_{k}\negthinspace\mathbf{W})\negthickspace-\negthickspace2\mathrm{Re}(\mathbf{w}_{k}^{\mathrm{H}}\mathbf{q}_{k}\negthinspace)\negthickspace-\negthickspace\mathsf{const}_{w,k}\negthinspace\bigr).\end{aligned}
\label{MR:W-Subproblem-MDA}
\end{equation}
Problem \eqref{MR:W-Subproblem-MDA} is a convex problem in variable
$\mathbf{s}$ with a simplex constraint ${\cal S}$, which can be
efficiently solved via many iterative algorithms. In this paper, we
adopt the mirror ascent algorithm (MAA) \cite{ben2001lectures,beck2003mirror}
outlined in the following. 
\begin{table}[H]
\noindent\fbox{\begin{minipage}[t]{1\columnwidth - 2\fboxsep - 2\fboxrule}%
\begin{center}
\uline{Mirror Ascent Algorithm (MAA)}
\par\end{center}
\textbf{Input:} function $h(\mathbf{s})$, initial feasible value
of $\mathbf{s}$.

\textbf{Repeat}

1. Calculate a subgradient $\mathbf{g}\in\partial h(\underline{\mathbf{s}})$
(the subdifferential of $h$ at $\underline{\mathbf{s}}$);

2. Update
\[
\mathbf{s}=\underset{}{\mathrm{arg}}\:\underset{\mathbf{s}\in\mathcal{S}}{\max}\bigl\{\mathbf{g}^{\mathrm{T}}\mathbf{s}-\frac{1}{\gamma}D_{\varphi}(\mathbf{s},\underline{\mathbf{s}})\bigr\},
\]
$\;$$\;$$\;$$\;$with $D_{\varphi}(\mathbf{s},\underline{\mathbf{s}})=\varphi(\mathbf{s})-\varphi(\underline{\mathbf{s}})-(\nabla\varphi(\underline{\mathbf{s}}))^{\mathrm{T}}(\mathbf{s}-\underline{\mathbf{s}});$

\textbf{Until} the value of the objective function $h(\mathbf{s})$
converges. \label{alg:MAA}%
\end{minipage}}
\end{table}
To solve Problem \eqref{MR:W-Subproblem-MDA} via MAA, we define input
function
\[
h(\mathbf{s})\negthinspace=\negthickspace\underset{\mathbf{W}\in\mathcal{W}}{\min}\sum_{k=1}^{K}\negthinspace s_{k}\negmedspace\left(\mathrm{tr}\negthinspace\left(\mathbf{W}^{\mathrm{H}}\mathbf{R}_{k}\negthinspace\mathbf{W}\right)\negthickspace-\negmedspace2\mathrm{Re}\left(\mathbf{w}_{k}^{\mathrm{H}}\mathbf{q}_{k}\right)\negthickspace-\negmedspace\mathsf{const}_{w,k}\right)\negthinspace,
\]
the subgradient $\mathbf{g}_{w}$ can be computed as
\[
[\mathbf{g}_{w}]_{k}=\mathrm{tr}\bigl(\mathbf{X}^{\mathrm{H}}\mathbf{R}_{k}\mathbf{X}\bigr)-2\mathrm{Re}\bigl(\bigl[\mathbf{X}\bigr]_{:,k}^{\mathrm{H}}\mathbf{q}_{k}\bigr)-\mathsf{const}_{w,k}
\]
with\textcolor{red}{{} }$\mathbf{X}\!=\underset{}{\mathrm{arg}}\underset{\mathbf{W}\in\mathcal{W}}{\min}\sum_{k=1}^{K}s_{k}\bigl(\mathrm{tr}(\mathbf{W}^{\mathrm{H}}\mathbf{R}_{k}\mathbf{W})-2\mathrm{Re}(\mathbf{w}_{k}^{\mathrm{H}}\mathbf{q}_{k})\bigr)$,
which is readily solved based on Lemma \ref{Prop: W-Subproblem-Solution}.
With the simplex space $\mathcal{S}$, we can choose
\[
\varphi(\mathbf{s})=\left\{ \begin{array}{ll}
\sum_{k=1}^{K}s_{k}\log s_{k} & \mathbf{s}\in\mathcal{S}\\
+\infty & \text{otherwise, }
\end{array}\right.
\]
and set stepsize $\gamma=r\frac{1}{\sqrt{t}}$ with constant $r>0$
at the $t$-th iteration, which leads to the following closed-form
update rule:
\begin{equation}
\mathbf{s}=c\frac{\underline{\mathbf{s}}\odot e^{-\gamma\mathbf{g}_{w}}}{\mathbf{1}^{\mathrm{T}}\left(\underline{\mathbf{s}}\odot e^{-\gamma\mathbf{g}_{w}}\right)}.\label{eq:MAA update}
\end{equation}

\subsection{The Update Step of $\boldsymbol{\Theta}$}

Given the iterate $\{\underline{\mathbf{W}},\underline{\boldsymbol{\Theta}}\}$,
the objective $f_{\mathsf{MR}}$ w.r.t. $\boldsymbol{\Theta}$ is
\[
f_{\mathsf{MR},\boldsymbol{\Theta}}(\boldsymbol{\Theta})=\underset{k}{\text{\ensuremath{\min}}}\,\log\bigl(1+\frac{\bigl|\underline{\mathbf{w}_{k}^{\mathrm{H}}}\mathbf{F}_{k}\boldsymbol{\theta}+\underline{\mathbf{w}_{k}^{\mathrm{H}}}\mathbf{h}_{k}^{\mathsf{d}}\bigr|^{2}}{\sum_{j,j\neq k}^{K}\bigl|\underline{\mathbf{w}_{j}^{\mathrm{H}}}\mathbf{F}_{k}\boldsymbol{\theta}+\underline{\mathbf{w}_{j}^{\mathrm{H}}}\mathbf{h}_{k}^{\mathsf{d}}\bigr|^{2}+\sigma^{2}}\bigr),
\]
where $\mathbf{F}_{k}=\mathbf{G}\mathrm{diag}\bigl(\mathbf{h}_{k}^{\mathrm{r}}\bigr)$.
Then based on Lemma \ref{Prop: Scalar MM}, a minorizing function
for $f_{\mathsf{MR},\boldsymbol{\Theta}}$ can be constructed as follows:
\[
\begin{aligned}f_{\mathsf{MR},\boldsymbol{\Theta}}^{\prime}(\boldsymbol{\Theta}) & =\underset{k}{\text{\ensuremath{\min}}}-\alpha_{k}\sum_{j=1}^{K}\bigl|\underline{\mathbf{w}_{j}^{\mathrm{H}}}\mathbf{F}_{k}\boldsymbol{\theta}+\underline{\mathbf{w}_{j}^{\mathrm{H}}}\mathbf{h}_{k}^{\mathsf{d}}\bigr|^{2}\\
 & \hspace{1cm}+2\mathrm{Re}(\beta_{k}\underline{\mathbf{w}_{k}^{\mathrm{H}}}\mathbf{F}_{k}\boldsymbol{\theta}+\beta_{k}\underline{\mathbf{w}_{k}^{\mathrm{H}}}\mathbf{h}_{k}^{\mathsf{d}})+\mathsf{const}_{w,k},
\end{aligned}
\]
which can be further rewritten as
\[
\begin{aligned} & f_{\mathsf{MR},\boldsymbol{\Theta}}^{\prime}(\boldsymbol{\Theta})=\underset{k}{\text{\ensuremath{\min}}}-\boldsymbol{\theta}^{\mathrm{H}}\mathbf{L}_{k}\boldsymbol{\mathbf{\theta}}-\alpha_{k}\sum_{j=1}^{K}\Bigl(2\mathrm{Re}\bigl(\boldsymbol{\theta}^{\mathrm{H}}\mathbf{F}_{k}^{\mathrm{H}}\underline{\mathbf{w}_{j}}\underline{\mathbf{w}_{j}^{\mathrm{H}}}\mathbf{h}_{k}^{\mathsf{d}}\bigr)\\
 & \hspace{0.55cm}+\bigl|\mathbf{h}_{\mathsf{d},k}^{\mathrm{H}}\underline{\mathbf{w}_{j}}\bigr|^{2}\Bigr)+2\mathrm{Re}(\beta_{k}\underline{\mathbf{w}_{k}^{\mathrm{H}}}\mathbf{F}_{k}\boldsymbol{\theta}+\beta_{k}\underline{\mathbf{w}_{k}^{\mathrm{H}}}\mathbf{h}_{k}^{\mathsf{d}})+\mathsf{const}_{w,k}
\end{aligned}
\]
with $\mathbf{L}_{k}\negthinspace=\negthinspace\alpha_{k}\sum_{j=1}^{K}\mathbf{F}_{k}^{\mathrm{H}}\underline{\mathbf{w}_{j}}\underline{\mathbf{w}_{j}^{\mathrm{H}}}\mathbf{F}_{k}.$
According to Lemma \ref{lem:quadratic_MM} and Lemma \ref{lem:Matrix Bound},
a piecewise linear minorizing function for $f_{\mathsf{MR},\boldsymbol{\Theta}}^{\prime}$
is further obtained as follows:
\[
f_{\mathsf{MR},\boldsymbol{\Theta}}^{\prime\prime}(\boldsymbol{\Theta})=\underset{k}{\text{\ensuremath{\min}}}\:-2\mathrm{Re}(\underline{\boldsymbol{\mathbf{\theta}}}^{\mathrm{H}}\mathbf{b}_{k})-\mathsf{const}_{\theta,k},
\]
where
\[
\mathbf{b}_{k}=\alpha_{k}\sum_{j=1}^{K}\mathbf{F}_{k}^{\mathrm{H}}\underline{\mathbf{w}_{j}}\underline{\mathbf{w}_{j}^{\mathrm{H}}}\mathbf{h}_{k}^{\mathsf{d}}-\beta_{k}^{*}\mathbf{F}_{k}^{\mathrm{H}}\underline{\mathbf{w}_{k}}+(\mathbf{L}-\lambda\mathbf{I})\underline{\boldsymbol{\mathbf{\theta}}}
\]
and $\mathsf{const}_{\theta,k}=N\lambda+\underline{\boldsymbol{\mathbf{\theta}}}^{\mathrm{H}}(\lambda\mathbf{I}-\mathbf{L})\underline{\boldsymbol{\mathbf{\theta}}}+\alpha_{k}\sum_{j=1}^{K}\bigl|\mathbf{h}_{\mathsf{d},k}^{\mathrm{H}}\underline{\mathbf{w}_{j}}\bigr|^{2}-2\mathrm{Re}(\beta_{k}\mathbf{h}_{\mathsf{d},k}^{\mathrm{H}}\underline{\mathbf{w}_{k}})-\mathsf{const}_{w,k}$
with $\lambda=\alpha_{k}\sum_{j=1}^{K}\bigl\Vert\underline{\mathbf{w}_{j}^{\mathrm{H}}}\mathbf{F}_{k}\bigr\Vert_{2}^{2}$.
Then the subproblem for $\boldsymbol{\theta}$ is given by
\[
\begin{aligned} & \underset{\boldsymbol{\Theta}\in\mathcal{C}}{\mathrm{minimize}} &  & \underset{k}{\max}\ 2\mathrm{Re}(\underline{\boldsymbol{\mathbf{\theta}}}^{\mathrm{H}}\mathbf{b}_{k})+\mathsf{const}_{\theta,k}.\end{aligned}
\]
As in the previous section, the discrete maximum form of the objective
can be transformed into a continuous maximum:
\begin{equation}
\begin{aligned} & \underset{\boldsymbol{\Theta}\in\mathcal{C}}{\mathrm{minimize}} &  & \negthinspace\negthinspace\underset{\mathbf{s}\in\mathcal{S}}{\max}\ \sum_{k=1}^{K}s_{k}\bigl(2\mathrm{Re}(\underline{\boldsymbol{\mathbf{\theta}}}^{\mathrm{H}}\mathbf{b}_{k})+\mathsf{const}_{\theta,k}\bigr).\end{aligned}
\label{MaxMin:Theta-Problem-Simplex}
\end{equation}

\begin{prop}
\label{Lemma: Relaxed Theta Constraint}A saddle point exists for
Problem \eqref{MaxMin:Theta-Problem-Simplex} and can be obtained
by solving the following relaxed problem
\begin{equation}
\begin{aligned} & \underset{\boldsymbol{\Theta}\in\mathcal{C}_{\mathsf{relaxed}}}{\mathrm{minimize}} &  & \negthinspace\negthinspace\underset{\mathbf{s}\in\mathcal{S}}{\max}\ \sum_{k=1}^{K}s_{k}\bigl(2\mathrm{Re}(\underline{\boldsymbol{\mathbf{\theta}}}^{\mathrm{H}}\mathbf{b}_{k})+\mathsf{const}_{\theta,k}\bigr),\end{aligned}
\label{MaxMin-Theta-Subproblem-Relaxed}
\end{equation}
where $\mathcal{C}_{\mathsf{relaxed}}\negmedspace=\negmedspace\bigl\{\boldsymbol{\theta}\mid\negmedspace\boldsymbol{\theta}\in\mathbb{C}^{N},\bigl|[\boldsymbol{\theta}]_{j}\bigr|\leq1,\forall j=1,\ldots,N\bigr\}.$
\end{prop}
\begin{IEEEproof}
The detailed proof is given in Appendix \ref{Appendix: Relaxed Theta Constraint}.
\end{IEEEproof}
We can swap the order of minimization and maximization as before,
and Problem \eqref{MaxMin-Theta-Subproblem-Relaxed} is equivalently
transformed to
\[
\underset{\mathbf{s}\in\mathcal{S}}{\mathrm{maximize}}\ \underset{\boldsymbol{\Theta}\in\mathcal{C}_{\mathsf{relaxed}}}{\min}\ \sum_{k=1}^{K}s_{k}\bigl(2\mathrm{Re}(\underline{\boldsymbol{\mathbf{\theta}}}^{\mathrm{H}}\mathbf{b}_{k})+\mathsf{const}_{\theta,k}\bigr),
\]
which can be solved via MAA with the input function
\[
h(\mathbf{s})=\underset{\boldsymbol{\Theta}\in\mathcal{C}_{\mathsf{relaxed}}}{\min}\ \sum_{k=1}^{K}s_{k}\bigl(2\mathrm{Re}(\underline{\boldsymbol{\mathbf{\theta}}}^{\mathrm{H}}\mathbf{b}_{k})+\mathsf{const}_{\theta,k}\bigr).
\]
Then the corresponding subgradient $\mathbf{g}_{\theta}$ can be calculated
as
\[
[\mathbf{g}_{\theta}]_{k}=2\mathrm{Re}(\mathbf{x}^{\mathrm{H}}\mathbf{b}_{k})+\mathsf{const}_{\theta,k},
\]
where $\mathbf{x}$ is computed from the following equation 
\begin{align*}
\mathbf{x} & =\mathrm{diag}\Bigl(\mathrm{arg}\underset{\boldsymbol{\Theta}\in\mathcal{C}_{\mathsf{relaxed}}}{\min}\ \sum_{k=1}^{K}2s_{k}\mathrm{Re}(\underline{\boldsymbol{\theta}}^{\mathrm{H}}\mathbf{b}_{k})\Bigr)\\
 & =e^{\mathfrak{j}\mathrm{ang}\left(-\sum_{k=1}^{K}s_{k}\mathbf{b}_{k}\right)},
\end{align*}
where the last line can be proved in a similar way as Lemma \ref{Prop: Theta-Subproblem-Solution}.
The update rules in MAA are chosen the same as \eqref{eq:MAA update}.

In summary, to solve the MR maximization problem via BMM, the two
variable blocks, i.e., $\mathbf{W}$ and $\boldsymbol{\Theta}$, will
be updated cyclically until some convergence criterion is met. The
overall algorithm is summarized in Algorithm \ref{MR:Algorithm} with
its convergence and complexity analyses given in Section \ref{sec:Convergence-and-Complexity}.

\begin{algorithm}[t]
\caption{The BMM Algorithm for Problem \eqref{MaxMin:Formulation}. \label{MR:Algorithm}}

\textbf{Input:} $\{\mathbf{h}_{i}^{\mathsf{r}}\}$, $\{\mathbf{h}_{i}^{\mathsf{d}}\}$,
$\mathbf{G}$, $P$, $\sigma^{2}$, initial feasible values of $\mathbf{W}$
and $\boldsymbol{\Theta}$.

\textbf{Repeat}

1. Update $\mathbf{W}$ by solving Prob. \eqref{MR:W-Subproblem-MDA}
via MAA;

2. Update $\boldsymbol{\Theta}$ by solving Prob. \eqref{MaxMin:Theta-Problem-Simplex}
via MAA;

\textbf{Until} the value of the objective function converges.
\end{algorithm}

\section{Sum-Rate Maximization for \protect \\
RIS-Aided MIMO Device-to-Device Networks \label{sec:SR}}

\subsection{System Model and Problem Formulation}

In this section, the RIS-aided MIMO D2D system is considered, where
$K$ transceiver pairs transmit multiple data streams. We assume the
$k$-th ($k=1,\ldots,K$) transmitter and receiver are equipped with
$M_{k}^{\mathsf{t}}$ and $M_{k}^{\mathsf{r}}$ antennas. Let $\mathbf{H}_{i,j}^{\mathsf{d}}\in\mathbb{C}^{M_{i}^{\mathsf{r}}\times M_{j}^{\mathsf{t}}}$,
$\mathbf{H}_{i}^{\mathsf{r}}\in\mathbb{C}^{M_{i}^{\mathsf{r}}\times N}$,
and $\mathbf{G}_{j}\in\mathbb{C}^{N\times M_{j}^{\mathsf{t}}}$ be
the direct channel between the $i$-th receiver and the $j$-th transmitter,
the channel in the reflection link between the RIS and the $i$-th
receiver, and the channel between the $j$-th transmitter and the
RIS, respectively. Other notations are defined the same as in previous
sections. Then the received signal at the $k$-th receiver is given
by
\begin{align*}
\mathbf{y}_{k} & =\underset{\text{desired signal}}{\underbrace{\left(\mathbf{H}_{k}^{\mathsf{r}}\bm{\Theta}\mathbf{G}_{k}+\mathbf{H}_{k,k}^{\mathsf{d}}\right)\mathbf{W}_{k}\mathbf{s}_{k}}}\\
 & \hspace{2.7cm}+\underset{\text{interference plus noise}}{\underbrace{\sum_{j,j\ne k}\left(\mathbf{H}_{k}^{\mathsf{r}}\bm{\Theta}\mathbf{G}_{j}+\mathbf{H}_{k,j}^{\mathsf{d}}\right)\mathbf{W}_{j}\mathbf{s}_{j}+\mathbf{e}_{k}}},
\end{align*}
where $\mathbf{W}_{k}\in\mathbb{C}^{M_{k}^{\mathsf{t}}\times d_{k}}$
denotes the beamforming matrix, $\mathbf{\bm{s}}_{k}\in\mathbb{C}^{d_{k}}$
is the symbol vector for the $k$-th transceiver pair, and $\mathbf{e}_{k}\sim\mathcal{CN}(\bm{0},\sigma^{2}\mathbf{I})$
represents the noise. 

The target is to maximize the SR of the system by jointly designing
the beamforming matrices $\left\{ \mathbf{W}_{i}\right\} $ and the
phase shift matrix $\boldsymbol{\Theta}$. The data rate at the $k$-th
receiver is defined as
\begin{align*}
\mathsf{R}_{k} & =\log\det\bigl(\mathbf{I}+\mathbf{W}_{k}^{\mathrm{H}}(\mathbf{H}_{k}^{\mathsf{r}}\bm{\Theta}\mathbf{G}_{k}+\mathbf{H}_{k,k}^{\mathsf{d}})^{\mathrm{H}}\mathbf{T}_{k}^{-1}\\
 & \hspace{4.4cm}\times(\mathbf{H}_{k}^{\mathsf{r}}\bm{\Theta}\mathbf{G}_{k}+\mathbf{H}_{k,k}^{\mathsf{d}})\mathbf{W}_{k}\bigr),
\end{align*}
where the interference-plus-noise term is
\[
\mathbf{T}_{k}\negthinspace=\negthinspace\negthinspace\sum_{j,j\ne k}\negthinspace\bigl(\mathbf{H}_{k}^{\mathsf{r}}\bm{\Theta}\mathbf{G}_{j}\negthinspace+\negthinspace\mathbf{H}_{k,j}^{\mathsf{d}}\bigr)\mathbf{W}_{j}\mathbf{W}_{j}^{\mathrm{H}}\bigl(\mathbf{H}_{k}^{\mathsf{r}}\bm{\Theta}\mathbf{G}_{j}\negthinspace+\negthinspace\mathbf{H}_{k,j}^{\mathsf{d}}\bigr)^{\mathrm{H}}\negthinspace+\negthinspace\sigma^{2}\mathbf{I}.
\]
 Then the SR maximization problem is formulated as follows:
\begin{equation}
\begin{aligned} & \underset{\{\mathbf{W}_{i}\},\bm{\Theta}}{\mathrm{maximize}} &  & f_{\mathsf{SR}}\left(\{\mathbf{W}_{i}\},\bm{\Theta}\right)=\sum_{k=1}^{K}\mathsf{R}_{k}\\
 & \mathrm{subject\ to} &  & \mathbf{W}_{i}\in\mathcal{W}_{i},\forall i=1,\ldots,K,\ \bm{\Theta}\in\mathcal{C},
\end{aligned}
\tag{SRMax}\label{SR:Formulation}
\end{equation}
where the transmit power constraint for the $i$-th transmitter is
\[
\mathcal{W}_{i}=\left\{ \mathbf{W}_{i}\mid\bigl\Vert\mathbf{W}_{i}\bigr\Vert_{\mathrm{F}}^{2}\le P_{i}\right\} .
\]
Problem \eqref{SR:Formulation} is nonconvex and NP-hard. In the following,
a BMM-based algorithm is developed for problem resolution.

\subsection{The Update Step of $\left\{ \mathbf{W}_{i}\right\} $}

Given the iterate $\{\{\underline{\mathbf{W}_{i}}\},\underline{\boldsymbol{\Theta}}\}$,
the $f_{\mathsf{SR}}$ w.r.t. $\left\{ \mathbf{W}_{i}\right\} $ is
\begin{align*}
f_{\mathsf{SR},\left\{ \mathbf{W}_{i}\right\} }(\left\{ \mathbf{W}_{i}\right\} ) & =\negthinspace\sum_{k=1}^{K}\log\det\left(\mathbf{I}+\mathbf{W}_{k}^{\mathrm{H}}\mathbf{H}_{k,k}^{\mathrm{H}}\mathbf{T}_{k}^{-1}\mathbf{H}_{k,k}\mathbf{W}_{k}\right),
\end{align*}
where $\mathbf{H}_{k,j}=\mathbf{H}_{k}^{\mathsf{r}}\underline{\bm{\Theta}}\mathbf{G}_{j}+\mathbf{H}_{k,j}^{\mathsf{d}}$. 
\begin{prop}
\label{Prop: Matrix MM} Function $\log\det\left(\mathbf{I}+\mathbf{X}^{\mathrm{H}}\mathbf{Y}^{-1}\mathbf{X}\right)$
with $\mathbf{X}\in\mathbb{C}^{M\times N}$ and $\mathbf{Y}\succ\mathbf{0}$
is minorized at $\left(\underline{\mathbf{X}},\underline{\mathbf{Y}}\right)$
as follows:
\[
\begin{aligned}\log\det(\mathbf{I}+\mathbf{X}^{\mathrm{H}}\mathbf{Y}^{-1}\mathbf{X})\geq-\mathrm{tr}\bigl((\underline{\mathbf{Y}}+\underline{\mathbf{X}}\underline{\mathbf{X}}^{\mathrm{H}})^{-1}\underline{\mathbf{X}}\hspace{1.7cm}\\
\times(\mathbf{I}+\underline{\mathbf{X}}^{\mathrm{H}}\underline{\mathbf{Y}}^{-1}\underline{\mathbf{X}})\underline{\mathbf{X}}^{\mathrm{H}}(\underline{\mathbf{Y}}+\underline{\mathbf{X}}\underline{\mathbf{X}}^{\mathrm{H}})^{-1}(\mathbf{Y}+\mathbf{X}\mathbf{X}^{\mathrm{H}})\bigr)\\
+2{\rm {\rm Re}}\bigl({\rm tr}\bigl((\mathbf{I}+\underline{\mathbf{X}}^{\mathrm{H}}\underline{\mathbf{Y}}^{-1}\underline{\mathbf{X}})\underline{\mathbf{X}}^{\mathrm{H}}(\underline{\mathbf{Y}}+\underline{\mathbf{X}}\underline{\mathbf{X}}^{\mathrm{H}})^{-1}\mathbf{X}\bigr)\bigr)\\
+\log\det(\mathbf{I}+\underline{\mathbf{X}}^{\mathrm{H}}\underline{\mathbf{Y}}^{-1}\underline{\mathbf{X}})-{\rm tr}(\underline{\mathbf{X}}^{\mathrm{H}}\underline{\mathbf{Y}}^{-1}\underline{\mathbf{X}}).
\end{aligned}
\]
\end{prop}
\begin{IEEEproof}
The proof is given in Appendix \ref{Appendix: Matrix MM Proof}.
\end{IEEEproof}
Based on Lemma \ref{Prop: Matrix MM}, taking $\mathbf{H}_{k,k}\mathbf{W}_{k}$
as $\mathbf{X}$ and $\mathbf{T}_{k}$ as $\mathbf{Y}$, a minorizing
function for $f_{\mathsf{SR},\left\{ \mathbf{W}_{i}\right\} }$ is
computed as follows:
\begin{align*}
f_{\mathsf{SR},\{\mathbf{W}_{i}\}}^{\prime}\bigl(\{\mathbf{W}_{i}\}\bigr)= & \sum_{k=1}^{K}\Bigl(-\mathrm{tr}\bigl(\mathbf{A}_{k}\sum_{j=1}^{K}\mathbf{H}_{k,j}\mathbf{W}_{j}\mathbf{W}_{j}^{\mathrm{H}}\mathbf{H}_{k,j}^{\mathrm{H}}\bigr)\\
 & \hspace{15bp}+2{\rm \mathrm{Re}}\bigl({\rm tr}(\mathbf{B}_{k}\mathbf{H}_{k,k}\mathbf{W}_{k})\bigr)\Bigr)+\mathsf{const}_{w},
\end{align*}
where $\mathbf{A}_{k}=(\mathbf{T}_{k}+\mathbf{H}_{k,k}\underline{\mathbf{W}_{k}}\underline{\mathbf{W}_{k}^{\mathrm{H}}}\mathbf{H}_{k,k}^{\mathrm{H}})^{-1}\mathbf{H}_{k,k}\underline{\mathbf{W}_{k}}\mathbf{B}_{k}$
with $\mathbf{B}_{k}=(\mathbf{I}+\underline{\mathbf{W}_{k}^{\mathrm{H}}}\mathbf{H}_{k,k}^{\mathrm{H}}\mathbf{T}_{k}^{-1}\mathbf{H}_{k,k}\underline{\mathbf{W}_{k}})\underline{\mathbf{W}_{k}^{\mathrm{H}}}\mathbf{H}_{k,k}^{\mathrm{H}}(\mathbf{T}_{k}+\mathbf{H}_{k,k}\underline{\mathbf{W}_{k}}\times$
$\underline{\mathbf{W}_{k}^{\mathrm{H}}}\mathbf{H}_{k,k}^{\mathrm{H}}){}^{-1}$,
and $\mathsf{const}_{w}=\sum_{k=1}^{K}\bigl(\underline{\mathsf{R}_{k}}-\mathrm{tr}(\underline{\mathbf{W}_{k}^{\mathrm{H}}}\mathbf{H}_{k,k}^{\mathrm{H}}\mathbf{T}_{k}^{-1}\times$
$\mathbf{H}_{k,k}\underline{\mathbf{W}_{k}})$ $-\sigma^{2}\mathrm{tr}(\mathbf{A}_{k})\bigr)$.
Ignoring the constant terms, we obtain the resultant subproblem for
$\mathbf{W}_{k}$ as follows:
\begin{equation}
\begin{aligned} & \underset{\mathbf{W}_{k}\in\mathcal{W}_{k}}{\mathrm{minimize}} &  & \mathrm{tr}(\mathbf{W}_{k}^{\mathrm{H}}\mathbf{R}_{k}\mathbf{W}_{k})-2\mathrm{Re}\bigl(\mathrm{tr}(\mathbf{W}_{k}^{\mathrm{H}}\mathbf{Q}_{k})\bigr),\end{aligned}
\label{SR:W-Subproblem}
\end{equation}
where $\mathbf{R}_{k}=\sum_{j=1}^{K}\mathbf{H}_{j,k}^{\mathrm{H}}\mathbf{A}_{k}\mathbf{H}_{j,k}$
and $\mathbf{Q}_{k}=\mathbf{H}_{k,k}^{\mathrm{H}}\mathbf{B}_{k}^{\mathrm{H}}$.
In Problem \eqref{SR:W-Subproblem}, $\mathbf{W}_{k}$ becomes separable.
Then $\left\{ \mathbf{W}_{i}\right\} $ can be updated separately
in parallel by solving \eqref{SR:W-Subproblem} via Lemma \ref{Prop: W-Subproblem-Solution}.

\subsection{The Update Step of $\boldsymbol{\Theta}$}

\begin{figure*}[tbh]
\begin{equation}
\begin{aligned}f_{\mathsf{SR},\boldsymbol{\Theta}}(\boldsymbol{\Theta}) & =\sum_{k=1}^{K}\log\det\Biggl(\mathbf{I}+\underline{\mathbf{W}_{k}^{\mathrm{H}}}\Bigl(\mathbf{H}_{k}^{\mathsf{r}}\odot\bigl(\boldsymbol{\theta}^{\mathrm{T}}\otimes\mathbf{1}\bigr)\mathbf{G}_{k}+\mathbf{H}_{k,k}^{\mathsf{d}}\Bigr)^{\mathrm{H}}\biggl(\sum_{j,j\ne k}\Bigl(\mathbf{H}_{k}^{\mathsf{r}}\odot\bigl(\boldsymbol{\theta}^{\mathrm{T}}\otimes\mathbf{1}\bigr)\mathbf{G}_{j}+\mathbf{H}_{k,j}^{\mathsf{d}}\Bigr)\underline{\mathbf{W}_{j}}\\
 & \hspace{2.9cm}\times\underline{\mathbf{W}_{j}^{\mathrm{H}}}\Bigl(\mathbf{H}_{k}^{\mathsf{r}}\odot\bigl(\boldsymbol{\theta}^{\mathrm{T}}\otimes\mathbf{1}\bigr)\mathbf{G}_{k}+\mathbf{H}_{k,k}^{\mathsf{d}}\Bigr)^{\mathrm{H}}+\sigma^{2}\mathbf{I}\biggr)^{-1}\Bigl(\mathbf{H}_{k}^{\mathsf{r}}\odot\bigl(\boldsymbol{\theta}^{\mathrm{T}}\otimes\mathbf{1}\bigr)\mathbf{G}_{j}+\mathbf{H}_{k,j}^{\mathsf{d}}\Bigr)\underline{\mathbf{W}_{k}}\biggr)\Biggr).
\end{aligned}
\label{SpanColumnFormulation:f_SR,Theta}
\end{equation}
\rule[0.5ex]{1\textwidth}{1pt}
\end{figure*}

\begin{figure*}[tbh]
\begin{equation}
\begin{aligned}f_{\mathsf{SR},\boldsymbol{\Theta}}^{\prime}(\boldsymbol{\Theta}) & =\sum_{k=1}^{K}\Biggl(-\mathrm{tr}\biggl(\mathbf{A}_{k}\sum_{j=1}^{K}\Bigl(\mathbf{H}_{k}^{\mathsf{r}}\odot\bigl(\boldsymbol{\theta}^{\mathrm{T}}\otimes\mathbf{1}\bigr)\mathbf{G}_{j}+\mathbf{H}_{kj}^{\mathsf{d}}\Bigr)\underline{\mathbf{W}_{k}}\underline{\mathbf{W}_{k}^{\mathrm{H}}}\Bigl(\mathbf{H}_{k}^{\mathsf{r}}\odot\bigl(\boldsymbol{\theta}^{\mathrm{T}}\otimes\mathbf{1}\bigr)\mathbf{G}_{j}+\mathbf{H}_{kj}^{\mathsf{d}}\Bigr)^{\mathrm{H}}\biggr)\\
 & \hspace{4.1cm}+2{\rm \mathrm{Re}}\biggl(\mathrm{tr}\Bigl(\mathbf{G}_{k}\underline{\mathbf{W}_{k}}\mathbf{B}_{k}\mathbf{H}_{k}^{\mathsf{r}}\odot\bigl(\boldsymbol{\theta}^{\mathrm{T}}\otimes\mathbf{1}\bigr)\Bigr)\biggr)+2{\rm \mathrm{Re}}\Bigl(\mathrm{tr}\bigl(\mathbf{B}_{k}\mathbf{H}_{kk}^{\mathsf{d}}\underline{\mathbf{W}_{k}}\bigr)\Bigr)\Biggr)+\mathsf{const}_{w}.
\end{aligned}
\label{SpanColumnFormulation:f^prime_SR,Theta}
\end{equation}
\rule[0.5ex]{1\textwidth}{1pt}
\end{figure*}
Given iterate $\{\{\underline{\mathbf{W}_{i}}\},\underline{\bm{\Theta}}\}$,
$f_{\mathsf{SR}}$ w.r.t. $\bm{\Theta}$ is given in Eq. \eqref{SpanColumnFormulation:f_SR,Theta}.
Then a minorizing function for $f_{\mathsf{SR},\boldsymbol{\Theta}}$
can be constructed based on Lemma \ref{Prop: Matrix MM} as in Eq.
\eqref{SpanColumnFormulation:f^prime_SR,Theta}, which is written
compactly as
\[
\begin{aligned}f_{\mathsf{SR},\boldsymbol{\Theta}}^{\prime}(\bm{\Theta})= & -\sum_{k=1}^{K}\mathrm{tr}\bigl(\mathbf{K}_{1}\odot(\boldsymbol{\theta}^{\mathrm{T}}\otimes\mathbf{1})\mathbf{K}_{2}\odot(\boldsymbol{\theta}^{\mathrm{T}}\otimes\mathbf{1})^{\mathrm{H}}\bigr)\\
 & \hspace{40bp}+2{\rm \mathrm{Re}}\Bigl(\mathrm{tr}\bigl(\mathbf{N}\odot(\boldsymbol{\theta}^{\mathrm{T}}\otimes\mathbf{1})\bigr)\Bigr)+\mathsf{const}_{\theta},
\end{aligned}
\]
where
\begin{align*}
 & \mathbf{K}_{1}=\mathbf{H}_{\mathsf{r},k}^{\mathrm{H}}\mathbf{A}_{k}\mathbf{H}_{\mathsf{r},k},\hspace{1.5cm}\mathbf{K}_{2}=\sum_{j=1}^{K}\mathbf{G}_{j}\underline{\mathbf{W}_{j}}\underline{\mathbf{W}_{j}^{\mathrm{H}}}\mathbf{G}_{j}^{\mathrm{H}},\\
 & \mathbf{N}\negthinspace=\negthinspace\sum_{k=1}^{K}\mathbf{G}_{k}\underline{\mathbf{W}_{k}}\mathbf{B}_{k}\mathbf{H}_{\mathsf{r},k}\negthinspace-\negthinspace\sum_{k=1}^{K}\sum_{j=1}^{K}\mathbf{G}_{j}\underline{\mathbf{W}_{j}}\underline{\mathbf{W}_{j}^{\mathrm{H}}}\mathbf{H}_{\mathsf{d},kj}^{\mathrm{H}}\mathbf{A}_{k}\mathbf{H}_{\mathsf{r},k},
\end{align*}
and $\mathsf{const}_{\theta}=\negthinspace\sum_{k=1}^{K}\Bigl(-\sum_{j=1}^{K}\mathrm{tr}\bigl(\mathbf{A}_{k}(\mathbf{H}_{\mathsf{d},kj}\underline{\mathbf{W}_{j}}\underline{\mathbf{W}_{j}^{\mathrm{H}}}\mathbf{H}_{\mathsf{d},kj}^{\mathrm{H}})\bigr)+2{\rm \mathrm{Re}}\bigl(\mathrm{tr}(\mathbf{B}_{k}\mathbf{H}_{\mathsf{d},kk}\underline{\mathbf{W}_{k}})\bigr)\Bigr)+\mathsf{const}_{w}$.
After some mathematical manipulation, the $f_{\mathsf{SR},\boldsymbol{\Theta}}^{\prime}(\bm{\Theta})$
can be further rewritten as
\[
f_{\mathsf{SR},\boldsymbol{\Theta}}^{\prime}(\bm{\Theta})=-\boldsymbol{\theta}^{\mathrm{H}}\mathbf{L}\boldsymbol{\theta}+2{\rm \mathrm{Re}}\bigl({\rm diag}\left(\mathbf{N}\right)^{\mathrm{T}}\boldsymbol{\theta}^{*}\bigr)+\mathsf{const}_{\theta}
\]
with $\mathbf{L}=\bigl(\sum_{k=1}^{K}\mathbf{K}_{1}\bigr)\odot\mathbf{K}_{2}^{\mathrm{T}}$.

Based on Lemma \ref{lem:quadratic_MM}, a linear minorizing function
for $f_{\mathsf{SR},\boldsymbol{\Theta}}^{\prime}$ can be further
obtained as follows:
\[
\begin{aligned}f_{\mathsf{SR},\boldsymbol{\Theta}}^{\prime\prime}(\bm{\Theta}) & =-2\mathrm{Re}\left(\boldsymbol{\theta}^{\mathrm{T}}\mathbf{b}\right)-N\lambda-\underline{\boldsymbol{\theta}}^{\mathrm{H}}(\lambda\mathbf{I}-\mathbf{L})\underline{\boldsymbol{\theta}}+\mathsf{const}_{\theta},\end{aligned}
\]
where $\mathbf{b}=\mathrm{diag}\bigl(\underline{\boldsymbol{\theta}}^{\mathrm{H}}(\mathbf{L}-\lambda\mathbf{I})^{\mathrm{H}}-\mathbf{N}\bigr)$
with $\lambda$ being the largest eigenvalue of $\mathbf{L}$. Then
the subproblem for $\boldsymbol{\Theta}$ is given by
\begin{equation}
\begin{aligned} & \underset{\boldsymbol{\Theta}\in\mathcal{C}}{\mathrm{minimize}} &  & \mathrm{Re}\bigl(\boldsymbol{\theta}^{\mathrm{T}}\mathbf{b}\bigr).\end{aligned}
\label{SR:Theta-Subproblem}
\end{equation}
Problem \eqref{SR:Theta-Subproblem} is separable over different elements
in $\boldsymbol{\theta}$ and can be solved in parallel via Lemma
\ref{Prop: Theta-Subproblem-Solution}.

In summary, based on BMM the variable blocks $\{\mathbf{W}_{i}\},\boldsymbol{\Theta}$
will be updated cyclically until some convergence criterion is met.
The overall algorithm is summarized in Algorithm \ref{SR:Algorithm}
with its convergence and complexity analyses given in Section \ref{sec:Convergence-and-Complexity}.
\begin{algorithm}[t]
\caption{The BMM Algorithm for Problem \eqref{SR:Formulation}. \label{SR:Algorithm}}

\textbf{Input:} $\{\mathbf{H}_{i}^{\mathsf{r}}\}$, $\{\mathbf{H}_{ij}^{\mathsf{d}}\}$,
$\{\mathbf{G}_{j}\}$, $\{P_{i}\}$, $\sigma^{2}$, initial feasible
values of $\{\mathbf{W}_{i}\}$ and $\boldsymbol{\Theta}$.

\textbf{Repeat}

1. $\negthinspace\negthinspace$Update $\negthinspace\{\mathbf{W}_{i}\}\negthinspace$
successively $\negthinspace$by $\negthinspace$solving $\negthinspace$Prob.$\,$\eqref{SR:W-Subproblem}$\negthinspace$
via $\negthinspace$Lemma $\negthinspace$\ref{Prop: W-Subproblem-Solution};

2. $\negthinspace\negthinspace$Update $\boldsymbol{\Theta}$ by solving
Prob. \eqref{SR:Theta-Subproblem} via Lemma \ref{Prop: Theta-Subproblem-Solution};

\textbf{Until} the value of the objective function converges.
\end{algorithm}

\section{Convergence and Complexity Analysis \label{sec:Convergence-and-Complexity}}

\subsection{Convergence Analysis}
\begin{thm}
\label{WSR: Theorem Convergence} Every limit point generated by Algorithm
\ref{WSR:Algorithm}, Algorithm \ref{MR:Algorithm}, or Algorithm
\ref{SR:Algorithm} is a stationary/KKT point of Prob. \eqref{WSR:Formulation},
Prob. \eqref{MaxMin:Formulation}, or Prob. \eqref{SR:Formulation},
respectively.
\end{thm}
\begin{IEEEproof}
The detailed proof is given in Appendix \ref{WSR:Proof-for-Convergence}.
\end{IEEEproof}

\subsection{Complexity Analysis}

For simplicity, we assume $M$, $\{M_{i}^{\mathsf{r}}\}$, $\{M_{i}^{\mathsf{t}}\}$,
$\{N_{i}\}$, and $\{d_{i}\}$ are of the same order, uniformly denoted
as $M$, and $L$ and $\{I_{i}\}$ are of the same order, uniformly
denoted as $L$. In the following, we analyze the per-iteration computational
complexities of the proposed BMM algorithms, where ``one iteration''
is defined when all variable blocks are updated once. 

\subsubsection{Complexity of the BMM algorithm for Prob. \eqref{WSR:Formulation}}

The complexity mainly comes from the calculation of matrices $\mathbf{R}$,
$\mathbf{L}_{i}$, as well as the inverse and eigendecomposition operations
of matrix $\mathbf{R}$, whose complexities are of orders $\mathcal{O}(LKM^{2})$,
$\mathcal{O}(LK^{2}M^{2}+L^{2}KM^{2}+L^{2}M^{3})$, and $\mathcal{O}(M^{3})$,
respectively. Therefore, the total complexity of the BMM algorithm
is $\mathcal{O}(LK^{2}M^{2}+L^{2}KM^{2}+L^{2}M^{3})$. Particularly,
for the single-hop case, i.e., $L=1$, the complexity reduces to $\mathcal{O}(K^{2}M^{2}+M^{3})$.

\subsubsection{Complexity of the BMM algorithm for Prob. \eqref{MaxMin:Formulation}}

The complexity mainly comes from the calculation of matrices $\mathbf{R}$,
$\mathbf{L}$, as well as the inverse and eigendecomposition operations
of matrix $\mathbf{R}$, whose complexities are $\mathcal{O}(KM^{2})$,
$\mathcal{O}(K^{2}M^{2})$, and $\mathcal{O}(M^{3})$, respectively.
Therefore, the total complexity of the BMM algorithm is $\mathcal{O}(\mathbb{I}_{w}KM^{2}+\mathbb{I}_{\theta}K^{2}M^{2}+\mathbb{I}_{w}M^{3})$,
where $\mathbb{I}_{w}$ and $\mathbb{I}_{\theta}$ are the iteration
times of the MAA algorithms for $\mathbf{W}$-block and $\boldsymbol{\Theta}$-block
respectively. 

\subsubsection{Complexity of the BMM algorithm for Prob. \eqref{SR:Formulation}}

The complexity mainly comes from the calculation of matrices $\mathbf{R}$
and $\mathbf{L}$, whose complexity are $\mathcal{O}(KM^{3})$ and
$\mathcal{O}(K^{2}M^{3})$. Therefore, the total complexity is $\mathcal{O}(K^{2}M^{3})$.

\section{Extensions and Generalizations\label{sec:Extensions-and-Generalizations}}

\subsection{A Further Majorization Step for The $\mathbf{W}$-Block}

In previous sections, solving the minimization problem for $\mathbf{W}$-block
relies on a one-dimensional line search method as given in Lemma \ref{Prop: W-Subproblem-Solution}.
Leveraging on the MM method, a closed-form solution free of line search
can also be obtained. In the following, we show this result by taking
the WSR maximization problem in Section \ref{sec:WSR} as an example,
while similar results hold for Problem \eqref{MaxMin:Formulation}
and Problem \eqref{SR:Formulation} as well. 

Further majorizing the objective in \eqref{WSR:W-Subproblem-vector-form}
based on Lemma \ref{lem:quadratic_MM} and Lemma \ref{lem:Matrix Bound},
then the $\mathbf{W}$-block subproblem in WSR maximization becomes
\begin{equation}
\begin{aligned} & \underset{\mathbf{W}\in\mathcal{W}}{\mathrm{minimize}} &  & \bigl\Vert\mathbf{W}-\boldsymbol{\Pi}\bigr\Vert_{\mathrm{F}}^{2},\end{aligned}
\label{eq:Further majorization for W}
\end{equation}
where
\[
\left[\boldsymbol{\Pi}\right]_{:,k}=\frac{[\mathbf{Q}]_{:,k}-(\mathbf{R}-\sum_{k=1}^{K}\omega_{k}\alpha_{k}\left\Vert \mathbf{h}_{k}\right\Vert _{2}^{2}\mathbf{I})\underline{\mathbf{w}_{k}}}{\sum_{k=1}^{K}\omega_{k}\alpha_{k}\left\Vert \mathbf{h}_{k}\right\Vert _{2}^{2}}.
\]
By solving the KKT system of Problem \eqref{eq:Further majorization for W},
the optimal solution can be obtained in closed-form as follows:
\[
\mathbf{W}^{\star}=\begin{cases}
\boldsymbol{\Pi} & \text{if }\bigl\Vert\boldsymbol{\Pi}\bigr\Vert_{\mathrm{F}}^{2}\leq P\\
\sqrt{\frac{P}{||\boldsymbol{\Pi}||_{\mathrm{F}}^{2}}}\mathbf{K} & \mathrm{\text{otherwise}}.
\end{cases}
\]
Therefore, with an additional majorization step, a cheaper result
is obtained leading to a lower per-iteration computational complexity.
When we apply the closed-form updating rule of $\mathbf{W}$, the
inverse and eigendecomposition operations of $\mathbf{R}$ are avoided
and the per-iteration complexity becomes $\mathcal{O}(K^{2}M^{2})$.
However, in practice, whether a further majorization step like this
brings better convergence property or not depends on the specific
problem and the characterization of the inner-loop residual error,
in that it solves a looser surrogate problem to trade for a closed-form
solution and, hence, it may requires more iterations for convergence
compared with its counterpart. 

\subsection{General Power Constraint for The $\mathbf{W}$-Block\label{subsec:General-Power-Constraint}}

Beyond the total power constraint, the proposed algorithm can also
handle more general power constraints \cite{wang2018optimal} like
\begin{equation}
\mathrm{tr}(\boldsymbol{\Omega}_{j}\mathbf{W}\mathbf{W}^{\mathrm{H}})\leq P_{j},\ \forall j=1,\ldots,J,\label{Extentions: General Power Constraints}
\end{equation}
where $\boldsymbol{\Omega}_{1},\ldots,\boldsymbol{\Omega}_{J}\succeq\mathbf{0}$
are application-oriented. This general power constraint reduces to
the total power limit considered considered in the previous sections
when $J=1$ and $\boldsymbol{\Omega}_{1}=\mathbf{I}$. It is also
commonly used to model a more realistic case that each antenna is
equipped with an independent power amplifier, where we can limit the
per-antenna power by setting $J=M$ and $\boldsymbol{\Omega}_{j}$
to be a diagonal matrix with the $j$-th diagonal element being one
while the other elements being zeros.
\begin{prop}
By solving the KKT system, the optimal solution to Problem \eqref{WSR:W-Subproblem}
with the general power constraints in \eqref{Extentions: General Power Constraints}
can be obtained in the following way
\[
\begin{aligned}\mathbf{W}^{\star}=\left\{ \begin{array}{ll}
\negthinspace\mathbf{R}^{-1}\mathbf{Q} & \text{if }\bigl\Vert\mathbf{R}^{-1}\mathbf{Q}\boldsymbol{\Omega}_{j}^{\frac{1}{2}}\bigr\Vert_{\mathrm{F}}^{2}\leq P_{j}\\
 & \hspace{1.7cm}\forall j=1,\ldots,J\\
\negthinspace(\mathbf{R}+\sum_{i=1}^{J}\gamma_{i}\boldsymbol{\Omega}_{i})^{-1}\mathbf{Q} & \text{otherwise},
\end{array}\right.\end{aligned}
\]
where the variables $\gamma_{1},\ldots,\gamma_{J}$ satisfies
\[
\Bigl\Vert(\mathbf{R}+\sum_{i=1}^{J}\gamma_{i}\boldsymbol{\Omega}_{i})^{-1}\mathbf{Q}\boldsymbol{\Omega}_{j}^{\frac{1}{2}}\Bigl\Vert_{\mathrm{F}}^{2}=P_{j},\ \forall j=1,\ldots,J.
\]
\end{prop}
\begin{IEEEproof}
It can be proved similarly as in Proposition \ref{Prop: W-Subproblem-Solution}.
\end{IEEEproof}

\subsection{A Serial Update Scheme for The $\boldsymbol{\Theta}$-Block}

Besides updating the elements in each phase shift matrix in parallel,
a serial update can also be conducted. Consider Problem \eqref{WSR:Formulation},
we can rewrite $f_{\mathsf{WSR},\boldsymbol{\Theta}_{l}}^{\prime}$
w.r.t. $[\boldsymbol{\theta}_{l}]_{j}$ as
\begin{align*}
 & f_{\mathsf{WSR},[\boldsymbol{\theta}_{l}]_{j}}^{\prime}([\boldsymbol{\theta}_{l}]_{j})=\negthinspace-2\mathrm{Re}\bigl([\boldsymbol{\theta}_{l}]_{j}^{*}[\mathbf{b}_{l}^{\mathsf{s}}]_{j}\bigr)\negthinspace-\negthinspace[\mathbf{L}_{l}]_{jj}\negthinspace-\negthinspace\underline{[\boldsymbol{\theta}_{l}]_{-j}^{\mathrm{H}}}\mathbf{L}_{l}\underline{[\boldsymbol{\theta}_{l}]_{-j}}\\
 & \negthinspace+\negthinspace\sum_{k=1}^{K}\negthinspace2\omega_{k}\mathrm{Re}\Bigl(\underline{[\boldsymbol{\theta}_{l}]_{-j}^{\mathrm{H}}}\mathbf{F}_{k,l}^{\mathrm{H}}\bigl(\beta_{k}^{*}\underline{\mathbf{w}_{k}}\negthinspace-\negthinspace\alpha_{k}\negthinspace\sum_{j=1}^{K}\underline{\mathbf{w}_{j}}\underline{\mathbf{w}_{j}^{\mathrm{H}}}\mathbf{h}_{k}^{\mathsf{d}}\bigr)\negthinspace\Bigr)\negthinspace\negthinspace+\negthinspace\mathsf{const}_{\theta,l},
\end{align*}
where
\[
\mathbf{b}_{l}^{\mathsf{s}}=\sum_{k=1}^{K}\omega_{k}\mathbf{F}_{k,l}^{\mathrm{H}}\bigl(\alpha_{k}\sum_{j=1}^{K}\underline{\mathbf{w}_{j}}\underline{\mathbf{w}_{j}^{\mathrm{H}}}\mathbf{h}_{k}^{\mathsf{d}}-\beta_{k}^{*}\underline{\mathbf{w}_{k}}\bigr)+\mathbf{L}_{l}\underline{[\boldsymbol{\theta}_{l}]_{-j}}.
\]
Then the subproblem w.r.t. $[\boldsymbol{\theta}_{l}]_{j}$ is given
by
\[
\begin{aligned} & \underset{\left|[\boldsymbol{\theta}_{l}]_{j}\right|=1}{\mathrm{minimize}} &  & \mathrm{Re}\bigl([\boldsymbol{\theta}_{l}]_{j}^{*}[\mathbf{b}_{l}^{\mathsf{s}}]_{j}\bigr),\end{aligned}
\]
which can be readily solved via Lemma \ref{Prop: Theta-Subproblem-Solution}. 

Based on the above formulation, the phase shift matrices $\{\boldsymbol{\Theta}_{i}\}$
will be updated in series with the elements in each phase shift matrix
updated also in series. Such update rule can also be applied to Problem
\eqref{MaxMin:Formulation} and Problem \eqref{SR:Formulation}.

\subsection{Discrete Phase for The $\boldsymbol{\Theta}$-Block}

Along the paper, we have taken a continuous-phase scheme for the phase
shift matrices, while for Problem \eqref{WSR:Formulation} and Problem
\eqref{SR:Formulation}, a discrete-phase scheme \cite{di2020hybrid},
which is more practical for hardware implementation, can also be considered
which is defined as follows:
\begin{align*}
\mathcal{D}_{i}= & \bigl\{\boldsymbol{\Theta}_{i}\mid\boldsymbol{\Theta}_{i}=\mathrm{diag}(\boldsymbol{\theta}_{i}),\ \boldsymbol{\theta}_{i}\in\mathbb{C}^{N_{i}},\ \bigl|[\boldsymbol{\theta}_{i}]_{j}\bigr|=1,\\
 & \hspace{3cm}\mathrm{ang}([\boldsymbol{\theta}_{i}]_{j})\in\Phi_{i},\ \forall j=1,\ldots,N_{i}\bigr\},
\end{align*}
where $\Phi_{i}$ denotes the set of fixed angles that is achievable
for the $i$-th RIS. Under the discrete-phase scheme, the closed-form
solutions given in Lemma \ref{Prop: Theta-Subproblem-Solution} should
be modified to be
\[
[\boldsymbol{\theta}_{i}^{\star}]_{j}=e^{\mathfrak{j}\mathrm{arg\,min}_{\boldsymbol{\phi}_{i}\in\Phi_{i}}||\boldsymbol{\phi}_{i}+\mathrm{ang}([\mathbf{b}_{i}]_{j})||_{2}},\ \forall j=1,\ldots,N_{i},
\]
which can be efficiently implemented on hardwares leveraging on a
look-up table.

\subsection{WSR Maximization for General-Topology Multi-Hop RIS-Aided Multi-User
MISO Cellular Networks\label{subsec:WSR-Maximization-for-general-topology}}

In Section \ref{sec:WSR}, a simplified model where only the direct
transmission paths from the BS to the users and the reflection transmission
paths through cascaded $I_{1},\ldots,I_{K}$ RISs to the users have
been considered. However, the developed BMM algorithm can be easily
extended for system designs with general-topology \cite{mei2021intelligent}.
To get a taste of it, let us first consider a feed-forward ``fully
connected'' double-RIS system (we neglect any feed-backward signals
which are in general much weaker when received). We adopt a similar
system setting as used in Section \ref{sec:WSR} and denote the channel
between the BS and the first RIS, the channel between the BS and the
second RIS, and the channel between the first RIS and the second RIS
as $\mathbf{G}_{0,1}$, $\mathbf{G}_{0,2}$, and $\mathbf{G}_{1,2}$,
respectively. Besides, we denote the channel between the first RIS
and the $k$-the user, the channel between the second RIS and the
$k$-the user, and the direct channel from the BS to the $k$-th user
as $\mathbf{h}_{k,1}^{\mathsf{r}}$, $\mathbf{h}_{k,2}^{\mathsf{r}}$,
and $\mathbf{h}_{k}^{\mathsf{d}}$, respectively. Then the SINR at
the $k$-th user is given by
\[
\mathsf{SINR}_{k}=\frac{\bigl|\mathbf{w}_{k}^{\mathrm{H}}\mathbf{h}_{k}\bigr|^{2}}{\sum_{j,j\neq k}^{K}\bigl|\mathbf{w}_{j}^{\mathrm{H}}\mathbf{h}_{k}\bigr|^{2}+\sigma^{2}},
\]
where
\begin{align*}
\mathbf{h}_{k}= & \underset{\text{double-reflection channel}}{\underbrace{\mathbf{G}_{0,1}\boldsymbol{\Theta}_{1}\mathbf{G}_{1,2}\boldsymbol{\Theta}_{2}\mathbf{h}_{k,2}^{\mathsf{r}}}}\\
 & \hspace{1.7cm}+\underset{\text{single-reflection channels}}{\underbrace{\mathbf{G}_{0,1}\boldsymbol{\Theta}_{1}\mathbf{h}_{k,1}^{\mathsf{r}}+\mathbf{G}_{0,2}\boldsymbol{\Theta}_{2}\mathbf{h}_{k,2}^{\mathsf{r}}}}+\underset{\text{direct channel}}{\underbrace{\mathbf{h}_{k}^{\mathsf{d}}}}.
\end{align*}
Following the general idea of the proposed BMM algorithms, given the
iterate $\{\underline{\mathbf{W}},\underline{\boldsymbol{\Theta}_{1}},\underline{\boldsymbol{\Theta}_{2}}\}$,
we optimize $\mathbf{W}$, $\boldsymbol{\Theta}_{1}$, and $\boldsymbol{\Theta}_{2}$
cyclically. It is easy to verify that $f_{\mathsf{WSR},\mathbf{W}}(\mathbf{W})$
takes exactly the same mathematical form as Eq. \eqref{WSR: W-Block WSR}
and hence is omitted. By defining $\mathbf{F}_{k,1}=\mathbf{G}_{0,1}\mathrm{diag}\bigl(\mathbf{G}_{1,2}\underline{\boldsymbol{\Theta}_{2}}\mathbf{h}_{k,2}^{\mathsf{r}}+\mathbf{h}_{k,1}^{\mathsf{r}}\bigr)$
and $\mathbf{h}_{k,1}^{\mathsf{d}}=\mathbf{G}_{0,2}\underline{\boldsymbol{\Theta}_{2}}\mathbf{h}_{k,2}^{\mathsf{r}}+\mathbf{h}_{k}^{\mathsf{d}}$,
we obtain the objective $f_{\mathsf{WSR}}$ w.r.t. $\boldsymbol{\Theta}_{1}$
given in the following way
\begin{align*}
 & f_{\mathsf{WSR},\boldsymbol{\Theta}_{1}}\left(\boldsymbol{\Theta}_{1}\right)\\
= & \sum_{k=1}^{K}\omega_{k}\log\bigl(1+\frac{\bigl|\underline{\mathbf{w}_{k}^{\mathrm{H}}}\mathbf{F}_{k,1}\boldsymbol{\theta}_{1}+\mathbf{w}_{k}^{\mathrm{H}}\mathbf{h}_{k,1}^{\mathsf{d}}\bigr|^{2}}{\sum_{j,j\neq k}^{K}\bigl|\underline{\mathbf{w}_{j}^{\mathrm{H}}}\mathbf{F}_{k,1}\boldsymbol{\theta}_{1}+\underline{\mathbf{w}_{j}^{\mathrm{H}}}\mathbf{h}_{k,1}^{\mathsf{d}}\bigr|^{2}+\sigma^{2}}\bigr),
\end{align*}
which shares the same form of expression as Eq. \eqref{WSR: Theta-Block WSR}.
Similar result applies to the objective $f_{\mathsf{WSR}}$ w.r.t.
$\boldsymbol{\Theta}_{2}$. Therefore, the problem of the feed-forward
``fully connected'' double-RIS system design can be addressed readily
following the same algorithmic procedure as introduced in Section
\ref{sec:WSR}.

Furthermore, we can extend our algorithm to the case of solving general-topology
multi-hop RIS-aided, a.k.a. multi-RIS, system designs equipped with
finite $L$ RISs. Assume there are $P_{k,n}$ reflection paths from
the BS to the $k$-th user that go through $n$ ($n=1,\ldots,L$)
RISs, within which we denote by $\mathcal{R}_{k,n}^{p}=\{p_{k,1},\ldots,p_{k,n}\}$
with $p=1,\ldots,P_{k,n}$ the $p$-th path, where the index of the
$i$-th intermediate RIS is denoted by $p_{k,i}$. (We also define
$p_{k,0}=0$ for any $k$ and $p$, which refers to the BS.) Note
that there is a chance that $P_{k,n}=0$ for some $n$ in practice.
The system model considered  in Section \ref{sec:WSR} can be seen
as a special case of this general-topology system where $P_{k,I_{k}}=1$
with $\mathcal{R}_{k,I_{k}}^{1}=\{1,\ldots,I_{k}\}$ and $P_{k,n}=0$
for $n\neq I_{k}$, $n=1,\ldots,L$. We denote by $\mathbf{G}_{0,i}$
the channel between the BS and the $i$-th RIS $(i=1,\ldots,L)$,
by $\mathbf{G}_{i,j}$ the channel between the $i$-th RIS and the
$j$-th RIS $(j=1,\ldots,L)$, by $\mathbf{h}_{k,i}^{\mathsf{r}}$
the channel in the reflection link between the $i$-th RIS and the
$k$-th user, respectively. Direct channels $\mathbf{h}_{k}^{\mathsf{d}}$
are defined as before. Then the SINR at the $k$-th user is computed
as
\[
\mathsf{SINR}_{k}=\frac{\bigl|\mathbf{w}_{k}^{\mathrm{H}}\mathbf{h}_{k}\bigr|^{2}}{\sum_{j,j\neq k}^{K}\bigl|\mathbf{w}_{j}^{\mathrm{H}}\mathbf{h}_{k}\bigr|^{2}+\sigma^{2}},
\]
with
\[
\begin{aligned}\mathbf{h}_{k}= & \sum_{n=1}^{L}\sum_{p=1}^{P_{n}}\biggl(\prod_{i=1}^{n}\mathbf{G}_{p_{i-1},p_{i}}\boldsymbol{\Theta}_{p_{i}}\biggr)\mathbf{h}_{k,p_{n}}^{\mathsf{r}}+\mathbf{h}_{k}^{\mathsf{d}}.\end{aligned}
\]
Already given the mathematical form above which resembles the aforementioned
double-RIS system, such a general-topology multi-hop RIS-aided system
designs can be addressed readily via the algorithm proposed in Section
\ref{WSR:Section-Theta-Block}.

\subsection{SINR Maximizations for RIS-Aided Wireless Networks}

Beyond the rate maximization problems, another commonly used system
performance metric is SINR \cite{zheng2021double}. The proposed BMM
method is also applicable in solving many SINR maximization problems
in RIS-aided wireless network designs, e.g., weighted sum-SINR maximizations,
minimum SINR maximizations, and sum-SINR maximizations. As for the
rate maximizations where the minorizing functions for the objectives
are constructed by minorizing individual rate functions for different
users, the construction of minorizing functions for the SINR-based
objectives can be done in a similar way. Taking the system considered
in Section \ref{sec:WSR} and Section \ref{sec:MR} as an example,
the minorizing functions for individual SINRs can be obtained based
on Lemma \ref{lem:SINR-lemma} (given in Appendix \ref{Appendix: Scalar MM Proof}).
And then the SINR maximization problems can be addressed by optimizing
the resultant surrogate problems following similar procedures as for
rate maximizations under the proposed BMM algorithmic framework.

\subsection{Acceleration Scheme for the BMM-Based Algorithms}

The BMM method may suffer from a slow convergence, in that its convergence
speed is dictated by the tightness of the minorizing function \cite{wu2010mm}.
To accelerate the algorithms, the SQUAREM method \cite{varadhan2008simple}
can be applied as an off-the-shelf accelerator, which was originally
proposed to accelerate the expectation-maximization algorithms and
later widely used for MM algorithm accelerations \cite{song2015optimization}.
It accelerates the convergence of a monotonic algorithm by squaring
the one-step update to be twice within each cycle of an extrapolation
scheme. Interested readers are referred to \cite{varadhan2008simple}
for details. 

\section{Numerical Simulations\label{sec:Numerical-Simulations}}

In this section, we provide numerical experiments to corroborate our
theoretical results with the codes available at 
\[
\textsf{https://github.com/zepengzhang/RateMaxRIS-BMM}.
\]
The simulations are performed in MATLAB on a personal computer with
a 3.3 GHz Intel Xeon W CPU.

\subsection{Simulations for RIS-Aided MISO Cellular Networks}

\subsubsection{System Settings\label{System-Settings}}

Under a three-dimensional Cartesian coordinate system, we consider
a multi-user MISO system where the BS located at $(0,0,10)\mathsf{m}$
communicates with $K$ users assisted by a RIS at $(d,0,10)\mathsf{m}$.
The $K$ users are randomly distributed in a circle centered at $(d,30,0)\mathsf{m}$
with radius of $10\mathsf{m}$. The antennas at the BS is arranged
as a uniform linear array with spacing of $\frac{\lambda}{2}$ and
the passive reflecting elements at the RIS is arranged as a uniform
planar array with spacing of $\frac{\lambda}{8}$, where $\lambda$
is the wavelength. We assume that the channel fading is frequency
flat and adopt the Rician fading model for all channels. To make an
example, the channel from the BS to the RIS is modeled as
\[
\mathbf{G}_{0,1}=\sqrt{\frac{\kappa_{\mathsf{G}}(d)}{K_{\mathsf{G}}+1}}(\sqrt{K_{\mathsf{G}}}\mathbf{G}_{0,1}^{\mathsf{LoS}}+\mathbf{G}_{0,1}^{\mathsf{NLoS}}),
\]
where $\kappa_{\mathsf{G}}(d)$ is the distance-dependent path loss,
$K_{\mathsf{G}}\in[0,\infty)$ is the Rician factor, $\mathbf{G}_{0,1}^{\mathsf{LoS}}$
is the line-of-sight (LoS) component, and $\mathbf{G}_{0,1}^{\mathsf{NLoS}}$
is the non-line-of-sight (NLoS) components. Specifically, the distance-dependent
path loss is computed as $\kappa_{\mathsf{G}}(d)=T_{0}(\frac{d}{d_{0}})^{-\varrho_{\mathsf{G}}}$
with the path-loss at the reference distance $d_{0}=1\mathsf{m}$
being $T_{0}=-30\mathsf{dB}$ and $\varrho_{\mathsf{G}}$ denoting
the path loss exponent, the LoS component is computed as the product
of the array responses at the two sides, and the NLoS component is
modeled by Rayleigh fading, with $[\mathbf{G}_{0,1}^{\mathsf{NLoS}}]_{ij}\sim\mathcal{CN}(0,1)$,
$i=1,\ldots,N_{1}$, $j=1,\ldots,M$. The other channels are modeled
similarly as for $\mathbf{G}_{0,1}$. The path loss exponents and
the Rician factors for channels $\{\mathbf{G}_{i-1,i}\}$, $\{\mathbf{h}_{i}^{\mathsf{d}}\}$,
and $\{\mathbf{h}_{i}^{\mathsf{r}}\}$ are set as $\varrho_{\mathsf{G}}=2.2$,
$K_{\mathsf{G}}=3$, $\varrho_{\mathsf{d}}=3.5$, $K_{\mathsf{d}}=0$,
and $\varrho_{\mathsf{r}}=2.8$, $K_{\mathsf{r}}=3$, respectively.
Besides, we have considered the noise power spectrum density of $-169\mathsf{dBm/Hz}$
and the transmission bandwidth of 240$\mathsf{kHz}$. In the following,
if not specified, we will assume $P=0\mathsf{dBm}$, $\sigma^{2}=1$,
$K=4$, $M=4$, $N=100$, and $d=200\mathsf{m}$. Moreover, all the
simulation curves are averaged over 100 independent channel realizations.

\subsubsection{Performance Evaluations}

Four benchmarks are considered for the WSR maximization as introduced
in Section \ref{sec:WSR}: i) WMMSE-based BCD method \cite{guo2020weighted}
in which case the $\mathbf{W}$-block is tackled by WMMSE \cite{shi2011iteratively}
and the $\boldsymbol{\mathbf{\Theta}}$-block is tackled via RCG;
ii) FP-based BCD method \cite{guo2020weighted} in which case the
problem is tackled via FP aided by the prox-linear update for the
$\mathbf{W}$-block and an MM step for the $\boldsymbol{\Theta}$-block;
iii) Random Phase in which case the phase shift matrix $\boldsymbol{\mathbf{\Theta}}$
is randomly initialized and $\mathbf{W}$ is optimized via MM; iv)
Without RIS, i.e., no RIS is used and $\mathbf{W}$ is optimized via
MM. 

\begin{figure}[t]
\begin{centering}
\includegraphics[width=1\columnwidth]{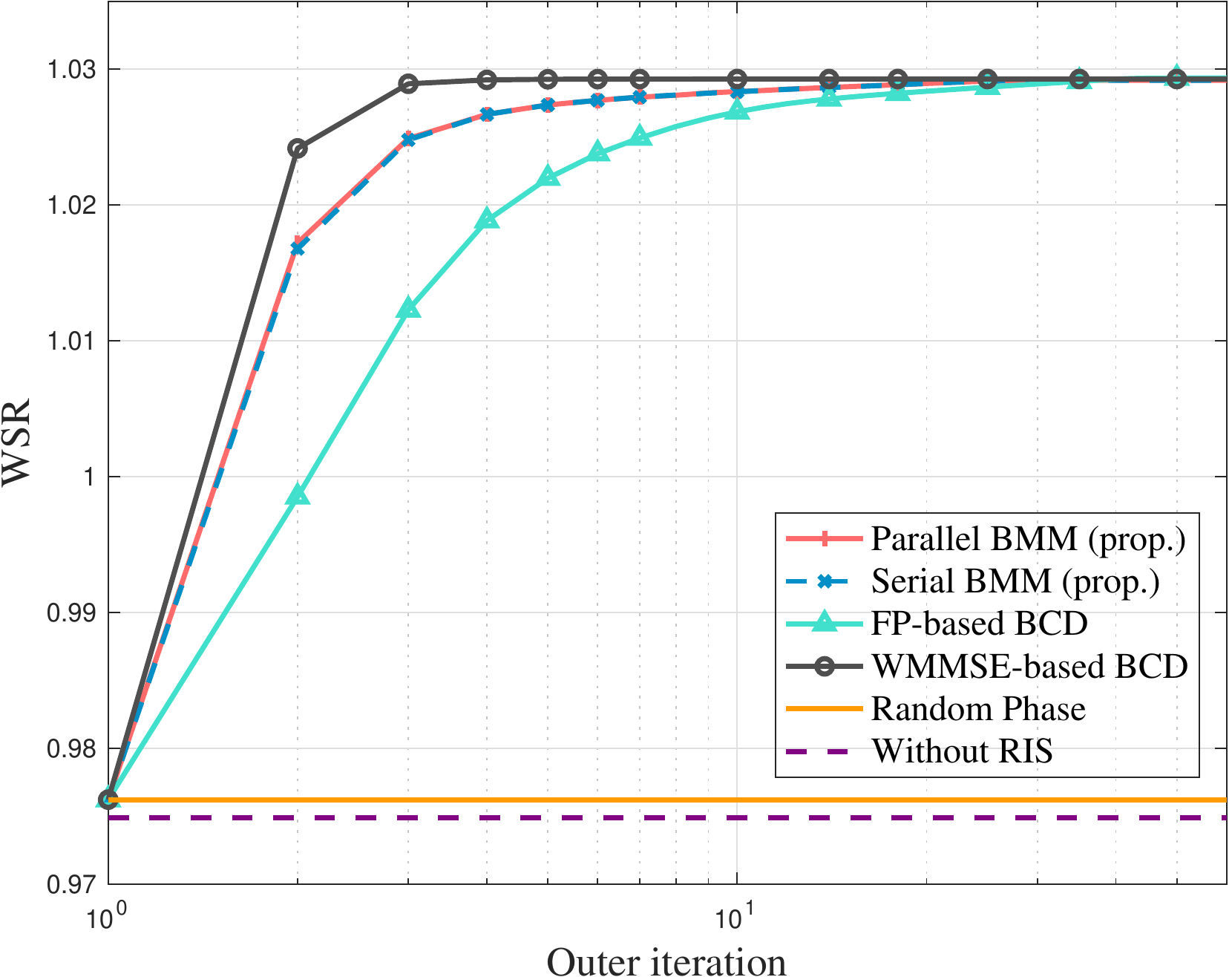}
\par\end{centering}
\caption{WSR versus outer iteration numbers.\label{Simulations:WSR-Convergence-Iterations}}
\end{figure}

\begin{figure}[t]
\begin{centering}
\includegraphics[width=1\columnwidth]{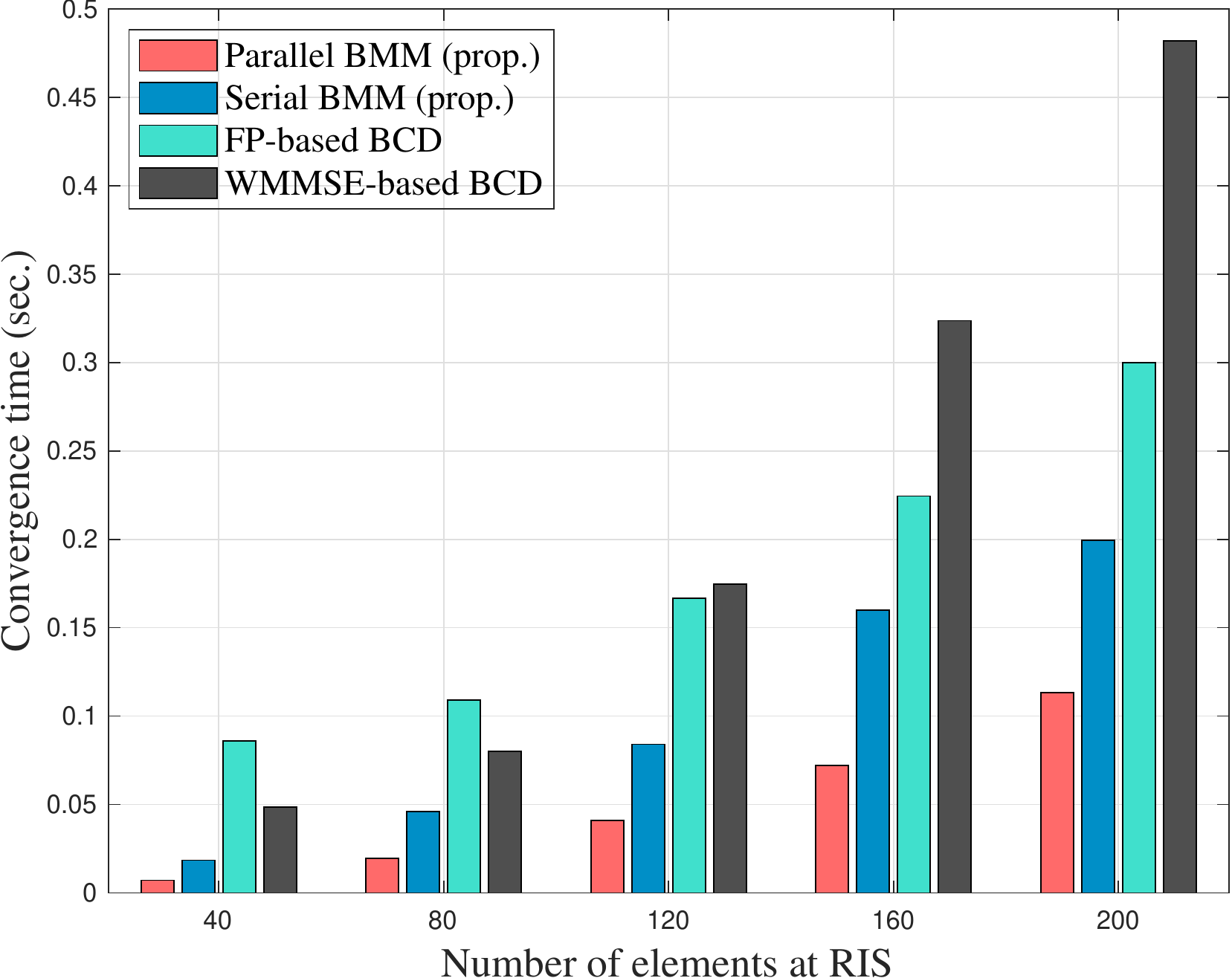}
\par\end{centering}
\caption{Convergence times versus number of elements at RIS.\label{Simulations:WSR-TIME-N}}
\end{figure}

\begin{figure}[t]
\begin{centering}
\includegraphics[width=1\columnwidth]{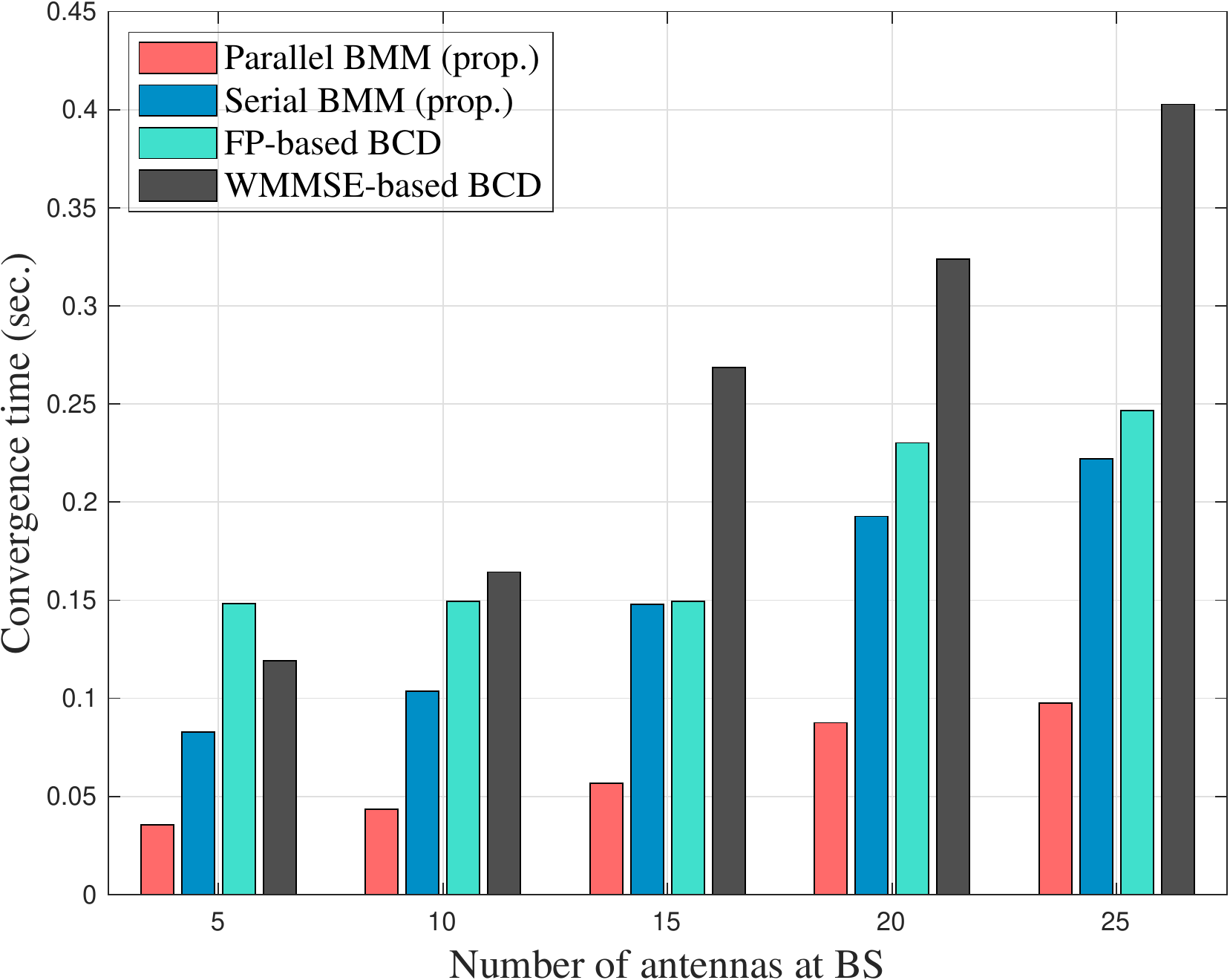}
\par\end{centering}
\caption{Convergence times versus number of antennas at BS.\label{Simulations:WSR-TIME-M}}
\end{figure}

The convergence behaviors of the proposed BMM algorithm and the above
benchmarks are investigated under the single-hop scenario. The comparisons
in terms of the outer iteration numbers are depicted in Fig. \ref{Simulations:WSR-Convergence-Iterations},
while the convergence times of different approaches with different
number of elements at RIS and those with different number of antennas
at BS are showcased in Fig. \ref{Simulations:WSR-TIME-N} and Fig.
\ref{Simulations:WSR-TIME-M}, respectively. It can be observed that
the serial BMM and the parallel BMM acquire similar outer iteration
numbers to converge, which is better or comparable w.r.t. the benchmark
methods. Moreover, the parallel BMM method consistently acquires the
lowest CPU computation times to converge in all the simulations cases. 


\begin{figure}[t]
\begin{centering}
\includegraphics[width=1\columnwidth]{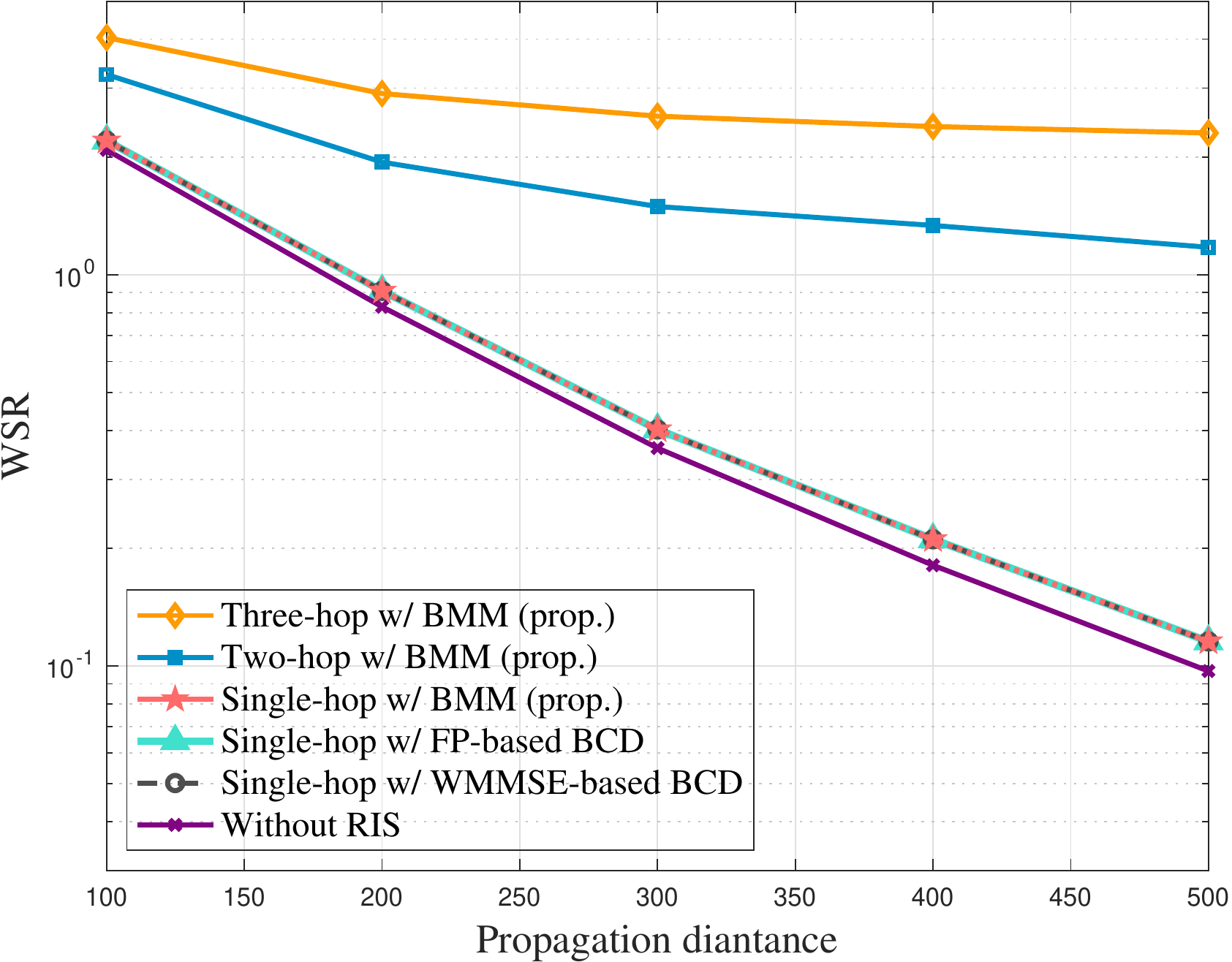}
\par\end{centering}
\caption{Achievable WSR versus distance.\label{Simulations:WSR-distance-multi-hop}}
\end{figure}

We further investigate the benefits of introducing multiple RISs,
especially in combating the large path loss in long-distance propagations.
As for the system setting, the second RIS at $(\frac{d}{2},0,10)\mathsf{m}$
is deployed in the two-hop system and an another RIS at $(\frac{d}{4},0,10)\mathsf{m}$
is included for the three-hop system. Besides, we assume the additional
RISs are also equipped with $N$ elements. The achievable WSRs for
different systems as distance increases are showcased in Fig. \ref{Simulations:WSR-distance-multi-hop},
in which the BMM approach (parallel BMM is used hereafter) and the
benchmarks converge to numerically similar WSR for the single-hop
scenario, and the benefits of introducing multiple RISs to improve
WSR is obvious. 

\begin{figure}[t]
\begin{centering}
\includegraphics[width=1\columnwidth]{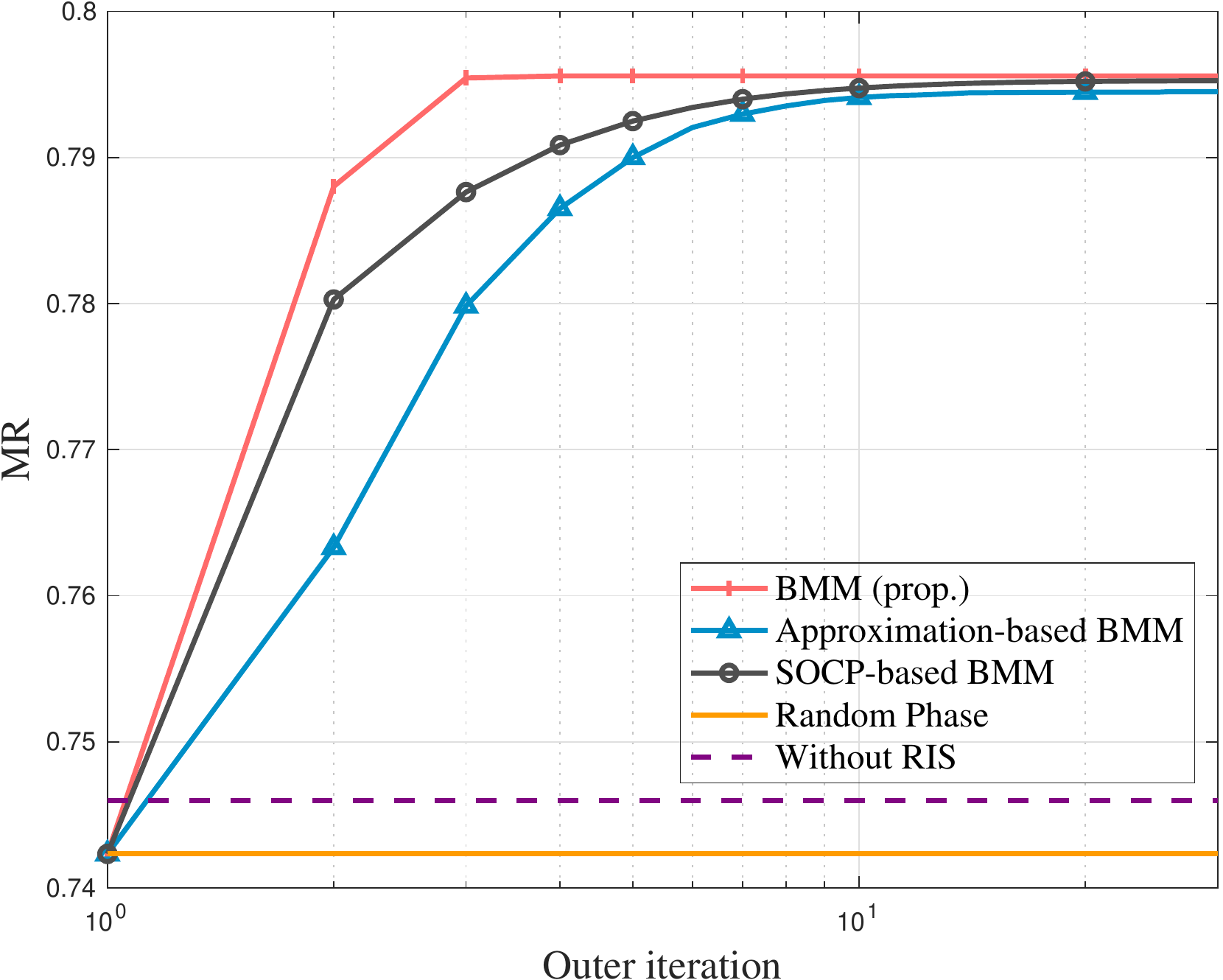}
\par\end{centering}
\caption{MR versus outer iteration numbers. \label{Simulation:MR-Iteration}}
\end{figure}

\begin{figure}[t]
\begin{centering}
\includegraphics[width=1\columnwidth]{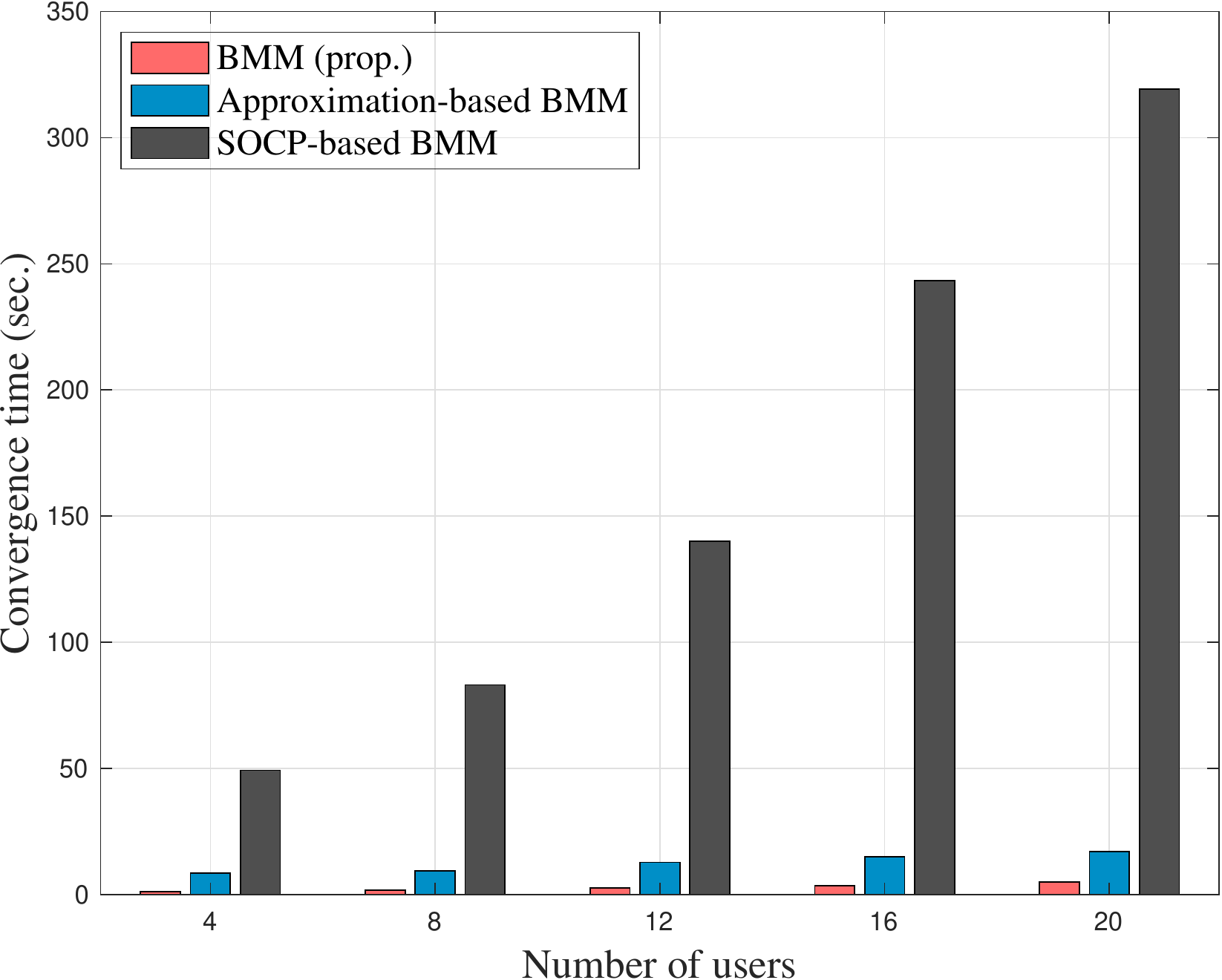}
\par\end{centering}
\caption{Convergence time versus number of users. \label{Simulation:MR-K}}
\end{figure}

For the MR maximization problems, two benchmarks are considered: i)
SOCP-based BMM method \cite{zhou2020intelligent} where the BMM method
is applied with the resultant SOCP subproblems solved via off-the-shelf
scripting language CVX \cite{grant2014cvx}; ii) Approximation-based
BMM method \cite{zhou2020intelligent} where the pointwise minimum
function is first smoothened with the log-sum-exp function and the
approximation problem is tackled via BMM. The convergence behaviors
of the proposed BMM algorithm along with these two benchmarks are
demonstrated in terms of outer iteration numbers in Fig. \ref{Simulation:MR-Iteration},
where the proposed BMM algorithm enjoys the fastest convergence speed
with the highest achievable MR among these three methods. Moreover,
the convergence time of different approaches with different number
of users is depicted in Fig. \ref{Simulation:MR-K}. We can easily
observe that the proposed BMM algorithm acquires the lowest CPU times
for convergence in all cases. 


\subsection{Simulations for RIS-Aided MIMO D2D Networks}

\subsubsection{System Settings}

We consider a MIMO D2D system where $K$ transceiver pairs transmit
multiple data streams with the assistance of a RIS located at $(d,30,0)\mathsf{m}$.
The transmitters and the receivers are randomly distributed in two
circles with radius of $10\mathsf{m}$ that are centered at $(0,0,10)\mathsf{m}$
and $(d,30,0)\mathsf{m}$, respectively. The antennas at the transmitters
and the receivers are arranged as uniform linear arrays with spacing
of $\frac{\lambda}{2}$. The number of antennas of each transmitter
and receiver is set uniformly to be $M$, and the dimension of symbol
vectors is set as $K$. The channels are modeled in a fashion similar
to that for the MISO system in Section \ref{System-Settings}, and
hence we omit the details here.

\subsubsection{Performance Evaluations}

\begin{figure}[t]
\begin{centering}
\includegraphics[width=1\columnwidth]{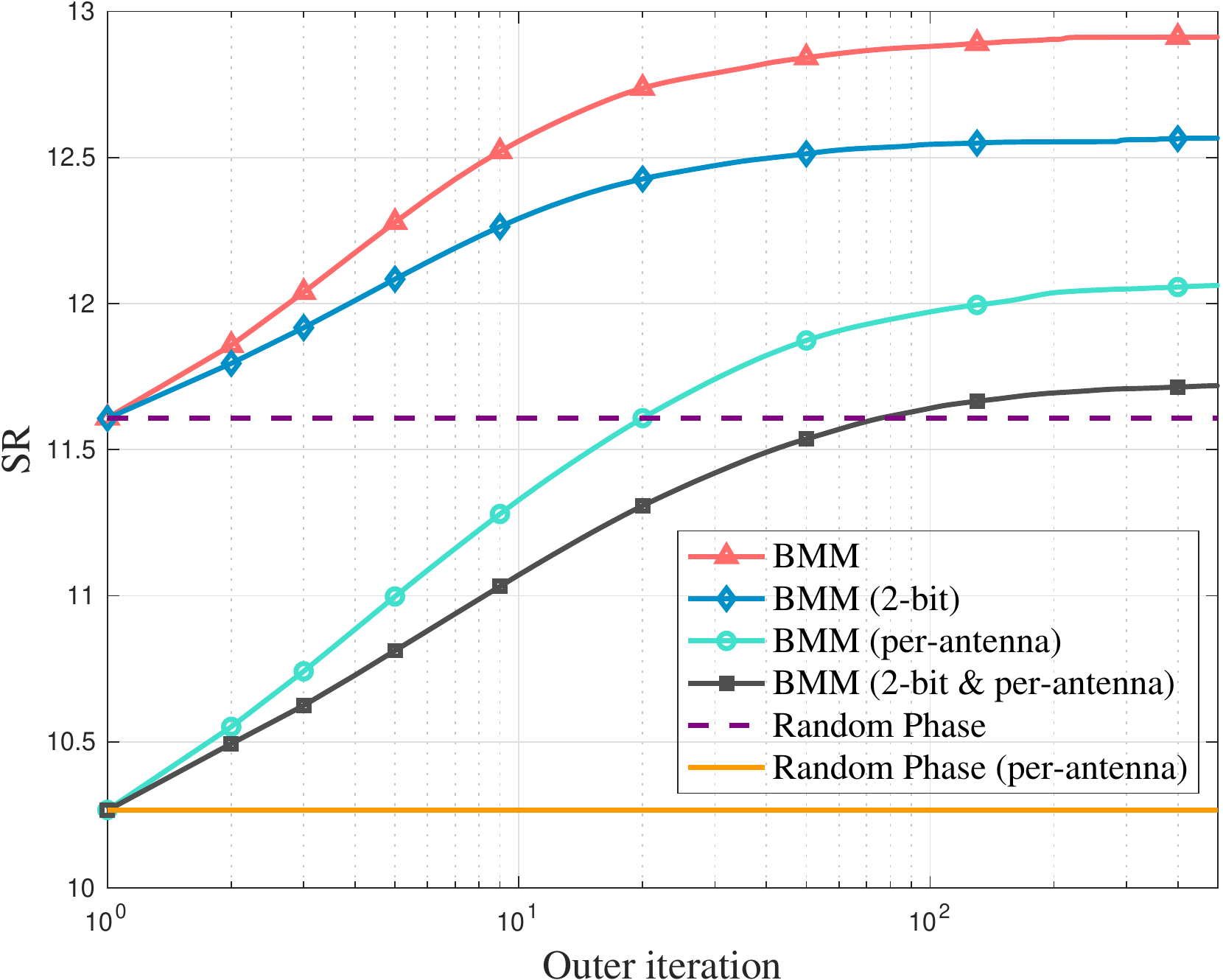}
\par\end{centering}
\caption{SR versus outer iteration numbers. \label{Simulation:SR}}
\end{figure}

\begin{figure}[t]
\begin{centering}
\includegraphics[width=1\columnwidth]{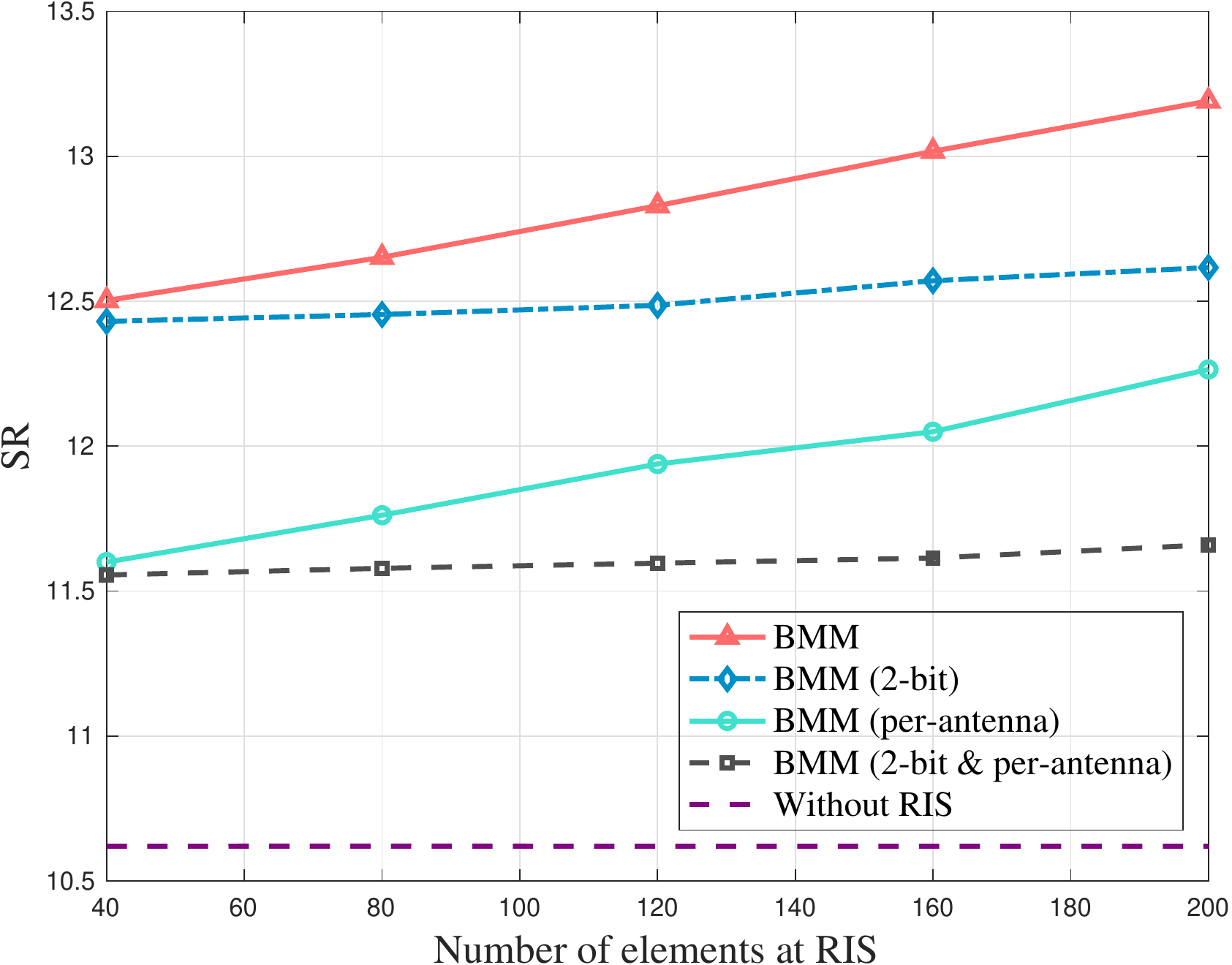}
\par\end{centering}
\caption{Achievable SR versus number of elements at RIS. \label{Simulation:SR-discrete}}
\end{figure}

In this section, we will investigate the performance loss of considering
more practical constraints in joint beamforming and reflecting design
for SR maximization in the specified MIMO D2D system. Particularly,
we consider the per-antenna power constraint introduced in Section
\ref{subsec:General-Power-Constraint}, which is more realistic due
to the fact that each antenna is usually equipped with an independent
power amplifier. Besides, considering hardware limitation, instead
of the continuous-phase shift, we further impose a 2-bit discrete-phase
constraint. Therefore, four variants of the proposed BMM algorithms
are implemented: i) BMM with total power limit and the continuous-phase
constraint; ii) BMM (2-bit) with total power limit and the 2-bit discrete-phase
constraint; iii) BMM (per-antenna) with per-antenna power limit and
the continuous-phase constraint; iv) BMM (2-bit \& per-antenna) with
per-antenna power limit and the 2-bit discrete-phase constraint. Convergence
behaviors of these four variants of BMM are demonstrated in Fig. \ref{Simulation:SR}
and performance comparisons by considering more practical constraints
is depicted in Fig. \ref{Simulation:SR-discrete}.


\section{Conclusions\label{sec:Conclusions}}

In this paper, we have considered the joint beamforming and reflecting
design problem for rate maximization problems in RIS-aided wireless
networks. A unified algorithmic framework based on the block minorization-maximization
(BMM) method has been developed. Merits of the algorithm have been
showcased via three system design problems, where problem-tailored
low-complexity and globally convergent algorithms are proposed. Benefits
of the BMM algorithms in comparison with existing algorithms have
been demonstrated\textcolor{red}{{} }numerically. We have also shown
that many more RIS-aided system design aspects can be addressed by
the unified BMM method, leaving much space to explore for future research.

\appendix{}

Refer to the attached Supplementary Materials.

\bibliographystyle{IEEEtran}
\bibliography{reference}

\newpage{}

\onecolumn
\begin{center}
{\LARGE{}Supplementary Materials for ``Rate Maximizations for Reconfigurable
Intelligent Surface-Aided Wireless Networks: A Unified Framework via
Block Minorization-Maximization''\vspace{5bp}
}{\LARGE\par}
\par\end{center}

\begin{center}
Zepeng~Zhang and Ziping~Zhao
\par\end{center}

\subsection{Proof for Proposition \ref{Prop: Scalar MM} \label{Appendix: Scalar MM Proof}}
\begin{IEEEproof}
The function $\log(z)$ with $z>0$ is concave and hence can be majorized
by its linear expansion around $\underline{z}$ as follows:
\[
\log(z)\leq\log(\underline{z})+\frac{1}{\underline{z}}(z-\underline{z}).
\]
By substituting $z$ with $\frac{y}{y+|x|^{2}}$, we get
\begin{equation}
\begin{aligned}\log\bigl(1+\frac{|x|^{2}}{y}\bigr) & \geq\log\bigl(1+\frac{|\underline{x}|^{2}}{\underline{y}}\bigr)-\frac{\underline{y}+|\underline{x}|^{2}}{\underline{y}}\bigl(\frac{y}{y+|x|^{2}}-\frac{\underline{y}}{\underline{y}+|\underline{x}|^{2}}\bigr)\\
 & =\log\bigl(1+\frac{|\underline{x}|^{2}}{\underline{y}}\bigr)-\frac{\underline{y}+|\underline{x}|^{2}}{\underline{y}}\bigl(\frac{|\underline{x}|^{2}}{\underline{y}+|\underline{x}|^{2}}-\frac{|x|^{2}}{y+|x|^{2}}\bigr)\\
 & =\log\bigl(1+\frac{|\underline{x}|^{2}}{\underline{y}}\bigr)-\frac{|\underline{x}|^{2}}{\underline{y}}+\bigl(1+\frac{|\underline{x}|^{2}}{\underline{y}}\bigr)\frac{|x|^{2}}{y+|x|^{2}},
\end{aligned}
\label{eq:scalar 1}
\end{equation}
where the equality is attained at $\left(x,y\right)=\left(\underline{x},\underline{y}\right)$. 
\begin{lem}
\label{lem:SINR-lemma}The function $\frac{\left|z_{1}\right|^{2}}{z_{2}}$
with $z_{1}\in\mathbb{C}$ and $z_{2}>0$ is minorized at $(\underline{z_{1}},\underline{z_{2}})$
as follows:
\[
\frac{|z_{1}|^{2}}{z_{2}}\geq\frac{2}{\underline{z_{2}}}\mathrm{Re}\bigl(\underline{z_{1}^{*}}z_{1}\bigr)-\frac{|\underline{z_{1}}|^{2}}{\underline{z_{2}^{2}}}z_{2}.
\]
\end{lem}
\begin{IEEEproof}
For $z_{1}\in\mathbb{C}$ and $z_{2}>0$, we have
\[
\bigl(\frac{z_{1}}{\sqrt{z_{2}}}-\frac{\underline{z_{1}}\sqrt{z_{2}}}{\underline{z_{2}}}\bigr)^{*}\bigl(\frac{z_{1}}{\sqrt{z_{2}}}-\frac{\underline{z_{1}}\sqrt{z_{2}}}{\underline{z_{2}}}\bigr)\geq0.
\]
Expanding the above formula and rearranging the terms lead to the
above result. 
\end{IEEEproof}
\begin{rem}
Lemma \ref{lem:SINR-lemma} reduces to the case of minorizing the
convex ``quadratic-over-linear'' function by the first order Taylor
expansion when $z_{1}\in\mathbb{R}$ and $z_{2}>0$ \cite{sun2016majorization}.
\end{rem}
Applying the results in Lemma \ref{lem:SINR-lemma} to the last line
of Eq. \eqref{eq:scalar 1} with $z_{1}=x$ and $z_{2}=y+\left|x\right|^{2}$,
we can conclude that
\begin{align*}
\log\bigl(1+\frac{|x|^{2}}{y}\bigr) & \geq\log\bigl(1+\frac{|\underline{x}|^{2}}{\underline{y}}\bigr)-\frac{|\underline{x}|^{2}}{\underline{y}}+\bigl(1+\frac{|\underline{x}|^{2}}{\underline{y}}\bigr)\Bigl(\frac{2}{\underline{y}+|\underline{x}|^{2}}\mathrm{Re}\bigl(\underline{x}^{*}x\bigr)-\frac{|x|^{2}}{\bigl(\underline{y}+|\underline{x}|^{2}\bigr)^{2}}\bigl(y+|x|^{2}\bigr)\Bigr)\\
 & =-\frac{|\underline{x}|^{2}}{\underline{y}\bigl(\underline{y}+|\underline{x}|^{2}\bigr)}\bigl(y+|x|^{2}\bigr)+\frac{2}{\underline{y}}\mathrm{Re}\bigl(\underline{x}^{*}x\bigr)+\log\bigl(1+\frac{|\underline{x}|^{2}}{\underline{y}}\bigr)-\frac{|\underline{x}|^{2}}{\underline{y}},
\end{align*}
where the equality is attained when $\left(x,y\right)=\left(\underline{x},\underline{y}\right)$. 
\end{IEEEproof}

\subsection{Proof for Proposition \ref{Lemma: Relaxed Theta Constraint} \label{Appendix: Relaxed Theta Constraint}}
\begin{IEEEproof}
The objective of Problem \eqref{MaxMin-Theta-Subproblem-Relaxed}
is concave-convex in $\boldsymbol{\Theta}$ and $\mathbf{s}$, and
constraint sets $\mathcal{C}_{\mathsf{relaxed}}$ and $\mathcal{S}$
are both nonempty compact and convex. Therefore, a saddle point $(\boldsymbol{\Theta}^{\star},\mathbf{s}^{\star})$
exists for Problem \eqref{MaxMin-Theta-Subproblem-Relaxed} \cite[Corollary 37.6.2]{rockafellar1970convex}.
In the following, we will verify that $\boldsymbol{\Theta}^{\star}\in\mathcal{C}$
always hold by contradiction, with which the proof is completed. 

Suppose $\boldsymbol{\Theta}^{\star}\notin\mathcal{C}$, i.e., in
the interior of $\mathcal{C}_{\mathsf{relaxed}}$, then there must
exist some element of $\boldsymbol{\Theta}^{\star}=\mathrm{diag}(\boldsymbol{\theta}^{\star})$,
say $\left[\boldsymbol{\theta}^{\star}\right]_{j}$, such that $|[\boldsymbol{\theta}^{\star}]_{j}|<1$.
If the $j$-th element of $\mathbf{b}_{k}$ is nonzero, we can always
reset the phase of $\left[\boldsymbol{\theta}\right]_{j}$ to be aligned
with the $j$-th element of $-\mathbf{b}_{k}$ and increase its modulus
by a small amount without violating feasibility. Then the objective
of Problem \eqref{MaxMin-Theta-Subproblem-Relaxed} will be pulled
down from the side of $\boldsymbol{\theta}$, which contradicts with
the saddle point nature of $(\boldsymbol{\Theta}^{\star},\mathbf{s}^{\star})$.
In case the $j$-th element of $\mathbf{b}_{k}$ is zero, the optimal
$\left[\boldsymbol{\theta}^{\star}\right]_{j}$ may be non-unique
(and thus the saddle point is non-unique), but we can always modulate
the modulus of $\left[\boldsymbol{\theta}\right]_{j}$ to find an
optimal $\left[\boldsymbol{\theta}^{\star}\right]_{j}$ on the boundary.
Therefore, we can conclude that the saddle point (or at least one
saddle point) of Problem \eqref{MaxMin-Theta-Subproblem-Relaxed}
naturally satisfies $\boldsymbol{\Theta}^{\star}\in\mathcal{C}$,
which means there must exist a saddle point for Problem \eqref{MaxMin:Theta-Problem-Simplex}
that can be obtained by solving Problem \eqref{MaxMin-Theta-Subproblem-Relaxed}.
\end{IEEEproof}

\subsection{Proof for Proposition \ref{Prop: Matrix MM} \label{Appendix: Matrix MM Proof}}
\begin{IEEEproof}
This proof is intrinsically parallel to that of Proposition \ref{Prop: Scalar MM},
based on which the result is extended to the matrix domain. The concave
function $\log\det\left(\mathbf{Z}\right)$ with $\mathbf{Z}\succ\mathbf{0}$
can be majorized by its linear expansion around $\underline{\mathbf{Z}}$
as follows:
\[
\log\det(\mathbf{Z})\leq\log\det(\underline{\mathbf{Z}})+{\rm tr}\bigl(\underline{\mathbf{Z}}^{-1}(\mathbf{Z}-\underline{\mathbf{Z}})\bigr).
\]
By substituting $\mathbf{Z}$ with $\bigl(\mathbf{I}+\mathbf{X}^{\mathrm{H}}\mathbf{Y}^{-1}\mathbf{X}\bigr)^{-1}$,
we get
\begin{align*}
\log\det\bigl(\mathbf{I}+\mathbf{X}^{\mathrm{H}}\mathbf{Y}^{-1}\mathbf{X}\bigr) & \geq\log\det\bigl(\mathbf{I}+\underline{\mathbf{X}}^{\mathrm{H}}\underline{\mathbf{Y}}^{-1}\underline{\mathbf{X}}\bigr)-\mathrm{tr}\Bigl(\bigl(\mathbf{I}+\underline{\mathbf{X}}^{\mathrm{H}}\underline{\mathbf{Y}}^{-1}\underline{\mathbf{X}}\bigr)\bigl(\mathbf{I}+\mathbf{X}^{\mathrm{H}}\mathbf{Y}^{-1}\mathbf{X}\bigr)^{-1}-\mathbf{I}\Bigr),
\end{align*}
where the equality is attained at $\left(\mathbf{X},\mathbf{Y}\right)=\left(\underline{\mathbf{X}},\underline{\mathbf{Y}}\right)$.
The Woodbury matrix identity gives
\begin{align*}
\bigl(\mathbf{I}+\mathbf{X}^{\mathrm{H}}\mathbf{Y}^{-1}\mathbf{X}\bigr)^{-1} & =\mathbf{I}^{-1}-\mathbf{I}^{-1}\mathbf{X}^{\mathrm{H}}\bigl(\mathbf{Y}+\mathbf{XI}\mathbf{X}^{\mathrm{H}}\bigr)^{-1}\mathbf{XI}^{-1}\\
 & =\mathbf{I}-\mathbf{X}^{\mathrm{H}}\bigl(\mathbf{Y}+\mathbf{X}\mathbf{X}^{\mathrm{H}}\bigr)^{-1}\mathbf{X},
\end{align*}
and hence
\begin{align*}
\log\det\bigl(\mathbf{I}+\mathbf{X}^{\mathrm{H}}\mathbf{Y}^{-1}\mathbf{X}\bigr) & \geq\log\det\bigl(\mathbf{I}+\underline{\mathbf{X}}^{\mathrm{H}}\underline{\mathbf{Y}}^{-1}\underline{\mathbf{X}}\bigr)-\mathrm{tr}\Bigl(\bigl(\mathbf{I}+\underline{\mathbf{X}}^{\mathrm{H}}\underline{\mathbf{Y}}^{-1}\underline{\mathbf{X}}\bigr)\bigl(\mathbf{I}-\mathbf{X}^{\mathrm{H}}\bigl(\mathbf{Y}+\mathbf{X}\mathbf{X}^{\mathrm{H}}\bigr)^{-1}\mathbf{X}\bigr)-\mathbf{I}\Bigr)\\
 & =\log\det\bigl(\mathbf{I}+\underline{\mathbf{X}}^{\mathrm{H}}\underline{\mathbf{Y}}^{-1}\underline{\mathbf{X}}\bigr)-\mathrm{tr}\bigl(\underline{\mathbf{X}}^{\mathrm{H}}\underline{\mathbf{Y}}^{-1}\underline{\mathbf{X}}\bigr)+\mathrm{tr}\Bigl(\bigl(\mathbf{I}+\underline{\mathbf{X}}^{\mathrm{H}}\underline{\mathbf{Y}}^{-1}\underline{\mathbf{X}}\bigr)\mathbf{X}^{\mathrm{H}}\bigl(\mathbf{Y}+\mathbf{X}\mathbf{X}^{\mathrm{H}}\bigr)^{-1}\mathbf{X}\Bigr).
\end{align*}
Since $\mathbf{I}+\underline{\mathbf{X}}^{\mathrm{H}}\underline{\mathbf{Y}}^{-1}\underline{\mathbf{X}}\succ\mathbf{0}$,
it admits a unique positive semidefinite square root $(\mathbf{I}+\underline{\mathbf{X}}^{\mathrm{H}}\underline{\mathbf{Y}}^{-1}\underline{\mathbf{X}})^{\frac{1}{2}}$
\cite[Theorem 7.2.6]{horn2012matrix}. Let $\mathbf{Z}_{1}=\mathbf{X}(\mathbf{I}+\underline{\mathbf{X}}^{\mathrm{H}}\underline{\mathbf{Y}}^{-1}\underline{\mathbf{X}})^{\frac{1}{2}}$
and $\mathbf{Z}_{2}=\mathbf{Y}+\mathbf{X}\mathbf{X}^{\mathrm{H}}$,
we have
\[
(\mathbf{Z}_{2}^{-\frac{1}{2}}\mathbf{Z}_{1}-\mathbf{Z}_{2}^{\frac{1}{2}}\underline{\mathbf{Z}_{2}^{-1}}\underline{\mathbf{Z}_{1}})^{\mathrm{H}}(\mathbf{Z}_{2}^{-\frac{1}{2}}\mathbf{Z}_{1}-\mathbf{Z}_{2}^{\frac{1}{2}}\underline{\mathbf{Z}_{2}^{-1}}\underline{\mathbf{Z}_{1}})\succeq\mathbf{0},
\]
which means
\[
\mathbf{Z}_{1}^{\mathrm{H}}\mathbf{Z}_{2}^{-1}\mathbf{Z}_{1}\succeq\underline{\mathbf{Z}_{1}^{\mathrm{H}}}\underline{\mathbf{Z}_{2}^{-1}}\mathbf{Z}_{1}+\mathbf{Z}_{1}^{\mathrm{H}}\underline{\mathbf{Z}_{2}^{-1}}\underline{\mathbf{Z}_{1}}-\underline{\mathbf{Z}_{1}^{\mathrm{H}}}\underline{\mathbf{Z}_{2}^{-1}}\mathbf{Z}_{2}\underline{\mathbf{Z}_{2}^{-1}}\underline{\mathbf{Z}_{1}}.
\]
Then we can conclude that
\begin{align*}
\log\det\bigl(\mathbf{I}+\mathbf{X}^{\mathrm{H}}\mathbf{Y}^{-1}\mathbf{X}\bigr) & \geq\log\det\bigl(\mathbf{I}+\underline{\mathbf{X}}^{\mathrm{H}}\underline{\mathbf{Y}}^{-1}\underline{\mathbf{X}}\bigr)-\mathrm{tr}\bigl(\underline{\mathbf{X}}^{\mathrm{H}}\underline{\mathbf{Y}}^{-1}\underline{\mathbf{X}}\bigr)+2\mathrm{Re}\Bigl(\mathrm{tr}\bigl((\mathbf{I}+\underline{\mathbf{X}}^{\mathrm{H}}\underline{\mathbf{Y}}^{-1}\underline{\mathbf{X}})\underline{\mathbf{X}}^{\mathrm{H}}(\underline{\mathbf{Y}}+\underline{\mathbf{X}}\underline{\mathbf{X}}^{\mathrm{H}})^{-1}\mathbf{X}\bigr)\Bigr)\\
 & \ \ -\mathrm{tr}\Bigl((\mathbf{I}+\underline{\mathbf{X}}^{\mathrm{H}}\underline{\mathbf{Y}}^{-1}\underline{\mathbf{X}})\underline{\mathbf{X}}^{\mathrm{H}}(\underline{\mathbf{Y}}+\underline{\mathbf{X}}\underline{\mathbf{X}}^{\mathrm{H}})^{-1}(\mathbf{Y}+\mathbf{X}\mathbf{X}^{\mathrm{H}})(\underline{\mathbf{Y}}+\underline{\mathbf{X}}\underline{\mathbf{X}}^{\mathrm{H}})^{-1}\underline{\mathbf{X}}\Bigr),
\end{align*}
where the equality is attained when $\left(\mathbf{X},\mathbf{Y}\right)=\left(\underline{\mathbf{X}},\underline{\mathbf{Y}}\right)$,
and the proof is completed by rearranging the terms.
\end{IEEEproof}

\subsection{Proof for Theorem \ref{WSR: Theorem Convergence} \label{WSR:Proof-for-Convergence}}
\begin{IEEEproof}
In the following, we will give the convergence proof for Algorithm
\ref{WSR:Algorithm} which is designed for Problem \eqref{WSR:Formulation},
and the convergence properties for Algorithm \ref{MR:Algorithm} and
Algorithm \ref{SR:Algorithm} can be proved similarly. Besides, it
can be verified that the convergence properties also hold for other
cases that BMM has been applied to in Section \ref{sec:Extensions-and-Generalizations}. 

The convergence proof partly hinges on the proof for BSUM in \cite{razaviyayn2013unified}.
The sequence $\{\mathbf{W}^{(t)},\{\boldsymbol{\Theta}_{i}^{(t)}\}\}_{t\in\mathbb{N}}$
generated by Algorithm \ref{WSR:Algorithm} lies in a compact set
and, hence, it has a limit point $\{\mathbf{W}^{(\infty)},\{\boldsymbol{\Theta}_{i}^{(\infty)}\}\}$.
Then we can get
\begin{equation}
f_{\mathsf{WSR},\mathbf{W}}^{\prime}\bigl(\mathbf{W}^{(\infty)},\mathbf{W}^{(\infty)},\{\boldsymbol{\Theta}_{i}^{(\infty)}\}\bigr)\geq f_{\mathsf{WSR},\mathbf{W}}^{\prime}\bigl(\mathbf{W},\mathbf{W}^{(\infty)},\{\boldsymbol{\Theta}_{i}^{(\infty)}\}\bigr),\ \ \forall\,\mathbf{W}\in\mathcal{W},\label{eq:convergence w}
\end{equation}
and
\begin{equation}
f_{\mathsf{WSR},\boldsymbol{\Theta}_{l}}^{\prime\prime}\bigl(\boldsymbol{\Theta}_{l}^{(\infty)},\mathbf{W}^{(\infty)},\{\boldsymbol{\Theta}_{i}^{(\infty)}\}\bigr)\geq f_{\mathsf{WSR},\boldsymbol{\Theta}_{l}}^{\prime\prime}\bigl(\boldsymbol{\Theta}_{l},\mathbf{W}^{(\infty)},\{\boldsymbol{\Theta}_{i}^{(\infty)}\}\bigr),\ \ \forall\,\boldsymbol{\Theta}_{l}\in\mathcal{C}_{l},\ \ l=1,\ldots,L.\label{eq:convergence theta}
\end{equation}

We first define the real counterparts of $\mathbf{W}$ and $\mathbf{h}_{k}$
in \eqref{WSR: W-Block WSR} as follows:
\[
\tilde{\mathbf{W}}=\left[\tilde{\mathbf{w}}_{k},\ldots,\tilde{\mathbf{w}}_{K}\right]=\left[\begin{array}{c}
\mathrm{Re}\left(\mathbf{W}\right)\\
\mathrm{Im}\left(\mathbf{W}\right)
\end{array}\right],\ \ \ \ \tilde{\mathbf{h}}_{k}=\left[\begin{array}{cc}
\mathrm{Re}\left(\mathbf{h}_{k}\right) & \mathrm{Im}\left(\mathbf{h}_{k}\right)\\
\mathrm{Im}\left(\mathbf{h}_{k}\right) & -\mathrm{Re}\left(\mathbf{h}_{k}\right)
\end{array}\right],\ \ k=1,\ldots,K,
\]
and then the constraint set for $\tilde{\mathbf{W}}$ derived from
${\cal W}$ is
\[
\tilde{\mathcal{W}}=\left\{ \tilde{\mathbf{W}}\mid\bigl\Vert\tilde{\mathbf{W}}\bigr\Vert_{\mathrm{F}}^{2}\leq P\right\} .
\]
With the above definitions, the counterpart functions of $f_{\mathsf{WSR},\mathbf{W}}$
and $f_{\mathsf{WSR},\mathbf{W}}^{\prime}$ with real variables can
be constructed as follows:
\begin{align*}
 & g_{\tilde{\mathbf{W}}}\bigl(\tilde{\mathbf{W}}\bigr)=\sum_{k=1}^{K}\omega_{k}\log\bigl(1+\frac{\bigl\Vert\tilde{\mathbf{w}}_{k}^{\mathrm{T}}\tilde{\mathbf{h}}_{k}\bigr\Vert_{2}^{2}}{\sum_{j,j\neq k}^{K}\bigl\Vert\tilde{\mathbf{w}}_{j}^{\mathrm{T}}\tilde{\mathbf{h}}_{k}\bigr\Vert_{2}^{2}+\sigma^{2}}\bigr),\\
 & g_{\tilde{\mathbf{W}}}^{\prime}\bigl(\tilde{\mathbf{W}}\bigr)=\sum_{k=1}^{K}\omega_{k}\bigl(-\alpha_{k}\sum_{j=1}^{K}\bigl\Vert\tilde{\mathbf{w}}_{j}^{\mathrm{T}}\tilde{\mathbf{h}}_{k}\bigr\Vert_{2}^{2}+2\beta_{k}\bigl\Vert\tilde{\mathbf{w}}_{k}^{\mathrm{T}}\tilde{\mathbf{h}}_{k}\bigr\Vert_{2}\bigr)+\mathsf{const}_{w},
\end{align*}
where $\alpha_{k}$ and $\mathsf{const}_{w}$ are defined as in \eqref{eq:f_WSR_W^prime}
while $\beta_{k}=\frac{\underline{\mathsf{SINR}}}{||\underline{\tilde{\mathbf{w}}_{k}^{\mathrm{T}}}\tilde{\mathbf{h}}_{k}||_{2}}$.
Given $(\mathbf{W}^{(\infty)},\{\boldsymbol{\Theta}_{i}^{(\infty)}\})$,
it is straightforward that the Problem \eqref{WSR:Formulation} w.r.t.
$\mathbf{W}$ given by
\begin{equation}
\begin{aligned} & \underset{\mathbf{W}\in\mathcal{W}}{\mathrm{maximize}} &  & f_{\mathsf{WSR},\mathbf{W}}\bigl(\mathbf{W},\mathbf{W}^{(\infty)},\{\boldsymbol{\Theta}_{i}^{(\infty)}\}\bigr)\end{aligned}
\label{Appendix: W subproblem}
\end{equation}
is equivalent to the following problem with real variable $\tilde{\mathbf{W}}$:
\begin{equation}
\begin{aligned} & \underset{\tilde{\mathbf{W}}\in\tilde{\mathcal{W}}}{\mathrm{maximize}} &  & g_{\tilde{\mathbf{W}}}\bigl(\tilde{\mathbf{W}},\mathbf{W}^{(\infty)},\{\boldsymbol{\Theta}_{i}^{(\infty)}\}\bigr),\end{aligned}
\label{Appendix: W subproblem real counterpart}
\end{equation}
and the corresponding surrogate problem for $\mathbf{W}$ given by
\begin{equation}
\begin{aligned} & \underset{\mathbf{W}\in\mathcal{W}}{\mathrm{maximize}} &  & f_{\mathsf{WSR},\mathbf{W}}^{\prime}(\mathbf{W},\mathbf{W}^{(\infty)},\{\boldsymbol{\Theta}_{i}^{(\infty)}\})\end{aligned}
\label{Appendix: W surrogate problem}
\end{equation}
is equivalent to the following problem with real variable $\tilde{\mathbf{W}}$:
\begin{equation}
\begin{aligned} & \underset{\tilde{\mathbf{W}}\in\tilde{\mathcal{W}}}{\mathrm{maximize}} &  & g_{\tilde{\mathbf{W}}}^{\prime}\bigl(\tilde{\mathbf{W}},\mathbf{W}^{(\infty)},\{\boldsymbol{\Theta}_{i}^{(\infty)}\}\bigr).\end{aligned}
\label{Appendix: W surrogate problem real counterpart}
\end{equation}
From Eq. \eqref{eq:convergence w}, we know that $\mathbf{W}^{(\infty)}$
is a global maximizer of Problem \eqref{Appendix: W surrogate problem},
then $\tilde{\mathbf{W}}^{(\infty)}$ is a global maximizer of Problem
\eqref{Appendix: W surrogate problem real counterpart} as well due
to the equivalence between Problem \eqref{Appendix: W surrogate problem real counterpart}
and Problem \eqref{Appendix: W surrogate problem}, indicating that
\[
\nabla g_{\tilde{\mathbf{W}}}^{\prime}(\tilde{\mathbf{W}},\mathbf{W}^{(\infty)},\{\boldsymbol{\Theta}_{i}^{(\infty)}\};\mathbf{d}_{w})\mid_{\tilde{\mathbf{W}}=\tilde{\mathbf{W}}^{(\infty)}}\leq0,\ \ \ \forall\mathbf{d}_{w}\ \text{ s.t. }\ \tilde{\mathbf{W}}^{(\infty)}+\mathbf{d}_{w}\in\tilde{\mathcal{W}}.
\]
It can be verified that $\nabla g_{\tilde{\mathbf{W}}}^{\prime}\bigl(\tilde{\mathbf{W}},\mathbf{W}^{(\infty)},\{\boldsymbol{\Theta}_{i}^{(\infty)}\}\bigr)=\nabla g_{\tilde{\mathbf{W}}}\bigl(\tilde{\mathbf{W}},\mathbf{W}^{(\infty)},\{\boldsymbol{\Theta}_{i}^{(\infty)}\}\bigr)$,
in that $g_{\tilde{\mathbf{W}}}^{\prime}\bigl(\tilde{\mathbf{W}}\bigr)$
is a minorizing function of $g_{\tilde{\mathbf{W}}}\bigl(\tilde{\mathbf{W}}\bigr)$
according to Proposition \ref{Prop: Scalar MM}. Therefore, we have
\[
\nabla g_{\tilde{\mathbf{W}}}(\tilde{\mathbf{W}},\mathbf{W}^{(\infty)},\{\boldsymbol{\Theta}_{i}^{(\infty)}\};\mathbf{d}_{w})\mid_{\tilde{\mathbf{W}}=\tilde{\mathbf{W}}^{(\infty)}}\leq0,\ \ \ \forall\mathbf{d}_{w}\ \text{ s.t. }\ \tilde{\mathbf{W}}^{(\infty)}+\mathbf{d}_{w}\in\tilde{\mathcal{W}},
\]
which means $\tilde{\mathbf{W}}^{(\infty)}$ is a stationary point
of Problem \eqref{Appendix: W subproblem real counterpart}. Then
$\mathbf{W}^{(\infty)}$ is also a stationary point of Problem \eqref{Appendix: W subproblem}
because of the equivalence between Problem \eqref{Appendix: W subproblem}
and Problem \eqref{Appendix: W subproblem real counterpart}.

We further define the real counterparts of $\tilde{\boldsymbol{\theta}}_{l}$,
$\underline{\mathbf{w}_{j}^{\mathrm{H}}}\mathbf{h}_{k}^{\mathsf{d}}$,
$\underline{\mathbf{w}_{j}^{\mathrm{H}}}\mathbf{F}_{k,l}$ in \eqref{WSR: Theta-Block WSR},
and $\mathbf{b}_{l}$ in \eqref{eq:f_WSR_theta_l^primeprime} as follows:
\[
\tilde{\boldsymbol{\theta}}_{l}\negthinspace=\negthinspace\left[\negthinspace\begin{array}{c}
\mathrm{Re}\left(\boldsymbol{\theta}_{l}\right)\\
\mathrm{Im}\left(\boldsymbol{\theta}_{l}\right)
\end{array}\negthinspace\right],\ \ \ \tilde{\mathbf{h}}_{j}^{\mathsf{d}}\negthinspace=\negthinspace\left[\negthinspace\begin{array}{c}
\mathrm{Re}\bigl(\underline{\mathbf{w}_{j}^{\mathrm{H}}}\mathbf{h}_{k}^{\mathsf{d}}\bigr)\\
\mathrm{Im}\bigl(\underline{\mathbf{w}_{j}^{\mathrm{H}}}\mathbf{h}_{k}^{\mathsf{d}}\bigr)
\end{array}\negthinspace\right],\ \ \ \tilde{\mathbf{f}}_{j}\negthinspace=\negthinspace\left[\negthinspace\begin{array}{cc}
\mathrm{Re}\bigl(\underline{\mathbf{w}_{j}^{\mathrm{H}}}\mathbf{F}_{k,l}\bigr) & \mathrm{Im}\bigl(\underline{\mathbf{w}_{j}^{\mathrm{H}}}\mathbf{F}_{k,l}\bigr)\\
-\mathrm{Im}\bigl(\underline{\mathbf{w}_{j}^{\mathrm{H}}}\mathbf{F}_{k,l}\bigr) & \mathrm{Re}\bigl(\underline{\mathbf{w}_{j}^{\mathrm{H}}}\mathbf{F}_{k,l}\bigr)
\end{array}\negthinspace\right],\ \ \ \tilde{\mathbf{b}}_{l}\negthinspace=\negthinspace\left[\negthinspace\begin{array}{c}
\mathrm{Re}\bigl(\mathbf{b}_{l}\bigr)\\
\mathrm{Im}\bigl(\mathbf{b}_{l}\bigr)
\end{array}\negthinspace\right],\ \ \ j\negthinspace=\negthinspace1,\ldots,K,
\]
and then the constraint set for $\tilde{\boldsymbol{\theta}}_{l}$
is
\[
\tilde{\mathcal{C}}_{l}=\left\{ \tilde{\boldsymbol{\theta}}_{l}\mid\tilde{\boldsymbol{\theta}}_{l}\in\mathbb{R}^{2N_{l}},\ [\boldsymbol{\theta}_{l}]_{j}^{2}+[\boldsymbol{\theta}_{l}]_{j+N_{l}}^{2}=1,\ j=1,\ldots,N_{i}\right\} .
\]
With the above definitions, the counterpart functions of $f_{\mathsf{WSR},\boldsymbol{\Theta}_{l}}$
and $f_{\mathsf{WSR},\boldsymbol{\Theta}_{l}}^{\prime\prime}$ with
real variable can be constructed as follows:
\begin{align*}
 & g_{\tilde{\boldsymbol{\theta}}_{l}}\bigl(\tilde{\boldsymbol{\theta}}_{l}\bigr)=\sum_{k=1}^{K}\omega_{k}\log\bigl(1+\frac{\bigl\Vert\tilde{\mathbf{f}}_{k}\tilde{\boldsymbol{\theta}}_{l}+\tilde{\mathbf{h}}_{k}^{\mathsf{d}}\bigr\Vert_{2}^{2}}{\sum_{j,j\neq k}^{K}\bigl\Vert\tilde{\mathbf{f}}_{j}\tilde{\boldsymbol{\theta}}_{l}+\tilde{\mathbf{h}}_{j}^{\mathsf{d}}\bigr\Vert_{2}^{2}+\sigma^{2}}\bigr)\\
 & g_{\tilde{\boldsymbol{\theta}}_{l}}^{\prime}\bigl(\tilde{\boldsymbol{\theta}}_{l}\bigr)=-2\tilde{\boldsymbol{\theta}}_{l}^{\mathrm{T}}\tilde{\mathbf{b}}_{l}+\mathsf{const}_{\theta,l}^{\prime},
\end{align*}
where $\mathsf{const}_{\theta,l}^{\prime}=\mathsf{const}_{\theta,l}-N_{l}\lambda_{l}-\underline{\boldsymbol{\theta}_{l}^{\mathrm{H}}}\bigl(\lambda_{l}\mathbf{I}-\mathbf{L}_{l}\bigr)\underline{\boldsymbol{\theta}_{l}}$
represents the constant parts in \eqref{eq:f_WSR_theta_l^primeprime}.
Given $(\mathbf{W}^{(\infty)},\{\boldsymbol{\Theta}_{i}^{(\infty)}\})$,
it is straightforward that the Problem \eqref{WSR:Formulation} w.r.t.
$\boldsymbol{\Theta}_{l}$ given by
\begin{equation}
\begin{aligned} & \underset{\boldsymbol{\Theta}_{l}\in\mathcal{C}_{l}}{\mathrm{maximize}} &  & f_{\mathsf{WSR},\boldsymbol{\Theta}_{l}}\bigl(\boldsymbol{\Theta}_{l},\mathbf{W}^{(\infty)},\{\boldsymbol{\Theta}_{i}^{(\infty)}\}\bigr)\end{aligned}
\label{Appendix: Theta subproblem}
\end{equation}
is equivalent to the following problem with real variable $\tilde{\boldsymbol{\theta}}_{l}$
\begin{equation}
\begin{aligned} & \underset{\tilde{\boldsymbol{\theta}}_{l}\in\tilde{\mathcal{C}}_{l}}{\mathrm{maximize}} &  & g_{\tilde{\boldsymbol{\theta}}_{l}}\bigl(\tilde{\boldsymbol{\theta}}_{l},\mathbf{W}^{(\infty)},\{\boldsymbol{\Theta}_{i}^{(\infty)}\}\bigr),\end{aligned}
\label{Appendix: Theta subproblem real counterpart}
\end{equation}
and the corresponding surrogate problem for $\boldsymbol{\Theta}_{l}$
given by
\begin{equation}
\begin{aligned} & \underset{\boldsymbol{\Theta}_{l}\in\mathcal{C}_{l}}{\mathrm{maximize}} &  & f_{\mathsf{WSR},\boldsymbol{\Theta}_{l}}^{\prime\prime}(\boldsymbol{\Theta}_{l},\mathbf{W}^{(\infty)},\{\boldsymbol{\Theta}_{i}^{(\infty)}\})\end{aligned}
\label{Appendix: Theta surrogate problem}
\end{equation}
is equivalent to the following problem with real variable $\tilde{\boldsymbol{\theta}}_{l}$
\begin{equation}
\begin{aligned} & \underset{\tilde{\boldsymbol{\theta}}_{l}\in\tilde{\mathcal{C}}_{l}}{\mathrm{maximize}} &  & g_{\tilde{\boldsymbol{\theta}}_{l}}^{\prime}(\tilde{\boldsymbol{\theta}}_{l},\mathbf{W}^{(\infty)},\{\boldsymbol{\Theta}_{i}^{(\infty)}\}).\end{aligned}
\label{Appendix: Theta surrogate problem real counterpart}
\end{equation}
From Eq. \eqref{eq:convergence theta}, we know that $\boldsymbol{\Theta}_{l}^{(\infty)}$
is a global maximizer of Problem \eqref{Appendix: Theta surrogate problem},
then $\tilde{\boldsymbol{\theta}}^{(\infty)}$ is a global maximizer
of Problem \eqref{Appendix: Theta surrogate problem real counterpart}
as well due to the equivalence between Problem \eqref{Appendix: Theta surrogate problem real counterpart}
and Problem \eqref{Appendix: Theta surrogate problem}, indicating
that
\[
\nabla g_{\tilde{\boldsymbol{\theta}}_{l}}^{\prime}(\tilde{\boldsymbol{\theta}}_{l},\mathbf{W}^{(\infty)},\{\boldsymbol{\Theta}_{i}^{(\infty)}\};\mathbf{d}_{\theta,l})\mid_{\tilde{\boldsymbol{\theta}}_{l}=\tilde{\boldsymbol{\theta}}_{l}^{(\infty)}}\leq0,\ \ \ \forall\mathbf{d}_{\theta,l}\in\mathcal{T}_{\tilde{\mathcal{C}}_{l}}(\tilde{\boldsymbol{\theta}}_{l}^{(\infty)}).
\]
It can be verified that $\nabla g_{\tilde{\boldsymbol{\theta}}_{l}}^{\prime}(\tilde{\boldsymbol{\theta}}_{l},\mathbf{W}^{(\infty)},\{\boldsymbol{\Theta}_{i}^{(\infty)}\})=\nabla g_{\tilde{\boldsymbol{\theta}}_{l}}(\tilde{\boldsymbol{\theta}}_{l},\mathbf{W}^{(\infty)},\{\boldsymbol{\Theta}_{i}^{(\infty)}\})$,
in that $g_{\tilde{\boldsymbol{\theta}}_{l}}^{\prime}\bigl(\tilde{\boldsymbol{\theta}}_{l}\bigr)$
is a minorizing function of $g_{\tilde{\boldsymbol{\theta}}_{l}}\bigl(\tilde{\boldsymbol{\theta}}_{l}\bigr)$
according to Proposition \ref{Prop: Scalar MM}, Lemma \ref{lem:quadratic_MM},
and Lemma \ref{lem:Matrix Bound}. Therefore, we have
\[
\nabla g_{\tilde{\boldsymbol{\theta}}_{l}}(\tilde{\boldsymbol{\theta}}_{l},\mathbf{W}^{(\infty)},\{\boldsymbol{\Theta}_{i}^{(\infty)}\};\mathbf{d}_{\theta,l})\mid_{\tilde{\boldsymbol{\theta}}_{l}=\tilde{\boldsymbol{\theta}}_{l}^{(\infty)}}\leq0,\ \ \ \forall\mathbf{d}_{\theta,l}\in\mathcal{T}_{\tilde{\mathcal{C}}_{l}}(\tilde{\boldsymbol{\theta}}_{l}^{(\infty)}),
\]
which means $\tilde{\boldsymbol{\theta}}_{l}^{(\infty)}$ is a stationary
point of Problem \eqref{Appendix: Theta subproblem real counterpart}.
Then $\boldsymbol{\Theta}_{l}^{(\infty)}$ is also a stationary point
of Problem \eqref{Appendix: Theta subproblem} because of the equivalence
between Problem \eqref{Appendix: Theta subproblem} and Problem \eqref{Appendix: Theta subproblem real counterpart}.

Similarly, by repeating the above argument for the other $\{\boldsymbol{\Theta}_{i}\}$
blocks, we can conclude that $(\mathbf{W}^{(\infty)},\{\boldsymbol{\Theta}_{i}^{(\infty)}\})$
is a coordinate-wise maximum of Problem \eqref{WSR:Formulation}.
Therefore, $(\mathbf{W}^{(\infty)},\{\boldsymbol{\Theta}_{i}^{(\infty)}\})$
is a stationary point of Problem \eqref{WSR:Formulation} as well
since the objective function of Problem \eqref{WSR:Formulation} is
regular at the limit point. 

Considering that the power limit constraint set $\mathcal{W}$ is
convex, Eq. \eqref{eq:convergence w} implies that $\mathbf{W}^{(\infty)}$
satisfies
\begin{equation}
\begin{aligned}0\leq\gamma\:\bot\:\|\mathbf{W}^{(\infty)}\|_{F}^{2}-P\leq0,\\
\nabla f_{\mathsf{WSR},\mathbf{W}}^{\prime}\bigl(\mathbf{W},\{\boldsymbol{\Theta}_{i}^{(\infty)}\}\bigr)\mid_{\mathbf{W}=\mathbf{W}^{(\infty)}}+\gamma\bigl(\|\mathbf{W}^{(\infty)}\|_{F}^{2}-P\bigr)=0,
\end{aligned}
\label{eq:convergence w KKT}
\end{equation}
where $\gamma$ is the Lagrange multiplier.
\begin{lem}
\label{lem:LICQ} Liner independence constraint qualification (LICQ)
\cite{peterson1973review} holds everywhere on set 
\[
\mathcal{C}_{i}=\bigl\{\boldsymbol{\Theta}_{i}=\mathrm{diag}\left(\boldsymbol{\theta}_{i}\right)\mid\boldsymbol{\theta}_{i}\in\mathbb{C}^{N_{i}},\bigl|[\boldsymbol{\theta}_{i}]_{j}\bigr|=1,\forall j=1,\ldots,N\bigr\},\ \ \ \forall i=1,\ldots,L
\]
and set
\[
\mathcal{D}_{i}=\bigl\{\boldsymbol{\Theta}_{i}=\mathrm{diag}\left(\boldsymbol{\theta}_{i}\right)\mid\boldsymbol{\theta}_{i}\in\mathbb{C}^{N_{i}},\bigl|[\boldsymbol{\theta}_{i}]_{j}\bigr|=1,\mathrm{ang}([\boldsymbol{\theta}_{i}]_{j})\in\Phi_{i},\forall j=1,\ldots,N_{i}\bigr\},\ \ \ \forall i=1,\ldots,L.
\]
\end{lem}
\begin{IEEEproof}
The constraint $\boldsymbol{\Theta}_{i}\in\mathcal{C}_{i}$ can be
rewritten as
\[
[\mathbf{c}_{i}(\boldsymbol{\theta}_{i})]_{j}=\left|[\boldsymbol{\theta}_{i}]_{j}\right|-1=0,\ j=1,\ldots,N_{i},
\]
while the constraint $\boldsymbol{\Theta}_{i}\in\mathcal{D}_{i}$
can be rewritten as
\[
[\mathbf{d}_{i}(\boldsymbol{\theta}_{i})]_{j}=\prod_{\phi\in\Phi_{i}}\bigl([\boldsymbol{\theta}_{i}]_{j}-e^{j\phi}\bigr)=0,\ j=1,\ldots,N_{i}.
\]
The gradients $\nabla[\mathbf{c}_{i}(\boldsymbol{\theta}_{i})]_{1},\ldots,\nabla[\mathbf{c}_{i}(\boldsymbol{\theta}_{i})]_{N_{i}}$
meet the LICQ evidently since $[\boldsymbol{\theta}_{i}]_{j}$ appears
only at the $j$-th entry of $\nabla[\mathbf{c}_{i}(\boldsymbol{\theta}_{i})]_{j}$,\footnote{The Wirtinger derivative is adopted for differentials of complex variables
\cite{hjorungnes2011complex}.} while the same goes for $\nabla[\mathbf{d}_{i}(\boldsymbol{\theta}_{i})]_{1},\ldots,\nabla[\mathbf{d}_{i}(\boldsymbol{\theta}_{i})]_{N_{i}}$,
through which the proof is completed.
\end{IEEEproof}
The constraint set $\mathcal{C}_{l}$ meet the LICQ according to Lemma
\ref{lem:LICQ}, then Eq. \eqref{eq:convergence theta} implies that
$\boldsymbol{\Theta}_{l}$ satisfies

\begin{equation}
\begin{aligned}\left[\boldsymbol{\Theta}_{l}\right]_{ij}=0,\ \ \left[\boldsymbol{\Theta}_{l}\right]_{ii}=1,\ \ \forall i\neq j,\ i,j=1,\ldots,N_{l},\\
\nabla f_{\mathsf{WSR},\boldsymbol{\Theta}_{l}}^{\prime\prime}(\boldsymbol{\Theta}_{l},\mathbf{W}^{(\infty)},\{\boldsymbol{\Theta}_{i}^{(\infty)}\})\mid_{\boldsymbol{\Theta}_{l}=\boldsymbol{\Theta}_{l}^{(\infty)}}+\sum_{i=1}^{N_{l}}\sum_{j=1,j\neq i}^{N_{l}}[\boldsymbol{\Gamma}]_{ij}\left[\boldsymbol{\Theta}_{l}\right]_{ij}+\sum_{i=1}^{N_{l}}[\boldsymbol{\Gamma}]_{ii}\left(\left[\boldsymbol{\Theta}_{l}\right]_{ii}-1\right)=0,
\end{aligned}
\label{eq:convergence Theat KKT}
\end{equation}
where $\boldsymbol{\Gamma}$ is the Lagrange multiplier. Similarly,
by repeating the above argument for the other $\{\boldsymbol{\Theta}_{i}\}$
blocks and putting them together with \eqref{eq:convergence w KKT}
and \eqref{eq:convergence Theat KKT}, we can conclude that $(\mathbf{W}^{(\infty)},\{\boldsymbol{\Theta}_{i}^{(\infty)}\})$
is a KKT point of Problem \eqref{WSR:Formulation}\textcolor{red}{}.
\end{IEEEproof}

\end{document}